%% file: MLD.tex
\documentclass[multphys,vecphys]{svmult}

\usepackage{makeidx}     % allows index generation
\usepackage{graphicx}    % standard LaTeX graphics tool
\usepackage{multicol}    % used for the two-column index
\makeindex             % used for the subject index
\def\half{\ensuremath{{\textstyle\frac{1}{2}}}}

\begin{document}

\title*{Hamiltonian monodromy as lattice defect}
\author{B. Zhilinskii}
 \authorrunning{Monodromy as lattice defect}
\institute{Universit\'e du Littoral, UMR du CNRS 8101, 59140 Dunkerque, France
\texttt{zhilin@univ-littoral.fr}}
\maketitle
        \begin{abstract}
The analogy between monodromy in dynamical (Hamiltonian) systems
and defect in crystal lattices is used in order to formulate some
general conjectures about possible types of qualitative features
of quantum systems which can be interpreted as a manifestation
of classical monodromy in quantum finite particle (molecular) problems.	
        \end{abstract}

\section{Introduction} \label{S:intro}
 The purpose of this paper is to demonstrate amazing similarity between
apparently different subjects: defects of regular periodic lattices,
monodromy of classical Hamiltonian integrable dynamical systems, and
qualitative features of joint quantum spectra of several commuting
observables for quantum finite-particle systems. First of all we
recall  why regular lattices and lattices with defects appear naturally
for classical integrable Hamiltonian systems and for their quantum analogs.
Then we describe several ``elementary dynamical'' defects using
tools and language developed in the theory of crystal defects. 
Comparison between defects arising in dynamical systems and crystal
defects leads to many interesting questions about possibility of
realization of certain defects in Hamiltonian dynamics and in crystals. 
 
\section{Integrable classical singular fibrations and monodromy} 
\label{S:classMon}
Let us start with the example of Liouville integrable classical 
Hamiltonian system with $N$ degrees of freedom 
\cite{Arnold}. This means that there exists a set
$F=\{F_1, \ldots, F_n\}$ of functions defined on $2n$-dimensional symplectic 
manifold $M$, which are functionally independent and mutually in involution.
The Hamiltonian $H$ can be locally represented as a function 
$H=f(F_1, \ldots, F_n)$. The mapping $F:M\rightarrow R^n$ defines the
integrable fibration. We call it a generalized energy-momentum map.
Each fiber is the union of connected component of inverse images
$F^{-1}(f)$ of points $f\in R^n$. If the differentials $\{dF_1,\ldots,dF_n\}$ of
functions from $F$ are linearly independent in each point the fibration is 
called regular. If moreover all fibers are compact, the fibration is toric.
We will be interested in integrable toric fibrations with singularities
of some simplest type.

\begin{figure}
\centering
\mbox{
\includegraphics[height=2.5cm]{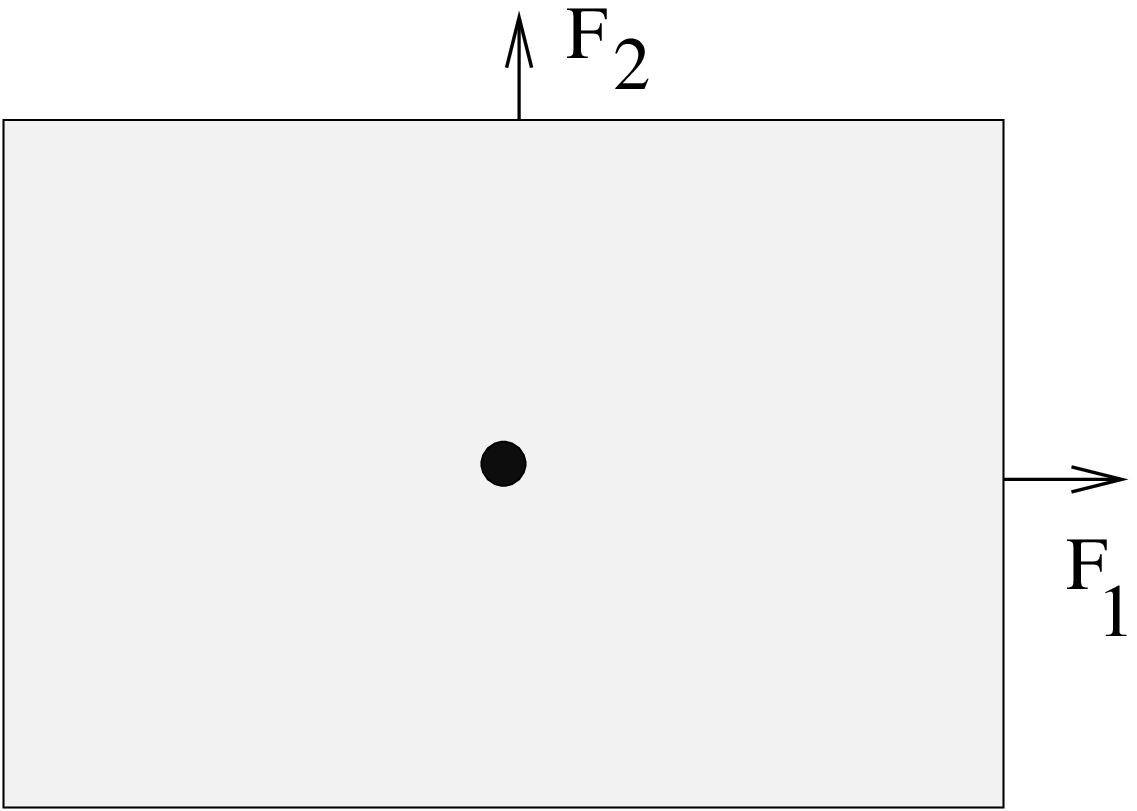} \hskip0.5cm
\includegraphics[height=2.5cm]{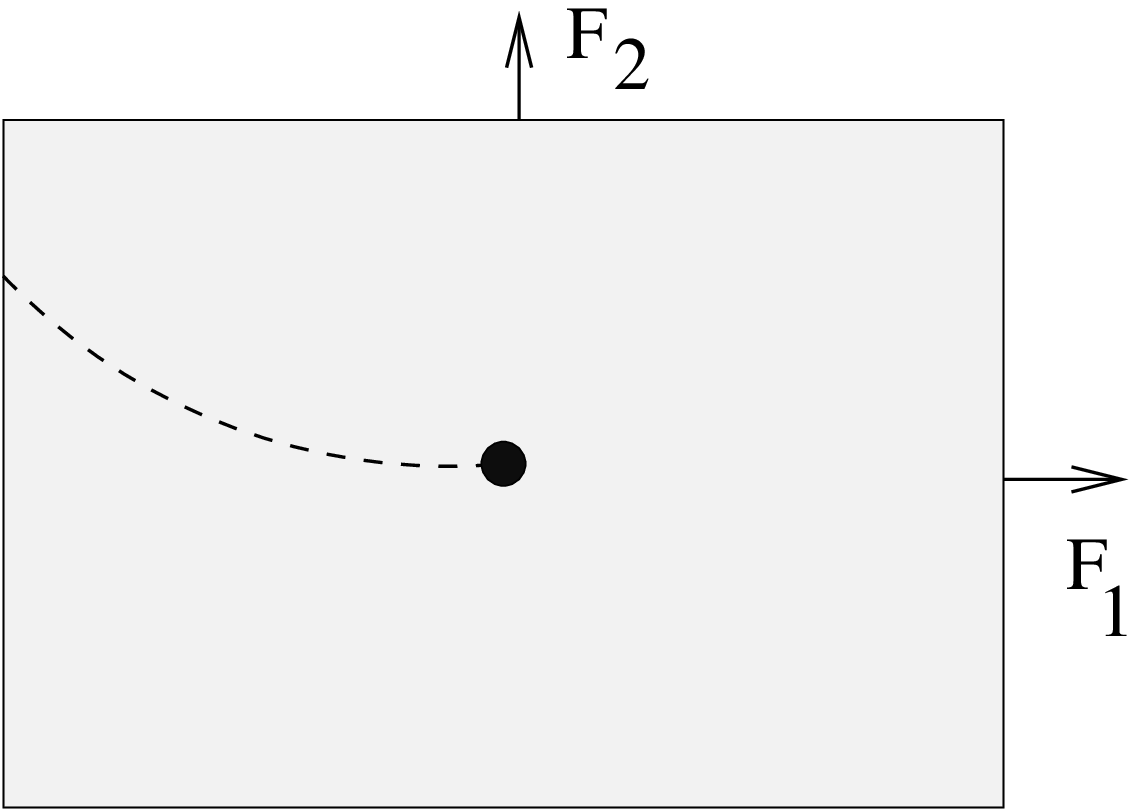} \hskip0.5cm
\includegraphics[height=2.5cm]{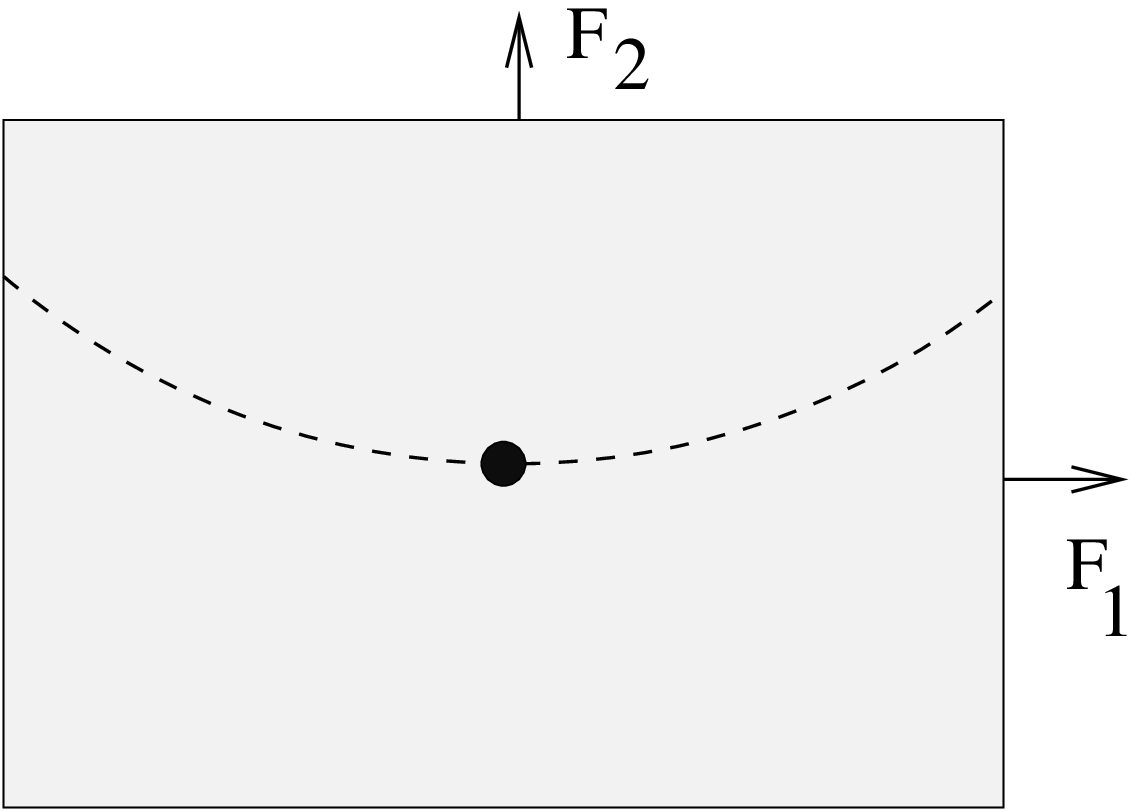} 
}
\caption{Examples of images of the energy momentum maps for singular toric
fibrations.}
\label{F:EMmaps}      
\end{figure}

\begin{figure}
\centering
\mbox{
\includegraphics[height=2.5cm]{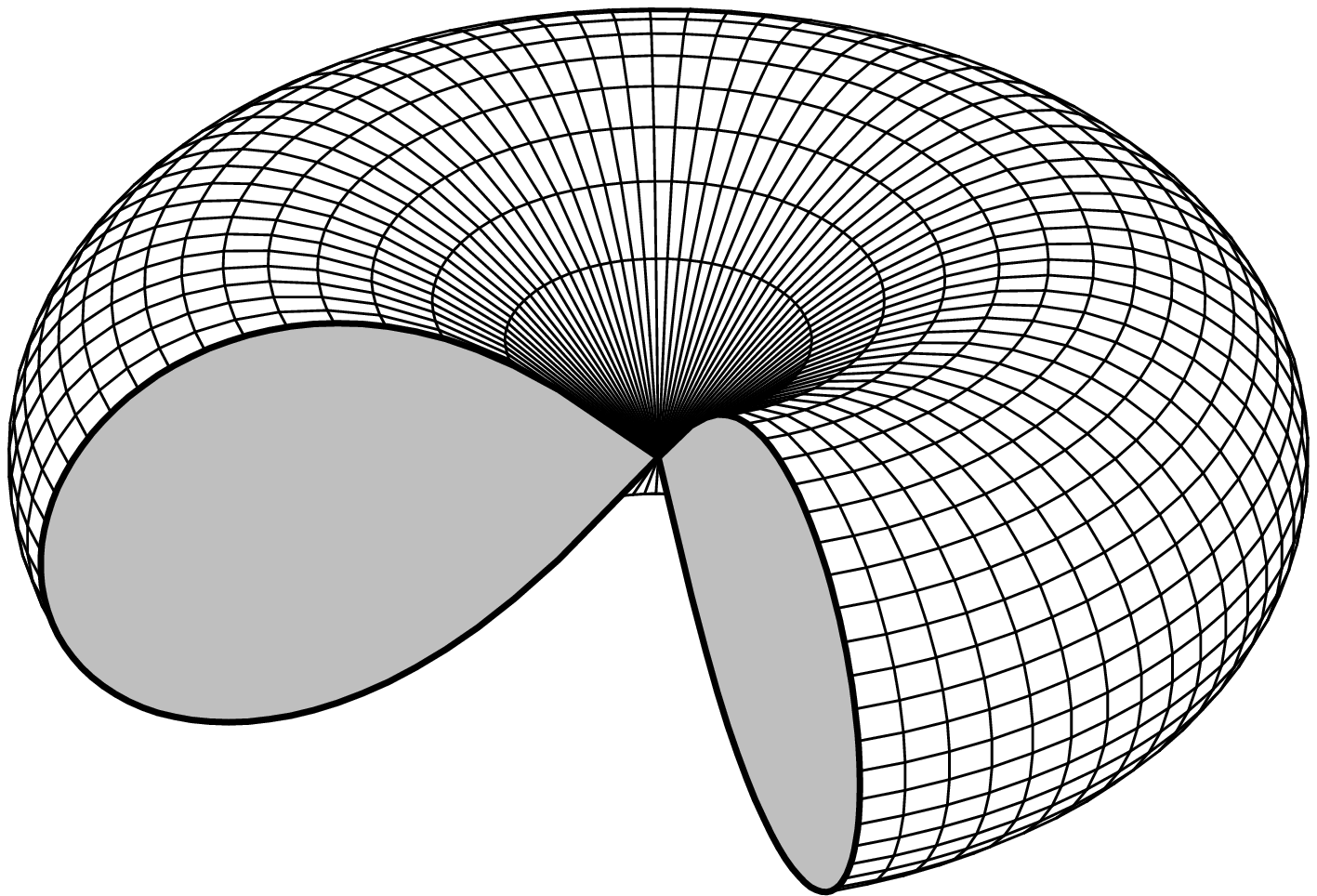} \hskip0.5cm
\includegraphics[height=2.5cm]{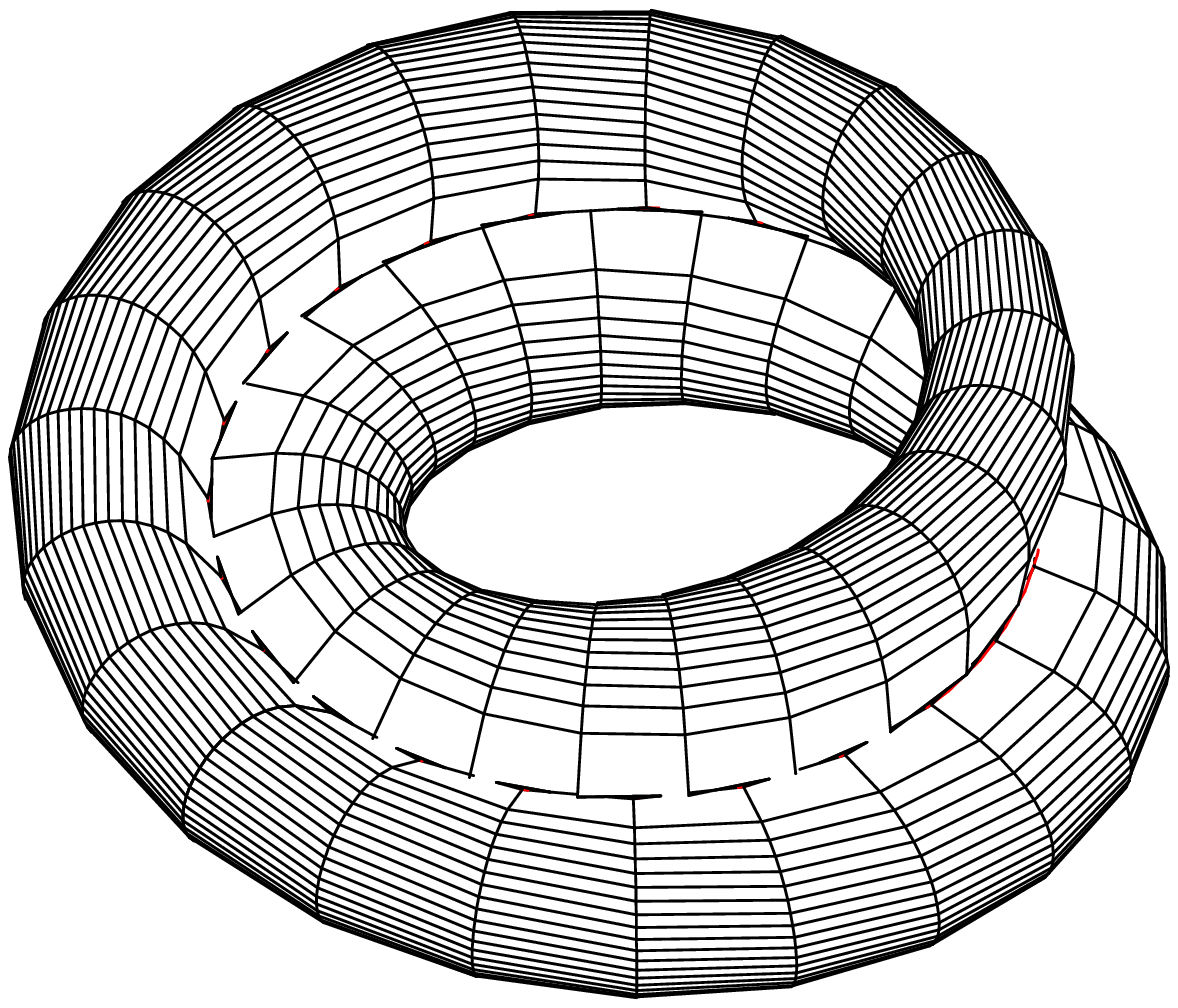} \hskip0.5cm
\includegraphics[height=2.5cm]{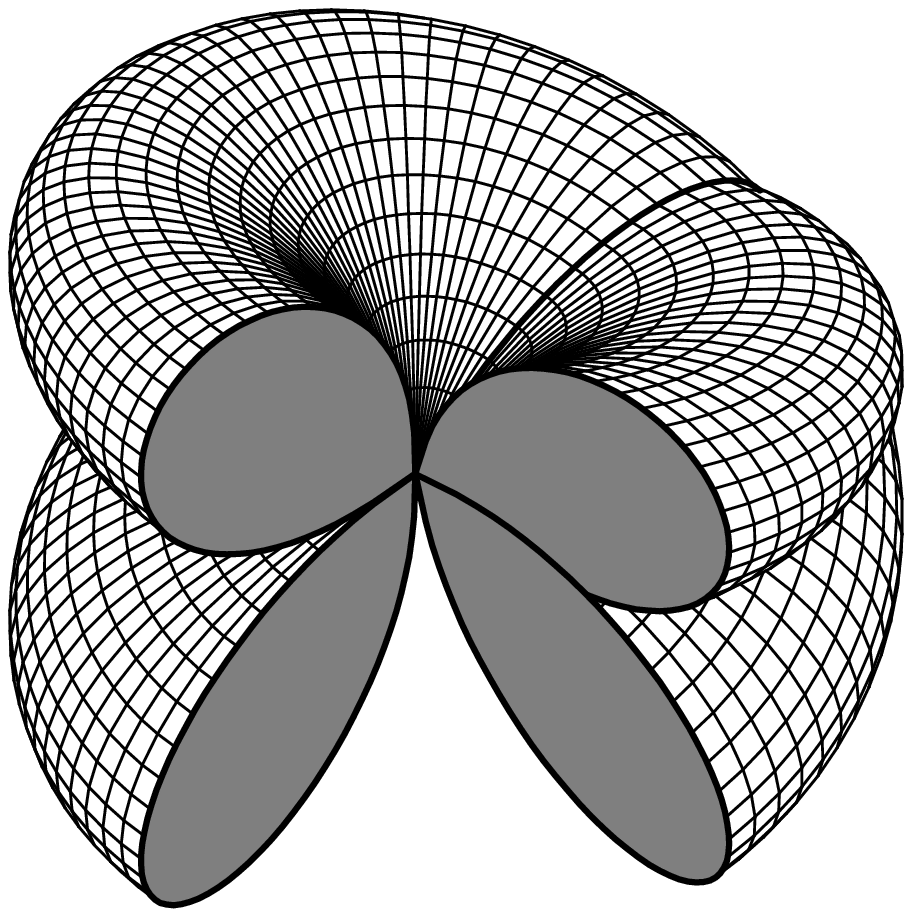} 
}
\caption{Singular fibers.  Pinched torus (left).  Curled torus (center).
Pinched curled torus (right).}
\label{F:SingF}      
\end{figure}

Let us restrict ourselves to systems with two degrees of freedom.
Typical examples of images of singular energy-momentum maps are shown
in Figure \ref{F:EMmaps}. The isolated critical value of the map $F$ (see Figure 
\ref{F:EMmaps}, left), also known as focus-focus singularity \cite{Lerman,zung97}, 
appears, for example, for such problems as spherical pendulum 
\cite{cushman83,Duist80,cushman-duistermaat88}, 
champagne bottle \cite{Bates,cushman-bates}, 
coupling of two angular momenta \cite{SadZhPhysLett}, etc. 
The singular fiber in this case is a
pinched torus (Figure \ref{F:SingF}, left) with one isolated critical 
point of rank 0. 

Presence of a half-line of critical values, together with end point, is typical
for nonlinear $1:(-k)$ resonant oscillator \cite{NekhSadZhil}. Each point on the 
singular half-line corresponds to singular ``curled torus'' (Figure \ref{F:SingF},
center shows curled torus for the case $k=2$) \cite{NekhSadZhil,ColinSan}, 
which differs from an ordinary torus due to presence of one
circular trajectory which covers itself $k$-times. This particular 
circular trajectory is formed by critical points of rank 1 of the map $F$.
The end point  (see Figure \ref{F:EMmaps}, center)  
corresponds to pinched curled torus with multiple circle shrinking to 
a point. This fiber has one critical point of rank 0 and 
is topologically equivalent to pinched torus but its
immersion into $4D$-space is different. A pinched curled torus for $k=2$ is
shown in Figure \ref{F:SingF}, right.

More general situation with two singular rays starting at one singular point
(as shown in Figure \ref{F:SingF}, right) corresponds to $k:(-l)$ 
resonant nonlinear
oscillator. Example of the integrable fibration corresponding to all 
shown in Figure  \ref{F:EMmaps} images of the energy-momentum maps with
two integrals $(F_1,F_2)$ in involution  can be
written as \cite{NekhSadZhil}
\begin{eqnarray}					\label{I1-def}
  F_1& = & m_1 \half
%{\ensuremath{{\textstyle\frac{m1}{2}}}}
 (p_1^2+q_1^2)  - m_2  \half(p_2^2+q_2^2), \\
  F_2& = & {\rm Im}\bigl[ (q_1+ip_1)^{m_2}(q_2+ip_2)^{m_1} \bigr]
            +  \bigl(m_1\half (p_1^2+q_1^2) + m_2\half  (p_2^2+q_2^2) \bigr)^s,
%  \quad {\rm with }\   s > m_1+m_2)/2 .		
	\label{I2-def}
\end{eqnarray}
with  $s > (m_1+m_2)/2 $, and $m_1,m_2$ positive integers.

All regular fibers are two-dimensional tori. Their fundamental groups are
abelian groups $Z^2$ with two generators, corresponding to two basic cycles on a 
torus. The fundamental groups for different regular tori are 
isomorphic among themselves and to $Z^2$ integer lattice. We can establish 
the correspondence between basic cycles defined on different tori by choosing
a continuous path in the $4D$-space which is transversal to fibers and by
deforming basic cycles continuously along this path. 
In particular, for a closed path passing only through  regular
tori we get the automorphism of the
fundamental group of a chosen regular torus.  The corresponding map of
basic cycles is the monodromy map. It is the same for all homotopy 
equivalent closed paths. If the path crosses singular lines similar
to those taking place for integrable fibration of the 
(\ref{I1-def},\ref{I2-def}) resonance oscillators only subgroup of
chains can be continuously deformed along the path and the monodromy
map in such a case can be defined only for a subgroup of fundamental group
\cite{NekhSadZhilinpress}.
Nevertheless this map can be linearly extended to a whole group.  In this
case the extended monodromy map is represented by a matrix with fractional
entries; while in the case of isolated critical values the monodromy map
is given by integer matrix $\mu\in SL(2,Z)$. 

%%%%%%%%%%%%%%%%%%%%%%%%%%%%%%%%%%%%%%%%%%%%%%%%%%%%%%%%%%%%%%%%%%%%%%%
\section{Quantum monodromy} \label{S:quantM}
In order to study the manifestation of classical monodromy in associated
quantum problems we first need to recall the existence of
local action-angle variables  \cite{Arnold,NNaav} and to replace the transformation of 
angles by corresponding transformation of actions. In the case
of locally regular integrable fibrations local action-angle variables
$\{I_k,\phi_k\}$ exist and the linear transformation of angles
$\phi^\prime = M\phi$, imposed by the monodromy map, corresponds to the
transformation of actions $I^\prime=\left(M^{-1}\right)^\dag I$.

\begin{figure}
\centering
\includegraphics[height=2.8cm]{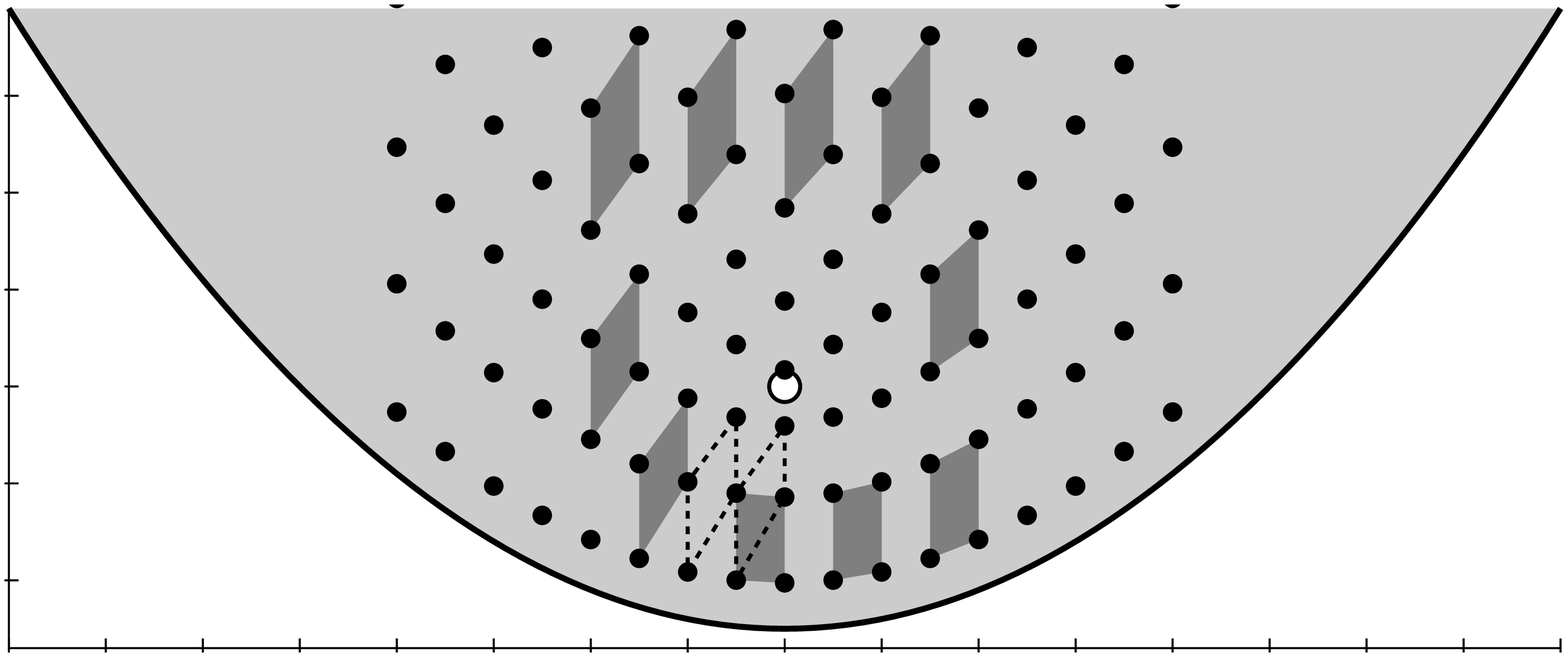} \hskip0.5cm
\includegraphics[height=2.8cm]{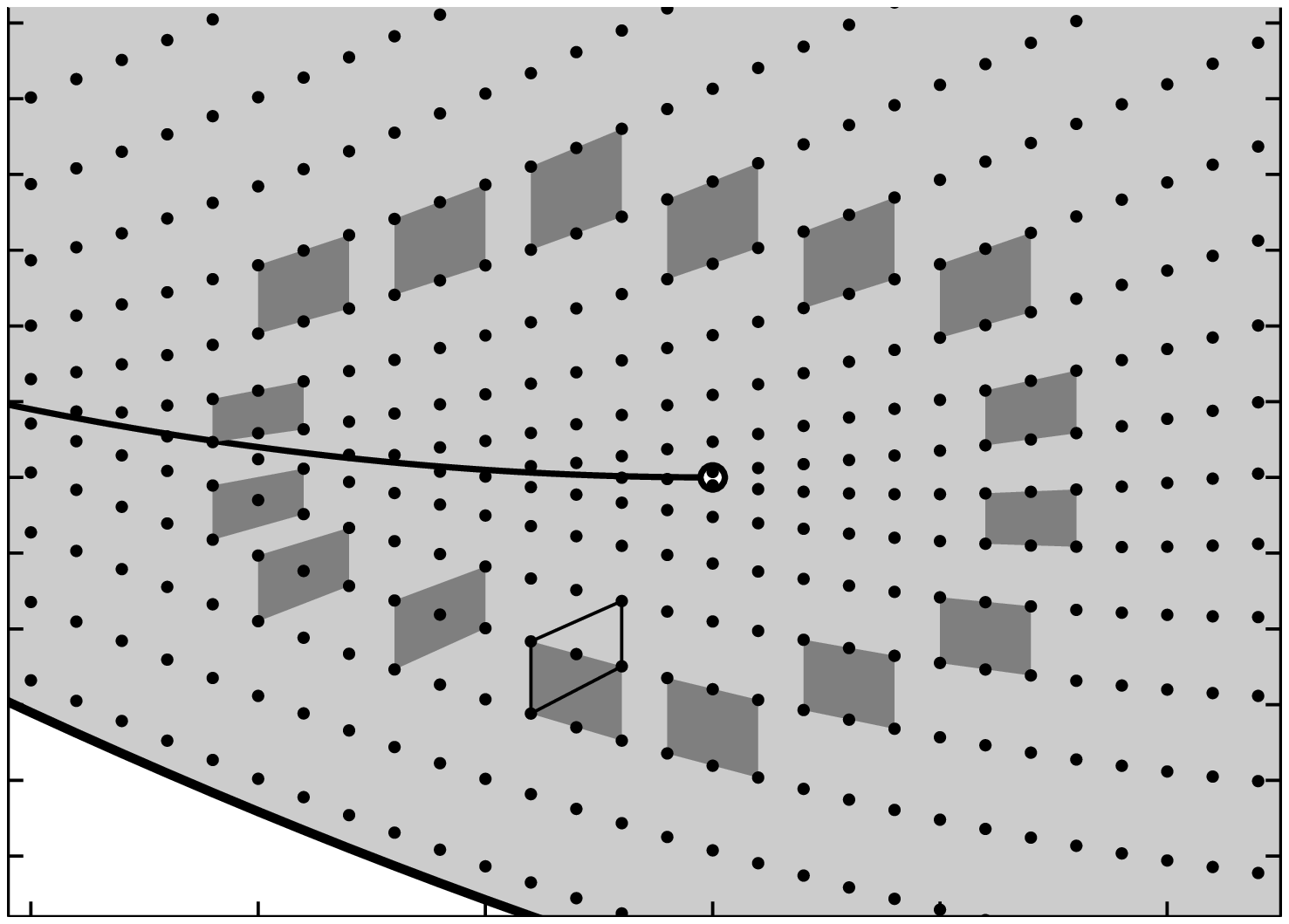} 
\caption{Example of the lattice of quantum states with monodromy.
Resonant oscillators (\ref{I1-def},\ref{I2-def}) with $m_1=m_2=1$ (left)
and $m_1=1$, $m_2=2$ (right).}
\label{F:Mon}      
\end{figure}

For quantum problems we are interested in the joint spectrum of commuting
operators, corresponding to classical integrals $\{F_1,F_2\}$
\cite{grondin-sadovskii-zhilinskii,Vu1,waalkens-dullin,child-weston-tennyson}. 
The collection  of joint
eigenvalues superimposed on the image of the energy-momentum map for
classical problem reveals locally the presence of a regular lattice
associated with integrality conditions imposed on local actions by
quantum mechanics. The lattice of quantum states for quantum problem
corresponding to classical oscillators with $1:(-1)$ and $1:(-2)$
resonances is represented in Fig. \ref{F:Mon} \cite{NekhSadZhil}.   

Due to existence of monodromy, lattice of quantum states can not be
regular globally. From figure  \ref{F:Mon}  it is clearly seen that
the transport of elementary cell of the locally regular part of the lattice
around the singularity shows nontrivial monodromy for a non-contractible
close path in the base space (in the space of $F_1,F_2$ values). 
The presence of quantum monodromy can be interpreted as a presence of 
defects of locally regular lattice of quantum states
\cite{SadZhPhysLett}. In the case of
isolated critical values of classical problem (Figure  \ref{F:Mon}, left)
the choice of elementary cell is arbitrary and the monodromy map is integer.
In the case of the presence of singular line at the image of the classical
energy momentum map, the dimension of the cell should be increased 
(doubled in the case of $1:(-2)$ resonance) in order
to ensure the unambiguous crossing of the singular line \cite{NekhSadZhil}.
In both cases the presence of singular fibers in classical problem
is reflected in  the appearance of 
some specific defects of the lattice of quantum states for
corresponding quantum problem. We want now to describe these specific
defects arising in the quantum theory of Hamiltonian systems using
methods and tools from defect theory of periodic lattices 
\cite{Kleman,Mermin,Michel,Kroner}.

%%%%%%%%%%%%%%%%%%%%%%%%%%%%%%%%%%%%%%%%%%%%%%%%%%%%%%%%%%%%%%%%%%%
\section{Elementary defects of lattices} \label{S:DefLat}
Let us play  with analogy between 2-D lattice of quantum numbers
(or of action variables in the classical limit) and the 2-D lattice
of regular solid with defects. More precisely the idea is  to see the correspondence
between defects of periodic solids and monodromy which is an
obstruction to the existence of global action-angle variables in Hamiltonian
dynamics (for integrable systems).

For 2-D  system each  quantum state
(or a site for a lattice formed by points) is characterized by two numbers, say
$(n_1,n_2)$. The existence of local order means that starting      
with some vertex (point of the lattice)
 one can  form two vectors, or
equivalently the elementary cell of the lattice by defining two vectors
as joining  respectively $(n_1,n_2)$ with $(n_1+1,n_2)$ and with
$(n_1,n_2+1)$. This corresponds to the choice of the elementary cell
with four vertices $\{(n_1,n_2), (n_1+1,n_2), (n_1,n_2+1), (n_1,n_2+1)\}$.
The choice of the elementary cell (or equivalently the choice of the
basis of the lattice) is not unique. It is defined only up to arbitrary
transformation with matrix $M\in SL(2,Z)$. But let us fix
some choice for a moment. The existence of local actions
in quantum-state lattice
language means that by elementary translations in two directions we can
label unambiguously all vertices by two numbers with difference in
numbers along each edge being 1 for one number and 0 for another.
This means that there is no defects (in the local region studied).

 Let us now analyze several different types of
defects which can be imagined  for periodic lattices in order to
find possible candidates to represent defects of lattices of
quantum numbers for quantum problems corresponding to classical
Hamiltonian systems with non-trivial (integer and fractional) monodromy.

\subsection{Vacations and linear dislocations} \label{sS:VacDisl}
\begin{figure}
\centering
\includegraphics[height=2.5cm]{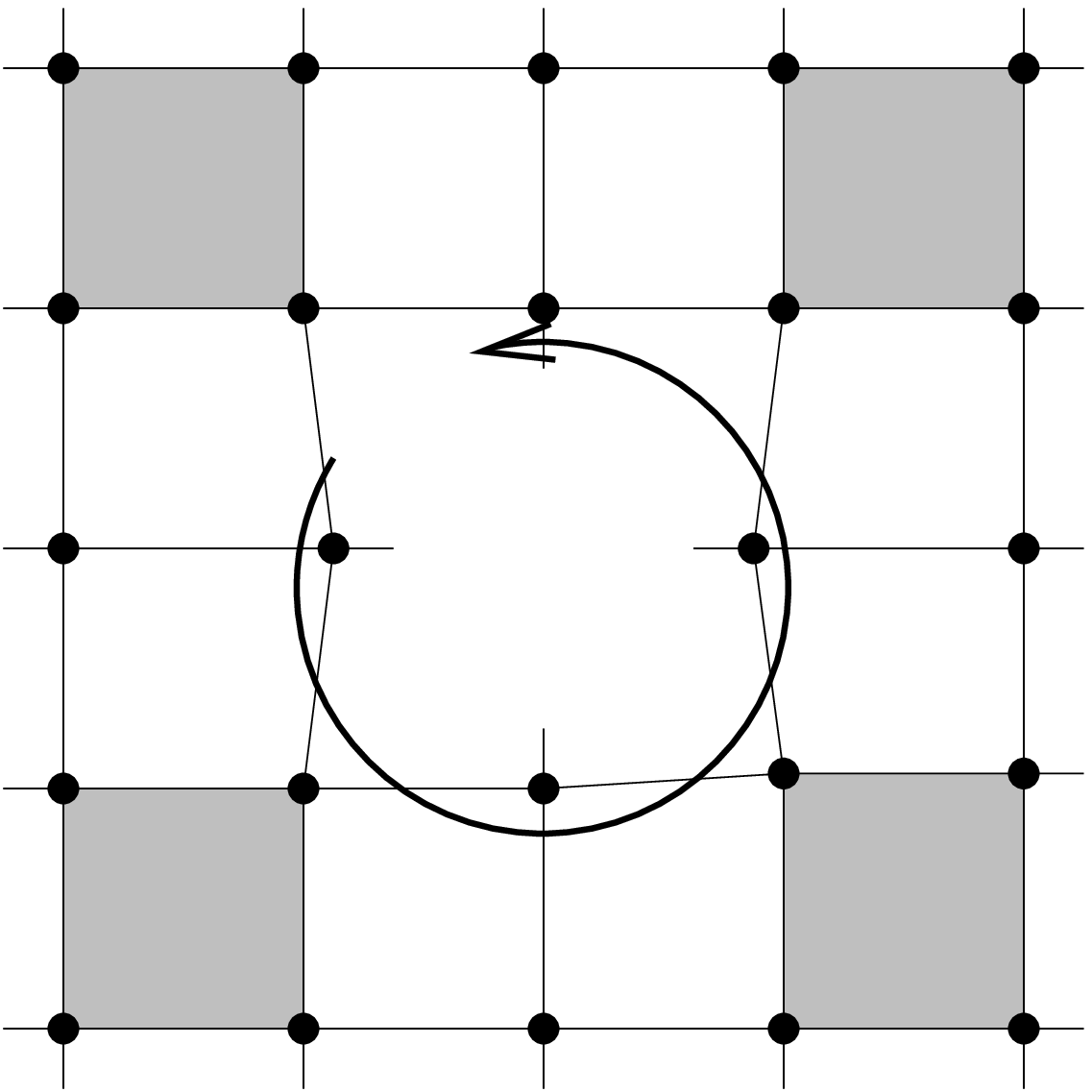} \hskip1.0cm
\includegraphics[height=2.5cm]{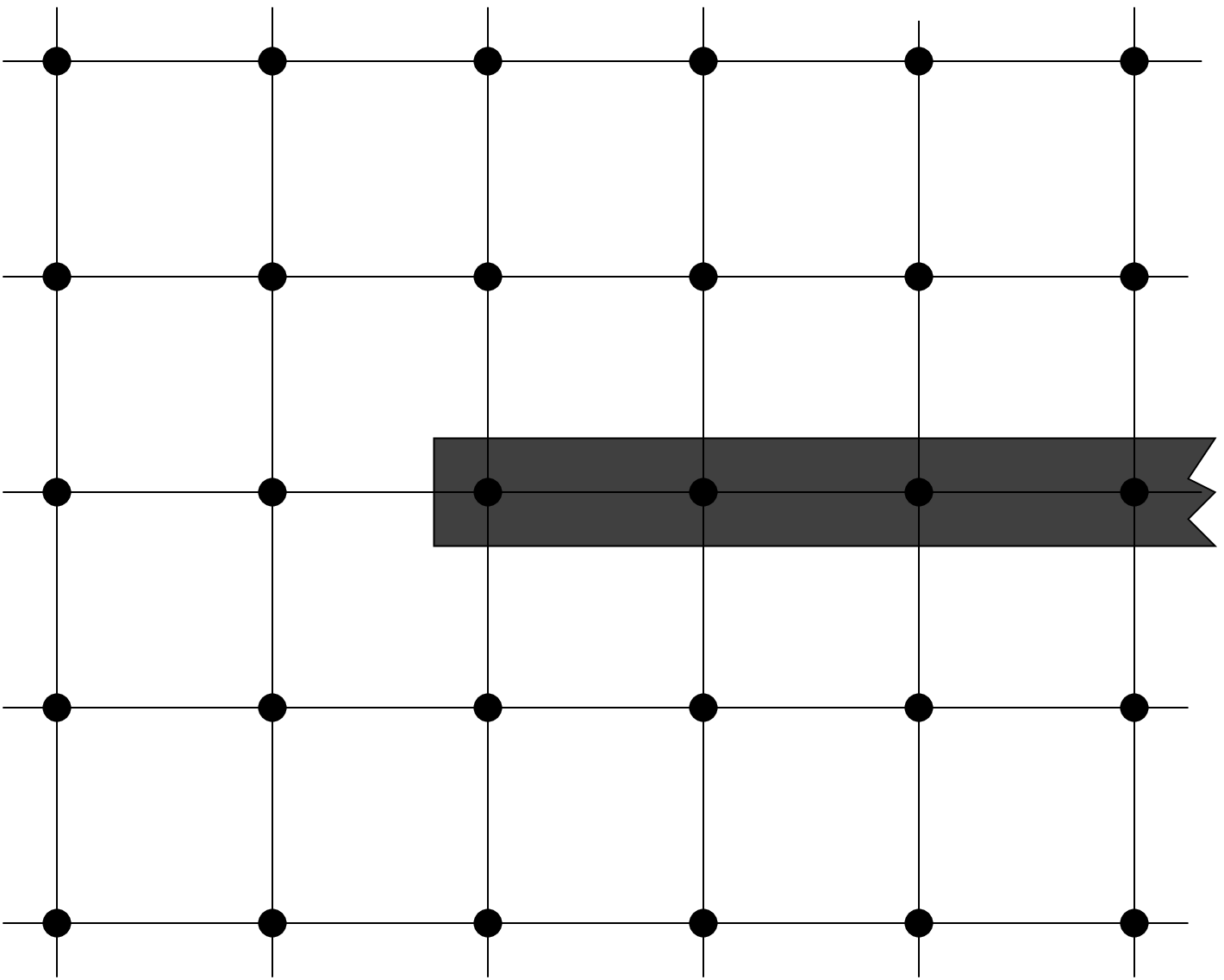} \hskip0.6cm
\includegraphics[height=2.5cm]{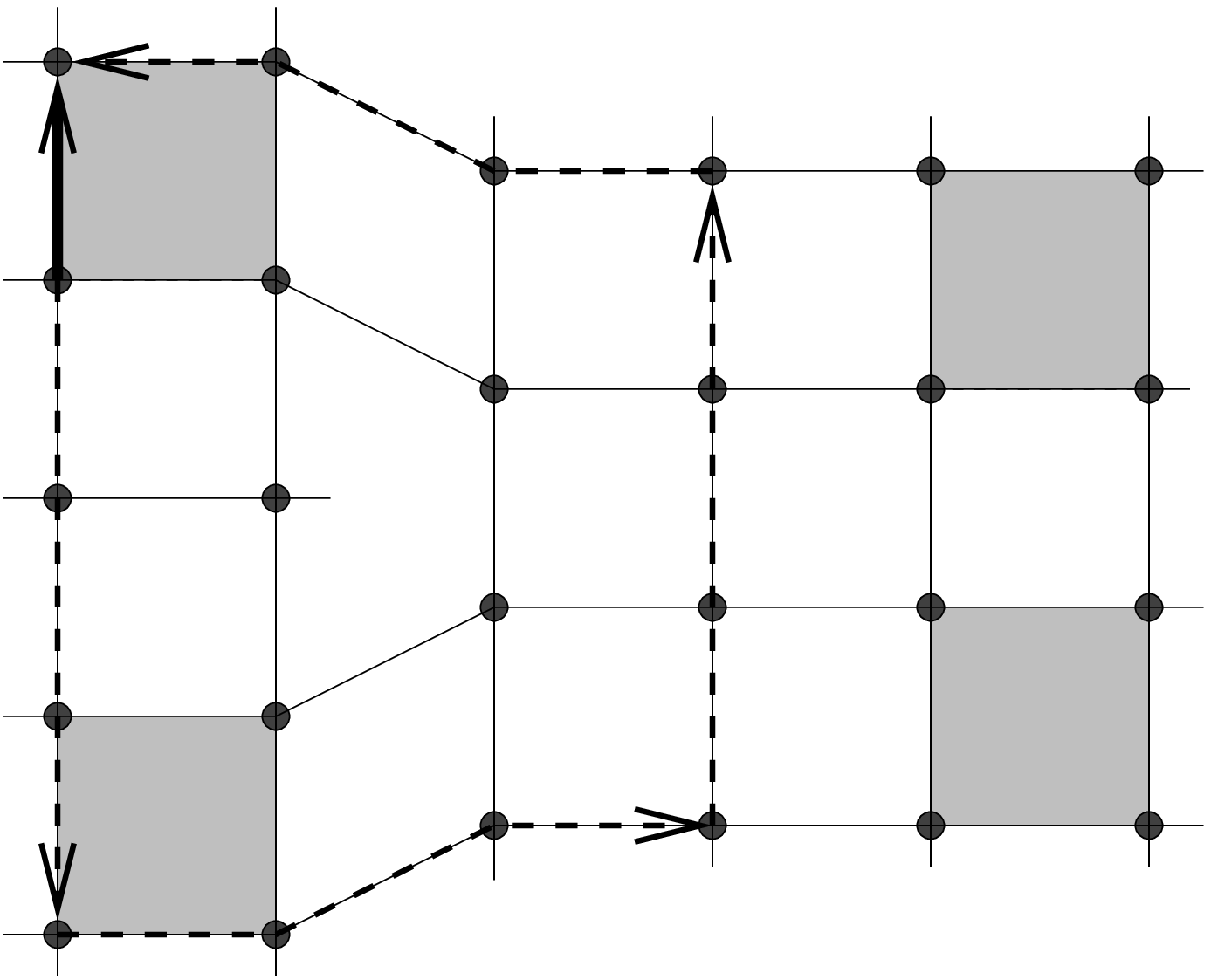}
\caption{ Lattice with vacation (left).  Construction of linear dislocation
(centre). Lattice with linear dislocation
(right). }
\label{F:VacLDisl}      
\end{figure}

The simplest point defect well known in solids is 
the absence of vertex (or the presence of additional vertex). This defect
does not  distort the system of edges not connected with the vacation.
The lattice is not deformed even slightly away from the point defect. The
elementary cell after a circular trip around the vacation has no modifications.
See Fig.  \ref{F:VacLDisl}, left. 

Linear dislocation  can be easily imagined to be formed through the 
following formal procedure. Let us remove all  vertices on the half-line
started at  a given vertex and join the vertices though the gap (see Fig.
\ref{F:VacLDisl}, center).
Equally, after making a cut along a line of vertices we can introduce
additional (one or even several) half-lines.
Now the circular path around this defect will show us the existence of the
defect. To observe this defect 
we should go around it by doing the same number of steps
in four directions (say, down, right, up, and left). If the final point will
not be the same as the initial point, there is a defect. The vector from
initial point to the final point (Burgers vector in solid state physics)
characterizes the dislocation. Observe that the elementary cell after the
round trip around the dislocation will return exactly to its initial place
(see Fig. \ref{F:VacLDisl}, right) because  
Burgers vector does not depend on the initial
point  and it is exactly the same
for all four vertices of the cell.   This means that  vacation and
linear dislocations can not be associated with monodromy type defects 
of regular lattices.

%%%%%%%%%%%%%%%%%%%%%%%%%%%%%%%%%%%%%%%%%%%%%%%%%%%%%%%%%%%%%%%%%%%%
\subsection{Angular dislocations as elementary monodromy defect}
Another general idea to form defect starting from the
regular lattice is to remove or to introduce ``the solid angle'' 
and to establish
in some way the regular correspondence between two boundaries
everywhere except one central point.

\label{sS:ElMdef}
\begin{figure}
\centering
\includegraphics[height=2.5cm]{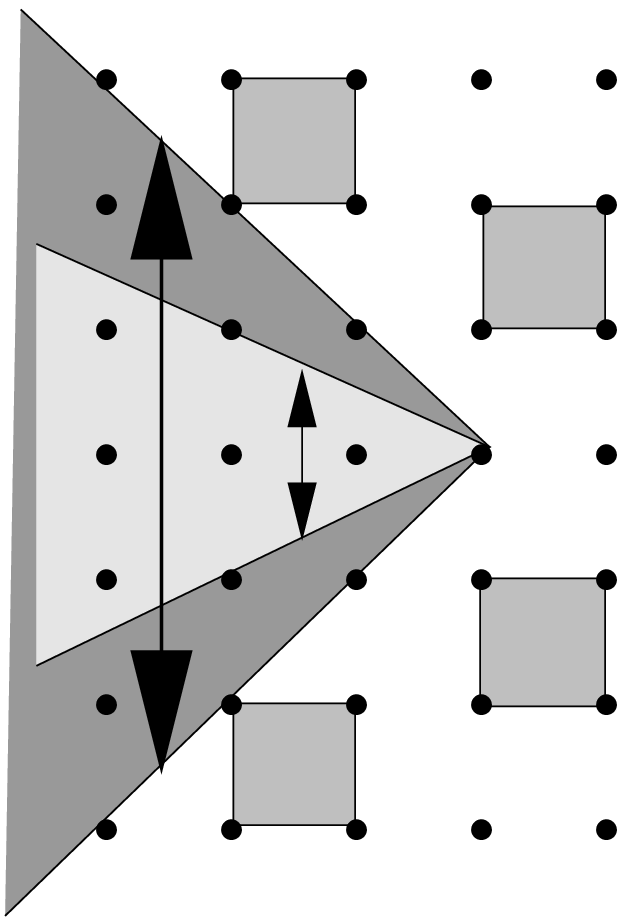} \hskip0.4cm
\includegraphics[height=2.5cm]{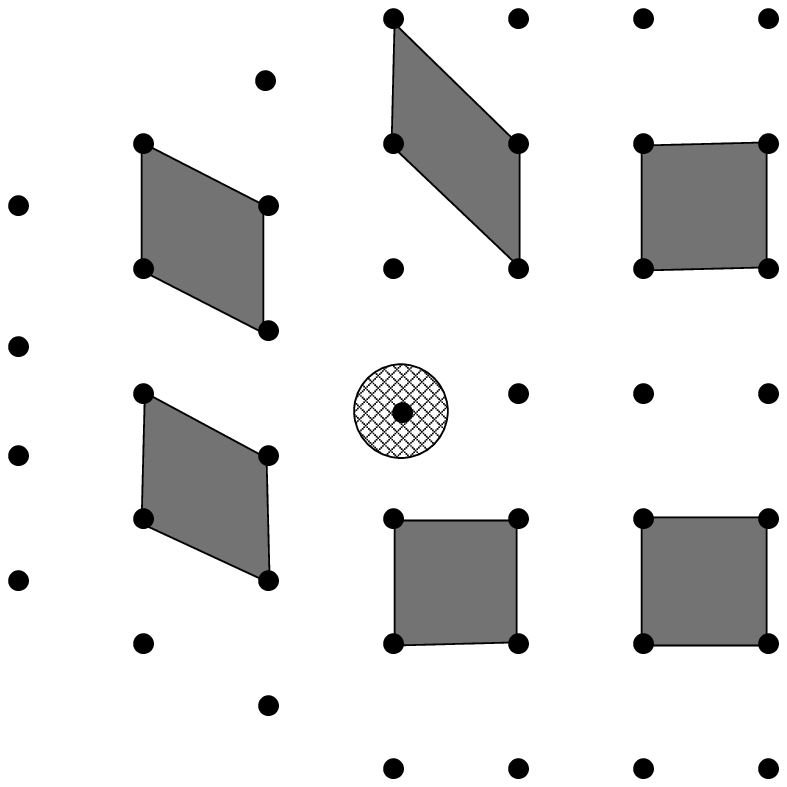}  \hskip0.2cm
\includegraphics[height=2.8cm]{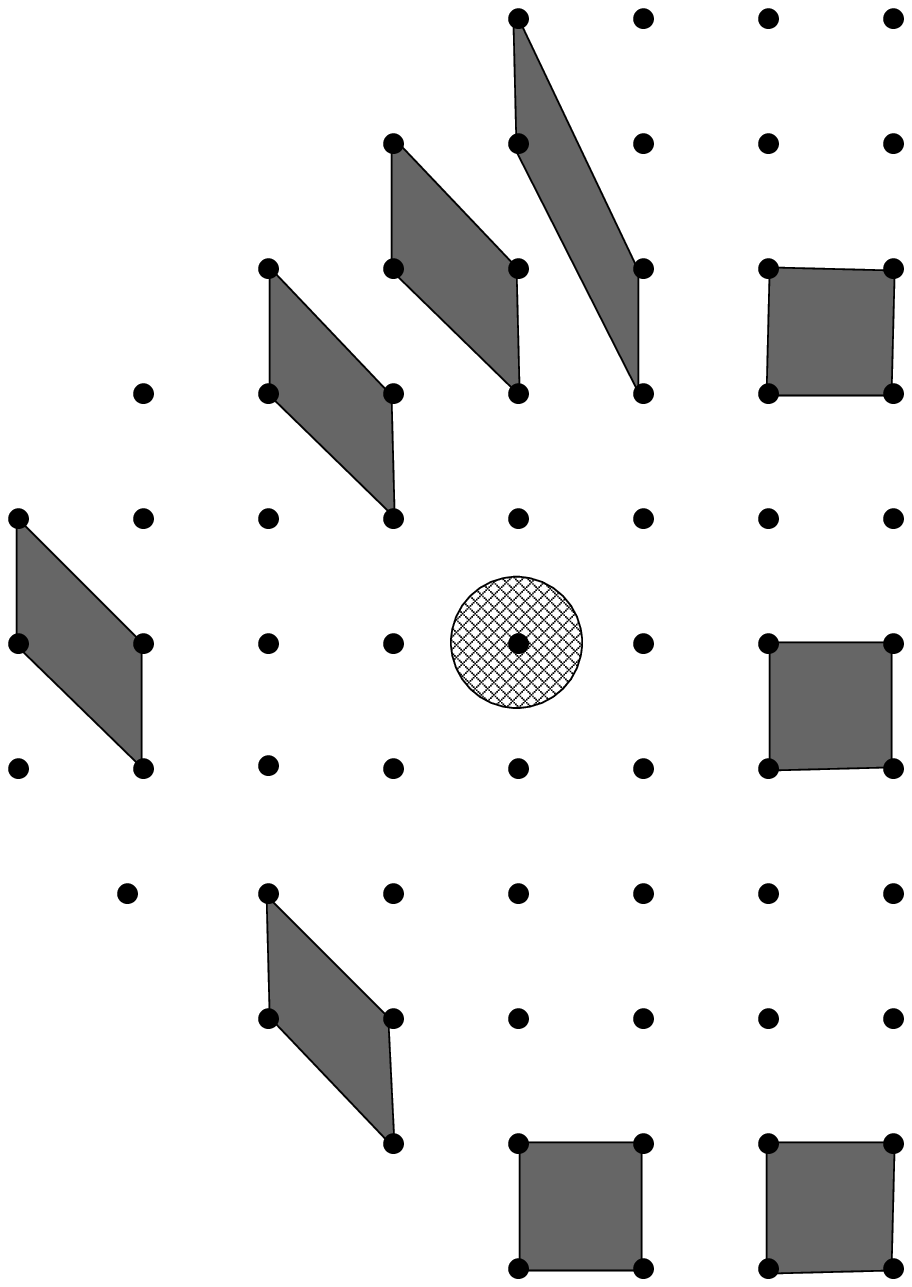}  \hskip0.2cm
\includegraphics[height=2.8cm]{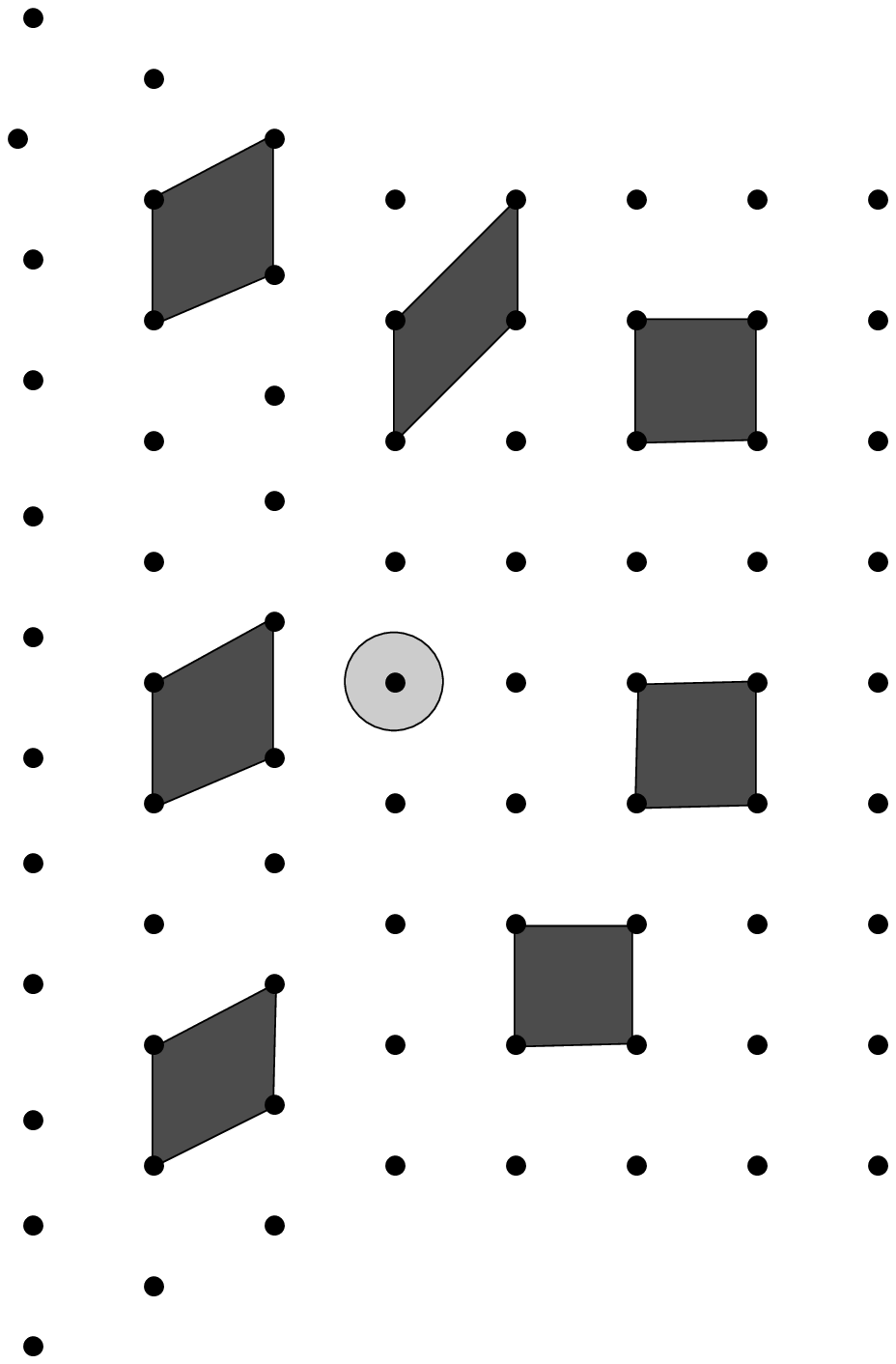}  \hskip0.2cm
\includegraphics[height=2.8cm]{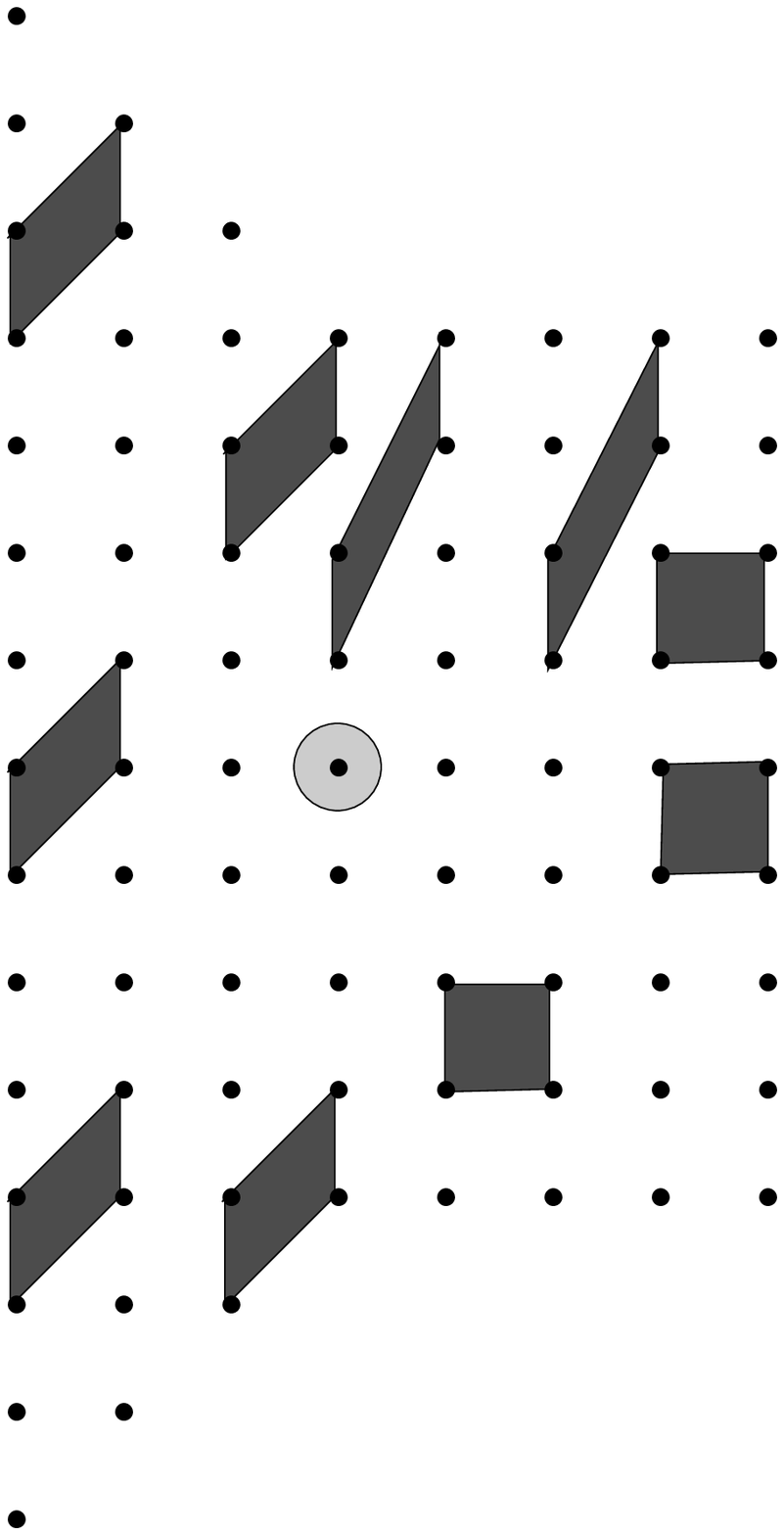} 
\caption{Construction of the angular dislocation by removing or adding
one of the solid angles shown on the left picture.  Reconstructed lattices
after removing or adding small or large sectors are shown together with
transport of elementary cell along a closed path around the defect on
the reconstructed lattice. The identification of boundaries after removing
or adding solid angle is done by the parallel shift of lattice points in
vertical direction.}
\label{F:AngDislPl}      
\end{figure}

\begin{figure}
\centering
\includegraphics[height=2.5cm]{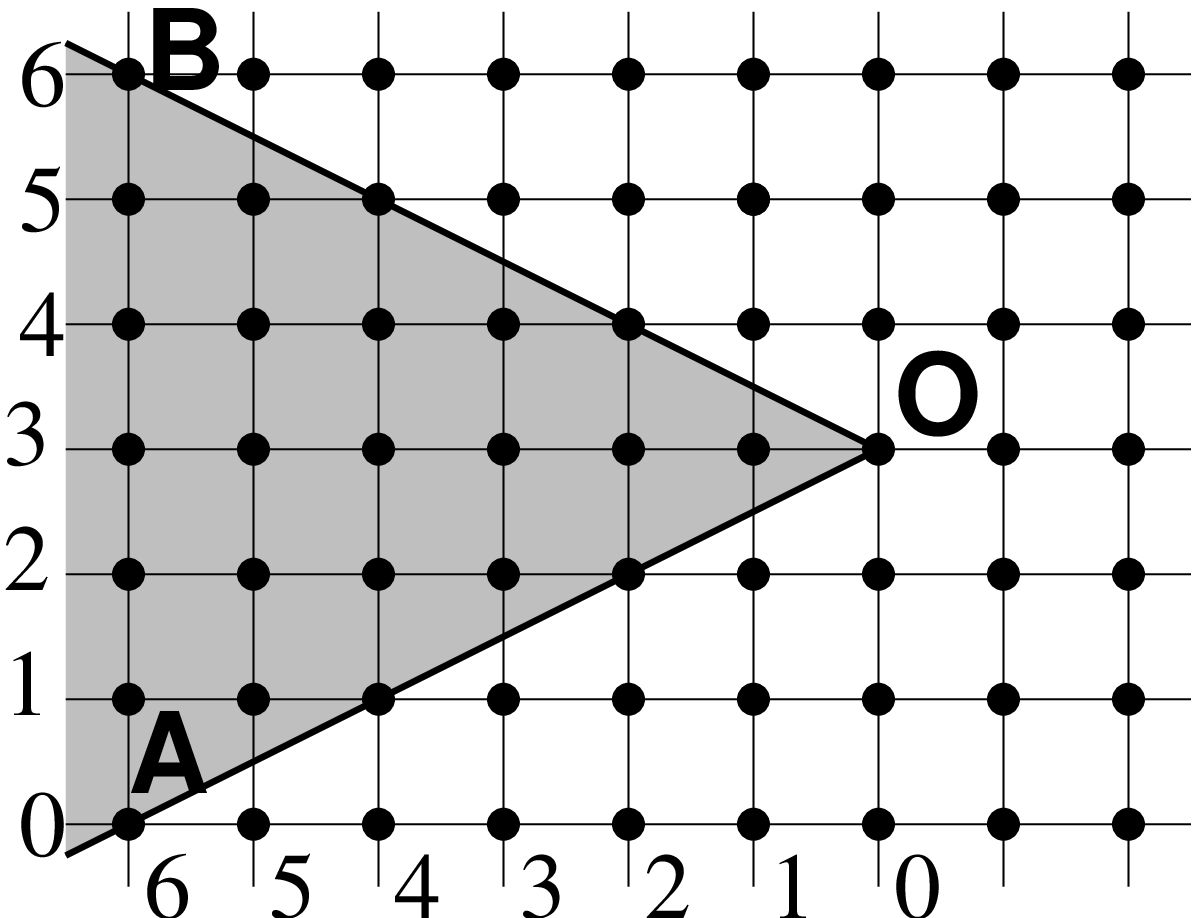} \hskip0.5cm
\includegraphics[height=2.5cm]{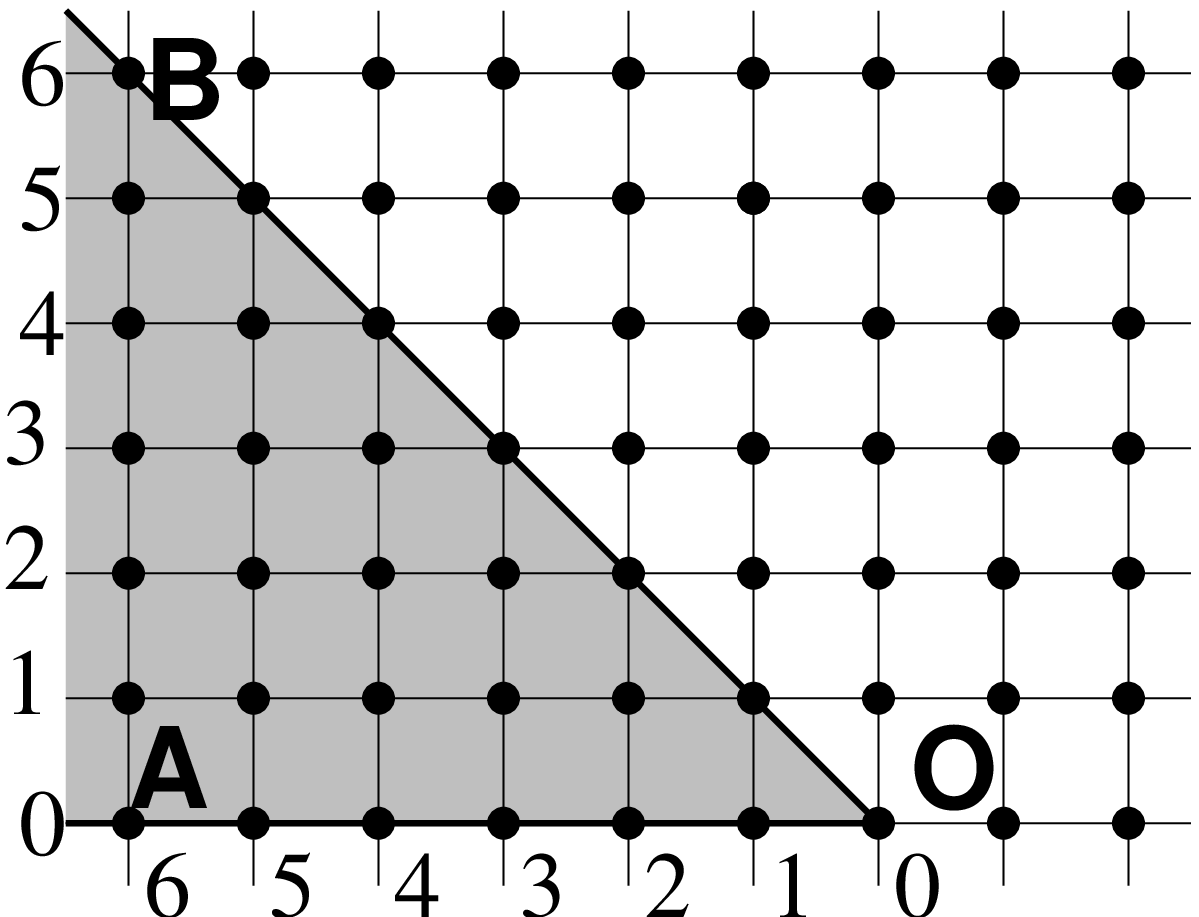}  \hskip0.5cm
\includegraphics[height=2.5cm]{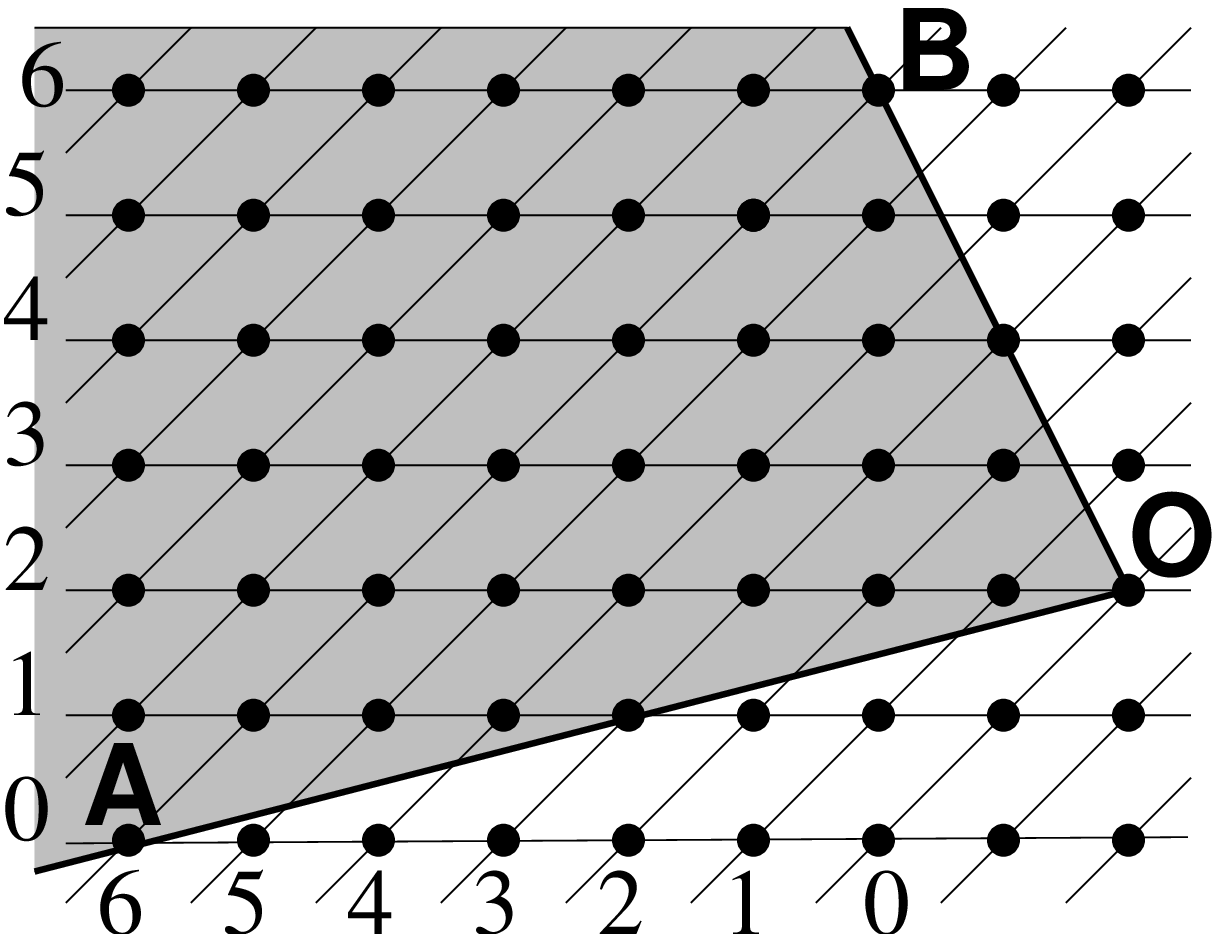} 
\caption{Alternative constructions of the same defect. }
\label{F:AngDislMi}      
\end{figure}

It is important to note that correspondence between two boundaries
should be imposed in order to reconstruct the lattice. We will 
look for different possibilities but let us start with the simplest one:
After removing (or introducing) the solid angle, the reconstruction is d.ne
by the parallel shift of lattice points in one chosen direction. 
The requirement for reconstructed
lattice to be well defined everywhere except singular point can be
satisfied only for some special values of removed or added angles.
Namely we should impose that the number of removed (added) points at each
vertical line is integer and varies linearly with distance from the
vertex of the solid angle. Figure \ref{F:AngDislPl}, left   shows 
examples of removed or added solid angles.  
Two different solid angles correspond respectively  to  removing (adding) of 
one or two additional points   
from vertical line at each step in the horizontal direction.
We can remove or add solid angles in different ways. Figure \ref{F:AngDislMi}
illustrates the construction of the removed angle. We start with
one chosen point $O$ of the lattice and two basis vectors corresponding
to
``horizontal'' and ``vertical'' directions. We put the first cut through the
vertex $A$ lying at the $k$-th vertical line counting from the vertex $O$.
($k=6$ on figure \ref{F:AngDislMi}). To construct the second cut 
we go from $A$ in vertical direction up to $k$-th horizontal line.
Two rays $OA$ and $OB$ show the sector to be removed. Observe that with this 
construction $s$ points are removed from $s$-th vertical line.
   
Figures \ref{F:AngDislPl} show graphically 
what happens  with
lattice after removing (or adding) solid angles. It is important to note
that just by looking on the deformation of elementary cell after the round 
trip on the reconstructed lattice we can easily find how big was
removed (added) solid angle and what transformation (removing or adding)
was exactly done. The absolute value of removed (added) solid angle
can be read directly by comparing the form of the initial and final
cell. It is sufficient   to write the transformation of two vectors
forming elementary cell in matrix form. This matrix is nothing else but
the monodromy matrix for actions. For two examples shown in 
Figure \ref{F:AngDislPl} this  monodromy matrix has the form
$\left(\begin{array}{cc} 1& p\\ 0 & 1\end{array}\right) $
or
$\left(\begin{array}{cc} 1& 0\\ p & 1\end{array}\right) $ 
with $p=\pm 1$ or $p=\pm 2$. One or another form of matrix and the sign of
$p$ depends on the choice of the first and second basis vector and on the
direction of the circular trip (clockwise or counterclockwise). At the
same time the absolute value of $|p|$ is unambiguously related with the absolute
value of the removed (added) solid angle. 

More subtle arguments are
needed to distinguish between adding and   removing solid
angle  with the same $|p|$.  We will denote below defects obtained
by removing solid angle by $(-)$ and by adding solid angle by
$(+)$. 

%%%%%%%%%%%%%%%%%%%%%%%%%%%%%%%%%%%%%%%%%%%%%%%%%%%%%%%%%%%%
\subsection{About the sign of the elementary monodromy defect}
\label{sS:SignDef}
The existence of the sign of Hamiltonian monodromy was conjectured by the
author on the basis of analogy between monodromy and 
$(+)$ and $(-)$ defects of lattices.
The proof was given by Cushman and Vu Ngoc \cite{CushmanSan}. 
We give here the characterization of the sign of defect in terms of
lattice transformation.
\begin{figure}
\centering
\includegraphics[height=2.3cm]{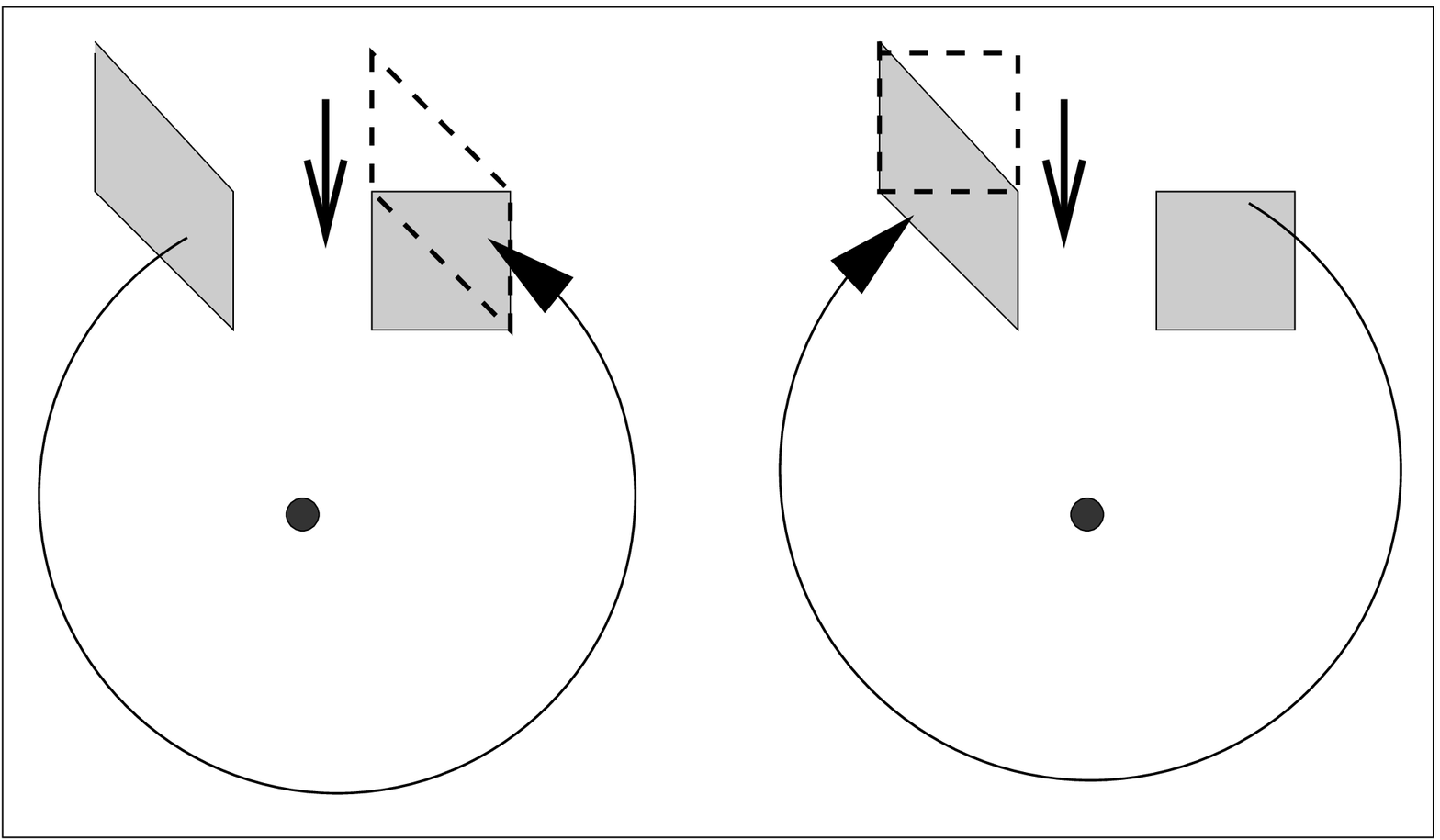} \hskip1.5cm
\includegraphics[height=2.3cm]{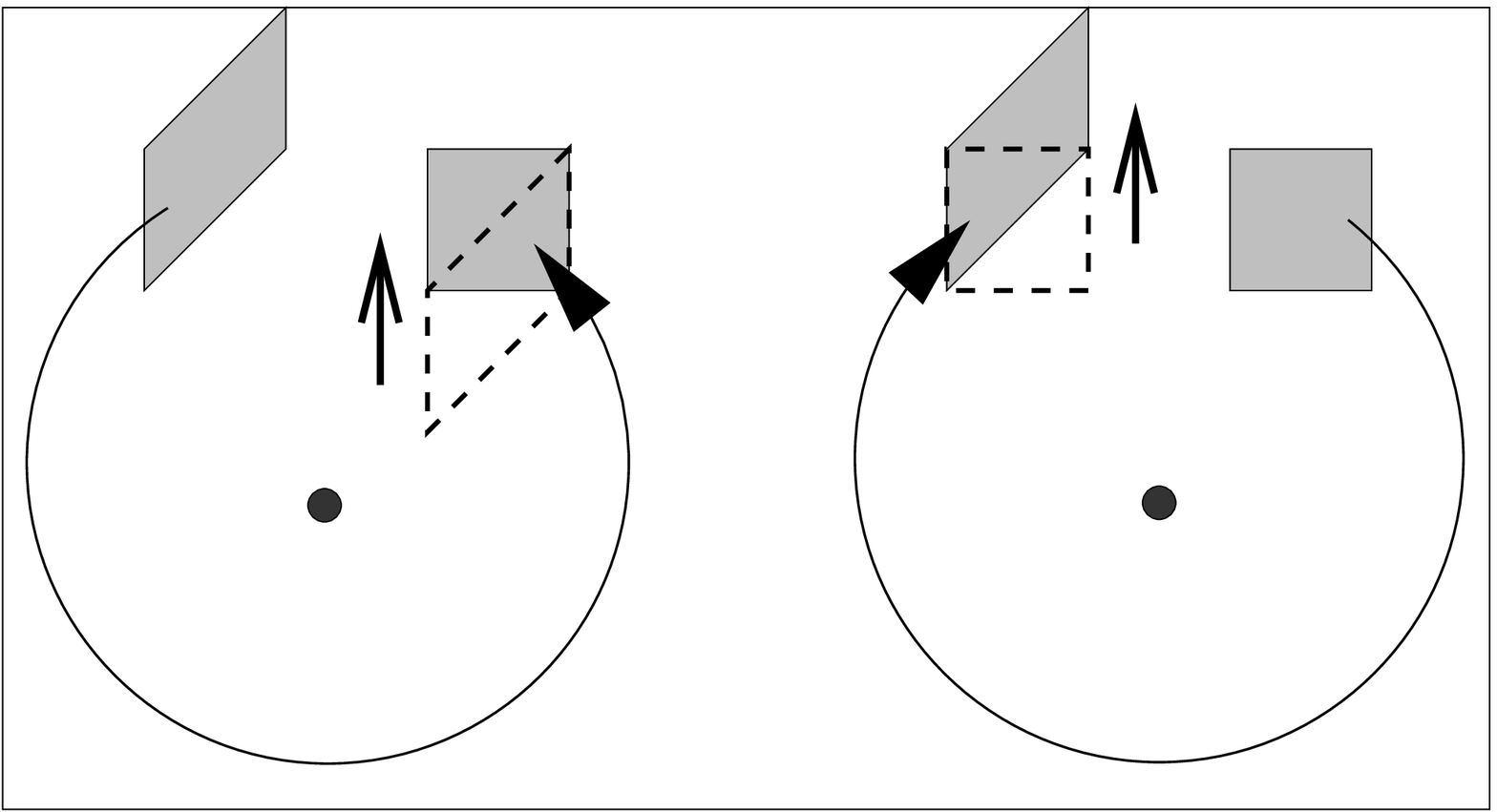}  
\caption{Comparison of initial and final cells after the circular path
around the singularity on the lattice reconstructed after removing
[$(-)$ defect, left] or adding [$(+)$ defect, right]
elementary solid angle. Both counterclockwise and 
 clockwise circular paths are shown for each type of defects.  }
\label{F:DownDef}      
\end{figure}

Let first compare initial and final cells for the same reconstructed
lattice (with $|p|=1$) obtained by removing simplest solid angle but for
both kinds of circular paths (clockwise and counterclockwise).
See Fig. \ref{F:DownDef}, left. The identification of initial cell with 
the final one
can be done only for two vertices. We choose the identified pair being
the back side of the cell in the final position (with respect to the 
direction given by the sense of rotation).
It is clearly seen from the figure  \ref{F:DownDef}, left
 that in order to deform the initial cell for $(-)$ defect
to the form of the final
cell we need to move the front side of the cell in the direction
inside the surrounded singularity. This result is unchanged if we
apply the same procedure to the clockwise or counterclockwise circular
path. 

A similar analysis can be done for lattice reconstructed after 
adding solid angle, i.e. for $(+)$ defect, 
(see Fig. \ref{F:DownDef}, right) shows that the deformation of the initial cell
after the round trip is now in the outside direction with respect to the
surrounded singularity. This result  remains again the same for
both clockwise and counterclockwise direction of the circular path.

Thus the simple geometrical analysis of the transformation of elementary
cell enables one to associate with elementary monodromy the specific
defect of the regular lattice. The defect obtained by
removing solid angle with $|p|=1$  will be called the elementary monodromy
defect. Exactly this defect appears in lattices of quantum states for
Hamiltonian systems corresponding to classical Hamiltonian systems with
focus-focus singularities. Observe that defects with $|p|>1$ appear naturally
in Hamiltonian systems with symmetries. One of the most interesting and
physically important systems of this kind is the integrable approximation
for hydrogen atom in crossed electric and magnetic fields  
\cite{cushman-sadovskii00}.  In classical systems 
monodromy with $|p|>1$ corresponds to presence of isolated singular fiber
which is $|p|$-times pinched torus \cite{matveev}. Cushman and
Vu Ngoc \cite{CushmanSan} have proved that only 
focus-focus singularities with the same sign
of monodromy can appear in a connected component of 
of the image of the generalised energy momentum map of an integrable Hamiltonian
system. In non-Hamiltonian systems monodromy of both sign can appear
simultaneously \cite{cushman-duistermaatNH}. 

%%%%%%%%%%%%%%%%%%%%%%%%%%%%%%%%%%%%%%%%%%%%%%%%%%%%%%%%%%%%%%%%%%%%%%%%
\subsection{Rational cuts and rational line defects} \label{sS:RatCut}
\begin{figure}
\centering
\includegraphics[height=2.5cm]{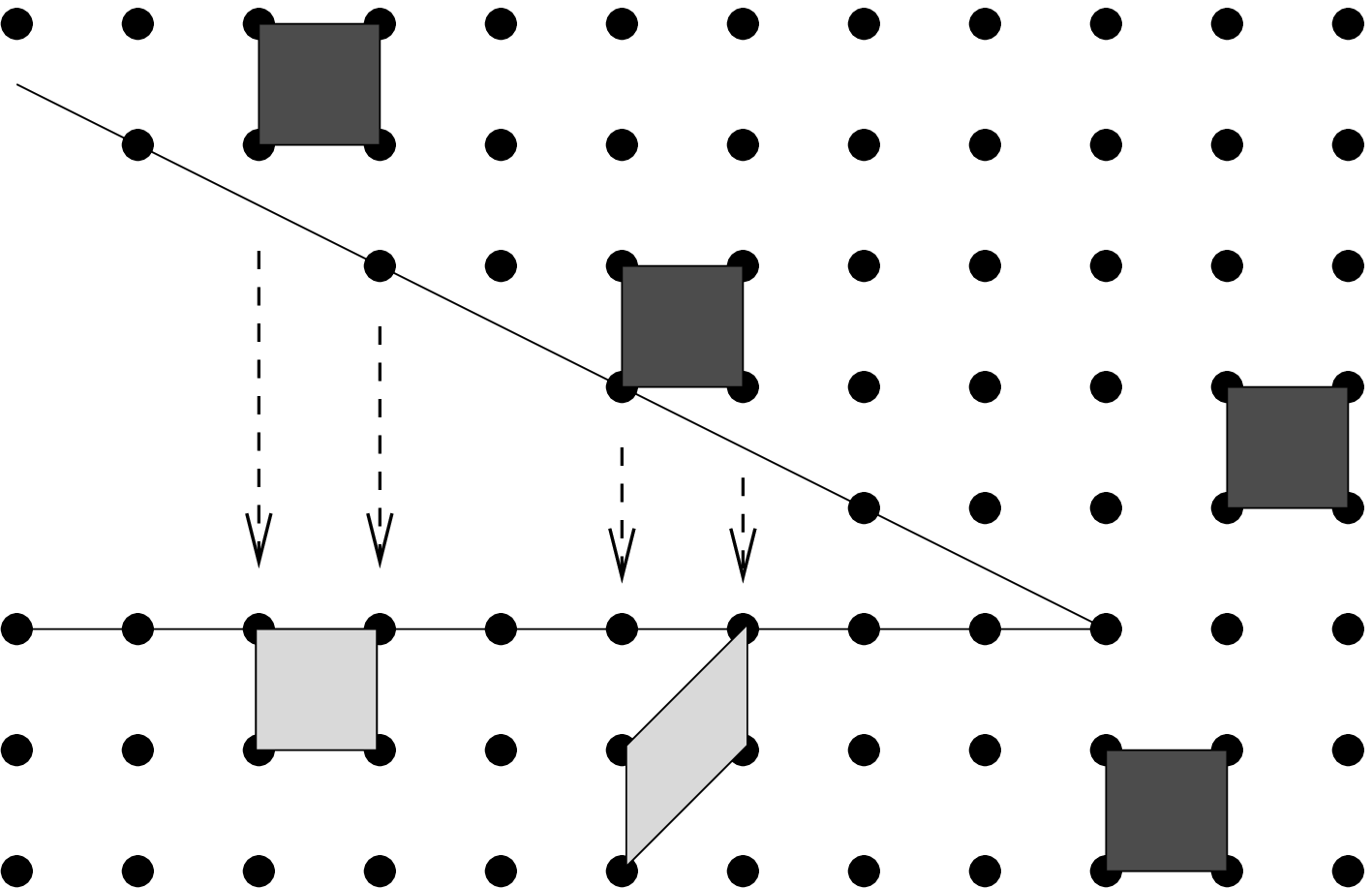} \hskip1.5cm
\includegraphics[height=2.5cm]{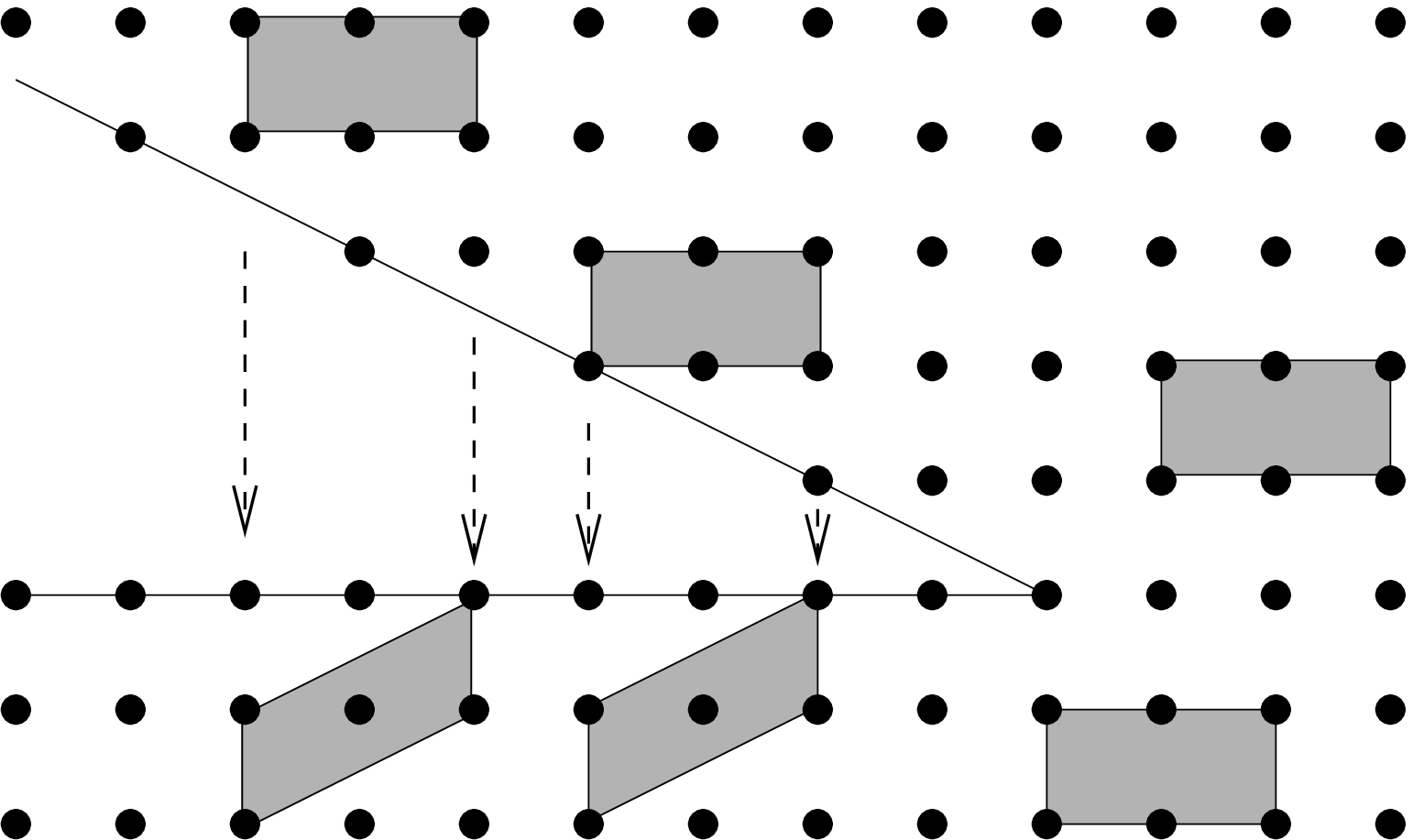}  
\caption{Construction of the $1:2$ rational lattice defect 
starting from the regular
square lattice. (Left) - Ambiguity in the transfer of $1\times 1$ cell through 
the cut.  (Right) -  
Unambiguous transfer of double cell through the cut. }
\label{F:1_2cut}      
\end{figure}

We have seen in previous section that only very special cuts together with
matching rules enables us to construct the point defects. Now we will
generalize the admissible cuts but keep the matching rule. Let us 
start with the example of $1:2$ rational cut which is defined as follows
(see Figure \ref{F:1_2cut}).

We cut out  half of the solid angle removed in the case of the
elementary monodromy defect. After removing this solid angle the two
boundaries are different. At one (lower boundary on Figure  \ref{F:1_2cut})
points are situated at each vertical line of the lattice. At the upper
boundary of the cut points are situated only at each second vertical line.
We keep the matching rules, i.e. identify the
boundaries by sliding points along the vertical lines. Naturally, the
reconstructed lattice is not homogeneous along the identified boundary. 
It is seen from the fact that the number of removed points from vertical
lines varies like $0,0,1,1,2,2,3,3, \ldots$ along the horizontal direction.
(Remark. The number of removed points can be represented in 
the form of the  sum of linear and oscillatory functions.) This means that the
reconstructed lattice has a line defect.   

 If we try to pass the 
elementary cell of the lattice through the cut the result depends on the
place where the cell goes through the boundary line. From Figure
\ref{F:1_2cut}, left,  it is clear that when the right side of the cell 
goes through the cut at even vertical
line (supposing the  vertical line going through the
vertex of the removed sector to be even) the form of the elementary cell 
remains unchanged. In contrast, when the right side of the cell 
goes through the cut at odd  vertical line, the form of the cell changes.
This ambiguity 
can be avoided if instead of elementary $1\times1$ cell we will use
larger cell. Namely, we double the dimension of the cell in the horizontal
direction. The double cell passes through the cut at any place in a similar
way. But the internal structure of the cell changes after crossing
the line defect. Cell transforms from 
``face centered'' to ``body centered'' in the crystallographic terminology.
But this modification is uniform along the cut. In some way, by
increasing the dimension of the cell we neglect the effects comparable 
with the dimension of the cell. This enables us to  define the transformation
of lattice vectors after traversing a closed path 
around the origin of the removed
sector. Putting $e_h$ and $e_v$ as horizontal and vertical basis vectors
of the square lattice shown in Figure \ref{F:1_2cut}
and $\{e_v,e_h^{double}=2e_h\}$ as vectors forming the double cell, 
the transformation of  vectors forming the double cell 
after a close path around the origin of the removed sector in
the counterclockwise direction is
\begin{equation}
\left(\begin{array}{c} e_v^\prime \\ (e_h^{double})^\prime\end{array} \right) 
= \left( \begin{array}{cc} 1&\ 0\\  1&\ 1 \end{array}\right) 
\left(\begin{array}{c} e_v \\ e_h^{double}\end{array} \right). 
\end{equation}
If we extend linearly this transformation to lattice vectors themselves
the transformation matrix takes the form 
\begin{equation}
\left(\begin{array}{c} e_v^\prime \\ e_h^\prime\end{array} \right) 
= \left( \begin{array}{cc} 1&\ 0\\ 1/2&\ 1 \end{array}\right) 
\left(\begin{array}{c} e_v \\ e_h\end{array} \right) .
\end{equation}
The so obtained matrix  with fractional entry coincides with the
fractional monodromy matrix for actions in the case of $1:(-2)$ resonant
classical oscillator and with quantum fractional monodromy for corresponding
quantum problem. 

\begin{figure}
\centering
\includegraphics[height=2.5cm]{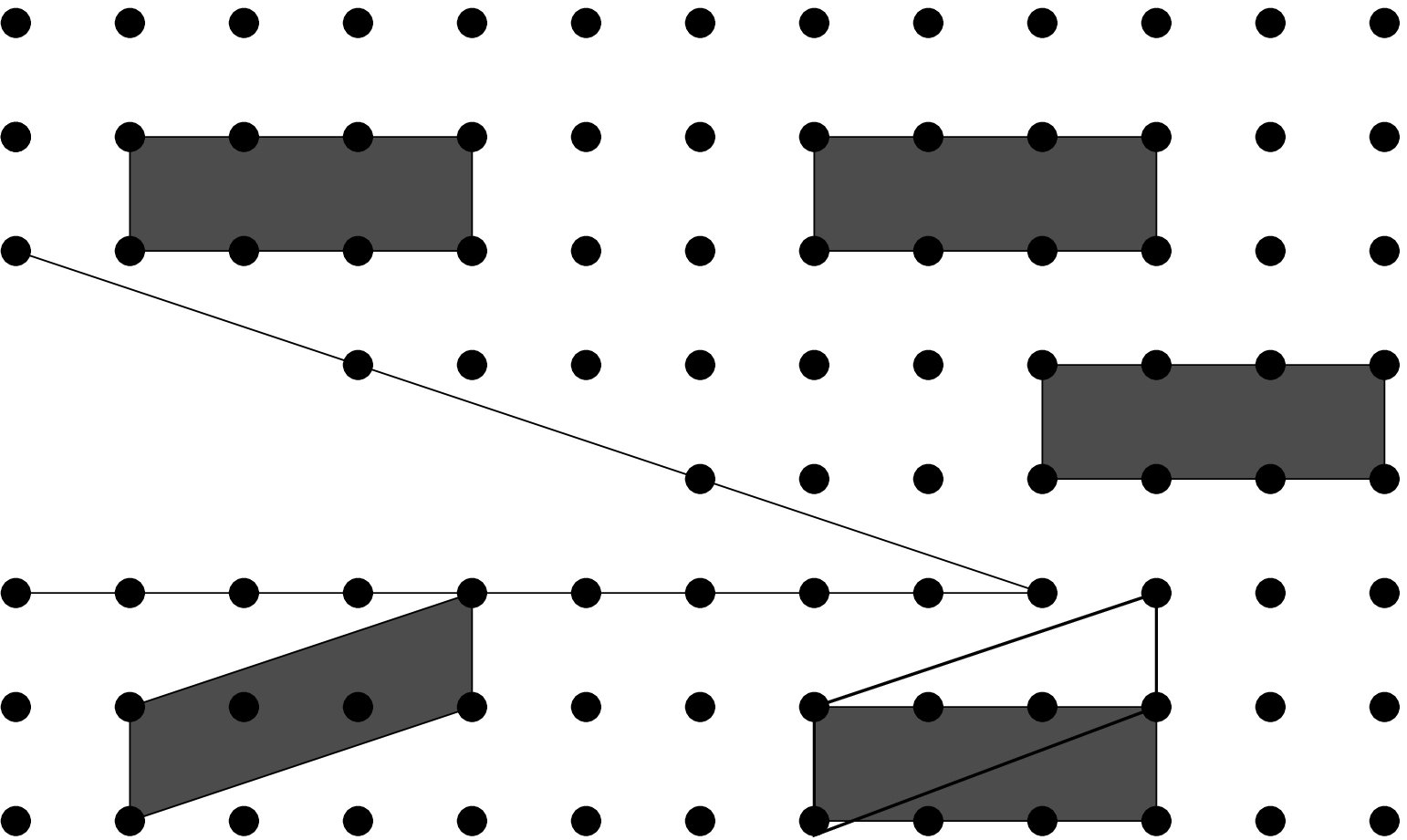} \hskip1.5cm
\includegraphics[height=2.5cm]{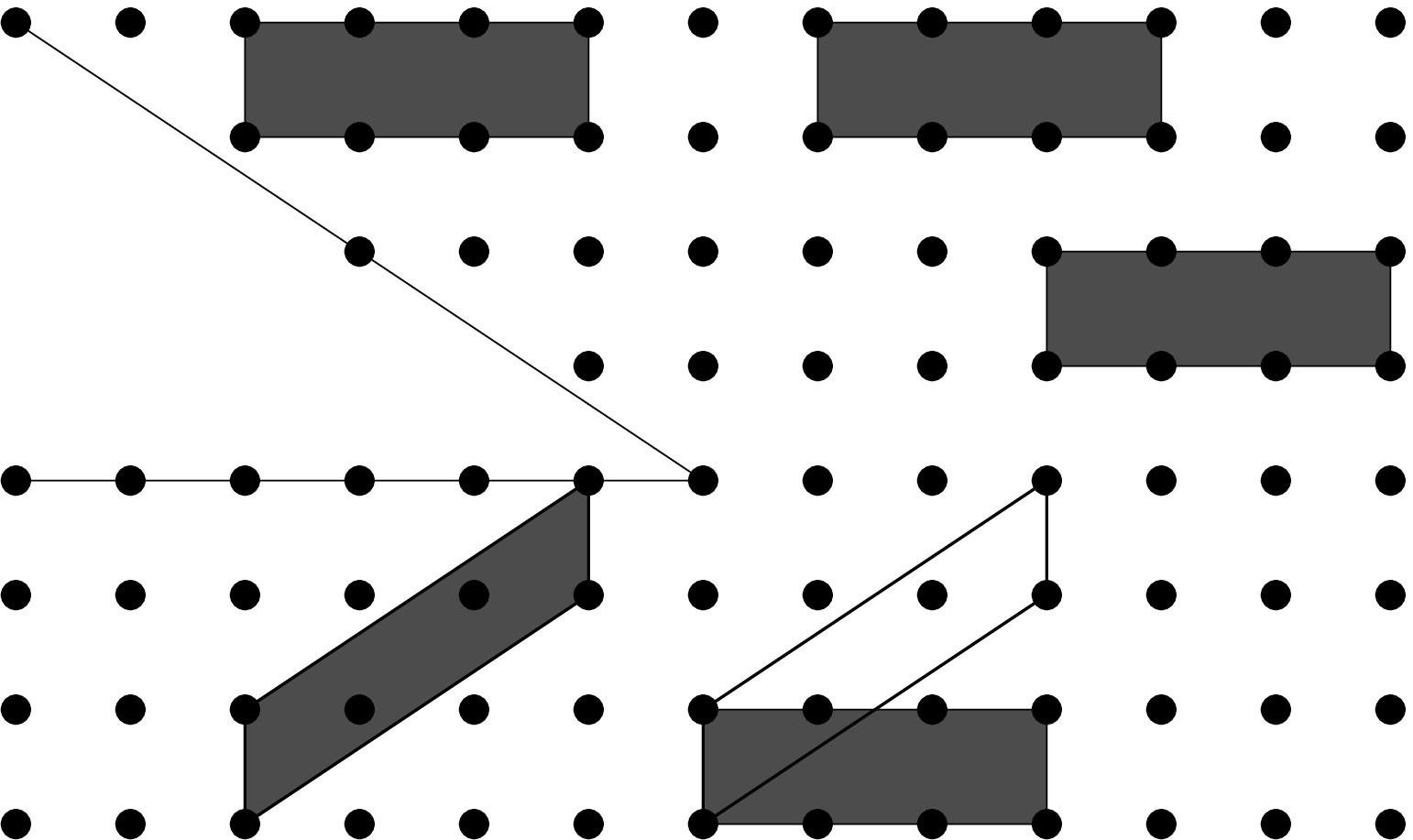}  
\caption{Construction of the $1:3$ and $2:3$ rational lattice defects 
starting from the regular
square lattice.}
\label{F:13Cuts}      
\end{figure}

Using the same principle we can construct, for example, line defects by
reconstructing lattice after $1:3$ or $2:3$ rational cuts shown in Figure
\ref{F:13Cuts}.   This notation means that we remove the solid angle 
$\varphi=\tan^{-1}(1/3)$ or $\tan^{-1}(2/3)$ respectively. 
We need to triple the dimension of cell in the horizontal direction in order
to get unambiguous transformation rules for the cell after crossing the
line defect on the 
reconstructed lattice. It is clear from the Figure \ref{F:13Cuts} that
the monodromy matrices for $1:3$ and $2:3$ rational defects have
respectively the form 
$\left( \begin{array}{cc} 1&0\\ 1/3&1 \end{array}\right)$
and  $\left( \begin{array}{cc} 1&0\\ 2/3&1 \end{array}\right)$.
Removing $2:3$ rational solid angle is equivalent to removing twice the
$1:3$ rational solid angle. 
Generalization to arbitrary rational cut with the same type of matching
rules for reconstruction of the lattice is quite obvious and leads to
half-line defect with fractional monodromy matrix.

\begin{figure}
\centering
\includegraphics[height=2.5cm]{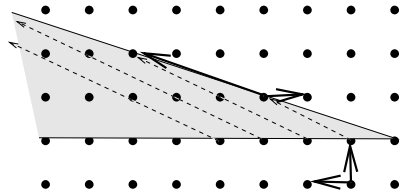}  
\caption{Matching rules for reconstruction of lattice after rational
cut. Example of $1:3$ cut. The monodromy of the resulting defect is the
Arnol'd cat map. }
\label{F:1_3Arn}      
\end{figure}

We can also suggest alternative matching rules after rational cuts with the
idea to obtain reconstructed lattice with only point rather than the line defect.
 Let us consider again as example the $1:3$ rational cut but with different
matching rules for two boundaries (see Figure \ref{F:1_3Arn}). 

 It is clear that
if we want to have on the reconstructed lattice only point defect
all vertices on two boundaries should be consecutively identified. This
identification imposes matching rule for one of the basis vectors of the
lattice. Another should be chosen in such a way that two new basis
vectors form elementary cell of the same volume, i.e. they should be 
related one to another with $SL(2,Z)$ transformation. 
%One can verify that for
%each rational cut it is possible to find the $SL(2,Z)$ matrix, which
%corresponds to matching rule for going over the cut. Otherwise for each
%$SL(2,Z)$ matrix one can construct a rational cut such that matching rules
%for this cut are expressed in terms of a chosen matrix. 
In fact this matrix
is precisely the monodromy matrix of the point defect just by the
construction. In the case of $1:3$ rational cut shown in
 Fig. \ref{F:1_3Arn} the resulting monodromy matrix has the form
$\left( \begin{array}{cc}0& -1\\1&\ \ 3 \end{array} \right )$.
This matrix is known as Arnol'd cat map
\cite{ArnAv}. A lot of different examples of 
point defects can be constructed in a similar way. But we will take 
as elementary point defect only $(-)$ defects corresponding to
elementary monodromy matrix. We will demonstrate now that all other defects
can be considered as more complicated objects composed in some way from
several elementary ones.  

%%%%%%%%%%%%%%%%%%%%%%%%%%%%%%%%%%%%%%%%%%%%%%%%%%%%%%%%%%%%%%%%%%
\section{Defects with arbitrary  monodromy} \label{sS:MultD}
We now turn to the description of defects which can be characterized by
arbitrary monodromy matrices. As  soon as
the choice of the basis of the lattice is ambiguous, the
matrix representation of the monodromy transformation is basis dependent.
For example the monodromy matrix
$M_a=\left(\begin{array}{cc} 1& 1\\ 0 & 1 \end{array}\right) $
after the transformation to another basis through the similarity
$M_a^\prime = A M_a A^{-1}$ with $A$ being arbitrary $SL(2,Z)$ matrix
takes the form
$$\left(\begin{array}{cc} a& b\\ c & d\end{array}\right)
   \left(\begin{array}{cc} 1& 1\\ 0 & 1\end{array}\right)
   \left(\begin{array}{cc} d& -b\\ -c & a\end{array}\right) =
   \left(\begin{array}{cc} 1-ac& a^2\\ -c^2 & 1+ac\end{array}\right)$$
From this family of equivalent matrices it is immediately clear that
matrices $\left(\begin{array}{cc} 1& 1\\ 0 & 1\end{array}\right)$
and $\left(\begin{array}{cc} 1& 0\\ -1 & 1\end{array}\right)$ are
equivalent but they are written in different frames. In contrast, matrix
$\left(\begin{array}{cc} 1& -1\\ 0 & 1\end{array}\right)$ 
is equivalent to
$\left(\begin{array}{cc} 1& 0\\ 1 & 1\end{array}\right)$ 
but is not equivalent to 
$\left(\begin{array}{cc} 1& 1\\ 0 & 1\end{array}\right)$ in spite of the fact
that these two matrices are mutually inversed.

In order to formulate precise statement about equivalence or in-equivalence
of different defects we should first establish equivalence of $SL(2,Z)$
matrices with respect to conjugation by elements of $SL(2,Z )$, i.e. to describe 
classes of conjugated elements of $SL(2,Z )$ group.

It is well known that the trace and the determinant of the matrix are
invariant with respect to similarity transformation. But these invariants
are not sufficient to completely characterize classes of conjugated elements.
Before looking for $SL(2,Z)$ matrices let us start with $SL(2,R)$ ones.

%%%%%%%%%%%%%%%%%%%%%%%%%%%%%%%%%%%%%%%%%%%%%%%%%%%%%%%%%%%%%%%%%%%%%%
\subsection{Topological description of unimodular matrices}
Let consider the subspace of $SL(2,R)$ matrices 
$M= \left(\begin{array}{cc} \alpha& \beta\\ \gamma & \delta\end{array}\right)$
with ${\rm Tr}\ M = K$. This means that four matrix elements
$\alpha, \beta, \gamma, \delta$ are related by two equations
\begin{equation}
\alpha\delta-\beta\gamma=1, \ \ \ \  \alpha+\delta = K.
\end{equation} 
Eliminating one parameter (say $\alpha$) we get the following relation
between three parameters $ \beta, \gamma, \delta$
\begin{equation}
-1+K\delta -\delta^2-\gamma\beta = 0.
\end{equation}
We can interpret this relation as the geometrical description of all 
$SL(2,R)$ matrices with given trace in the three dimensional space of
parameters $ \beta, \gamma, \delta$. The geometrical form of so obtained
surface depends on the value of $K$. Topologically there are three different
situations. 

\begin{figure}
\centering
\includegraphics[height=5cm]{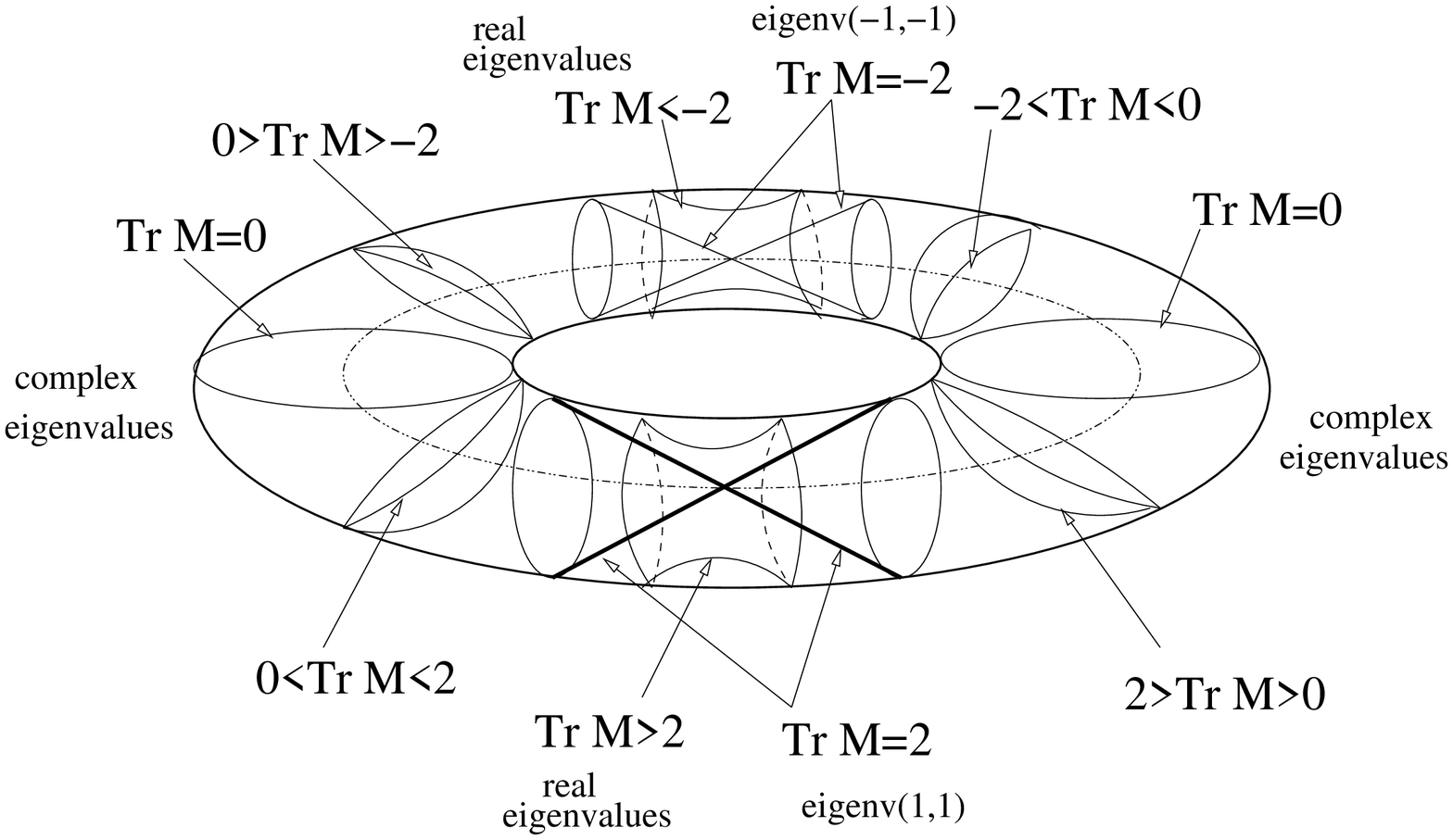}
\caption{The topological structure of the space of 2 by 2 matrices with
determinant 1. The solid tore in 3-D space is foliated by levels
corresponding to a given value of the trace of matrix $M$. 
\label{torL2R} }
\end{figure}

If $K=\pm2$ we have double cone with vertex corresponding to
$\pm \left(\begin{array}{cc} 1& 0\\ 0 & 1\end{array}\right) $ matrix.
If $|K|>2$ we have a  hyperboloid of one-sheet
and if $-2<K< 2$ we have two-sheeted hyperboloid. 
We can schematically represent the whole family of $SL(2,R)$ matrices
by filling the solid torus in three-dimensional space of parameters by
surfaces corresponding to all possible values of traces. This 
representation is given in Figure \ref{torL2R}.

%%%%%%%%%%%%%%%%%%%%%%%%%%
%%%%%%%%%%%%%%%%%%%%%%%%%%%%%%%%%%
The existence of two disjoint connected components for matrices with $-2\leq 
{\rm Tr} M \leq 2$ implies the existence of additional invariant which 
classifies matrices with the same trace into smaller subclasses of conjugate
elements. We need such description but only for matrices in $SL(2,Z)$.
The important difference between $SL(2,R)$ and $SL(2,Z)$ cases 
 is due to the fact that
there is only finite number of possible values of the trace 
in $SL(2,Z)$ which correspond to matrices with complex eigenvalues 
in $SL(2,R)$. In physical language this is the
consequence of the fact that only axes of second, third, fourth and sixth
orders are compatible with the existence of the lattice.
 
Formal proof: Characteristic polynomial for the $SL(2,Z)$ matrix $M$ has the form
$\lambda^2 - ({\rm Tr}\ M) \lambda +1=0$. It has complex eigenvalues only if
the discriminant $({\rm Tr}\ M)^2 -4 <0$. As soon as the trace is integer,
it is only possible that  $({\rm Tr}\ M)=0, \pm 1$.

Now we can return to the study of $SL(2,Z)$ case.

%%%%%%%%%%%%%%%%%%%%%%%%%%%%%%%%%%%%%%%%%%%%%%%%%%%%%%%%%%%%%%%%%%%%%%%
\subsection{Classes of conjugated elements and ''normal form'' of 
$SL(2,Z)$ matrices}
The matrices  $M\in SL(2,Z)$ will be named parabolic, elliptic, or
hyperbolic, depending on their trace. Parabolic matrices have trace
equal $\pm2$ and their eigenvalues are $\{+1, +1\}$ or  $\{-1, -1\}$.
Elliptic matrices have trace $\pm1$ or $0$. Their eigenvalues are complex
numbers. Hyperbolic matrices have $|{\rm Tr}\ M|>2$. Their eigenvalues
are real irrational numbers. 
Identity matrix 
$\left(\begin{array}{cc}  1 & 0\\  0 & 1 \end{array}\right)$ and minus identity
$\left(\begin{array}{cc} -1 & 0\\  0 &-1 \end{array}\right)$ commute with
all elements from $SL(2,Z)$ and each form a proper class of conjugate
elements consisting of one element. We will consider these classes
separately. In Fig, \ref{torL2R} these matrices correspond to vertices of
double cones of matrices with trace $\pm2$. 

Within each class of conjugate elements we can choose one matrix to be
the ``normal form''. All classes of conjugate elements together with
normal forms are listed in Table \ref{T:sl2zCE}.

\begin{table}
\centering
\caption{Classes of conjugated elements of $SL(2,Z)$ group together with
normal forms of matrices for each class.}
\label{T:sl2zCE}      
\begin{tabular}{l|ccccc}
\hline\noalign{\smallskip}
Trace    & $K, (|K|>2)$ & $\pm2$ & $\pm1$ & $0$  \\
\noalign{\smallskip}\hline\noalign{\smallskip}
Module    & $-$ & $p=0,\pm1,\ldots$ & $\varepsilon=\pm1$ & $\varepsilon=\pm1$  \\
\noalign{\smallskip}
Normal form & $\left(\begin{array}{cc} K& 1\\ -1 & 0\end{array}\right)$ 
& $\left(\begin{array}{cc} 1&  p \\  0 & 1\end{array}\right)$ &
$\left(\begin{array}{cc} \pm(1+\varepsilon)/2 & \varepsilon \\
           -\varepsilon & \pm(1-\varepsilon)/2 \end{array}\right)$ &
$\left(\begin{array}{cc} 0& \varepsilon \\ -\varepsilon & 0 
 \end{array}\right)$  \\
\noalign{\smallskip}\hline
\end{tabular}
\end{table}

%%%%%%%%%%%%%%%%%%%%%%%%%%%%%%%%%%%%%%%%%%%%%%%%%%%%%%%%%%%%%%%%%%%%%%
\subsection{Several elementary monodromy defects}\label{sS:MultElMD}
We have defined  the construction of elementary defects using one chosen
lattice basis. Both cuts and matching rules were precisely defined in that
basis. But the choice of the lattice basis is not unique. 
If there are two defects, characterized both by elementary monodromy matrix
but the choice of basis and the orientation of cuts for these two defects
are different, the global monodromy, corresponding to transformation of 
elementary cell after a circular path around both defects, depends on
relative orientation of two removed solid angles. Let us again start with some
particular examples of systems with several elementary defects. 

%%%%%%%%%%%%%%%%%%%%%%%%%%%%%%%%%%%%%%%%%%%%%%%%%%%%%%%%%%%%%%%%%%%%%%%%%
\subsubsection{Disclinations as a composition of elementary 
monodromy defects}\label{ssS:Discl}
\begin{figure}
\centering
\includegraphics[height=4.5cm]{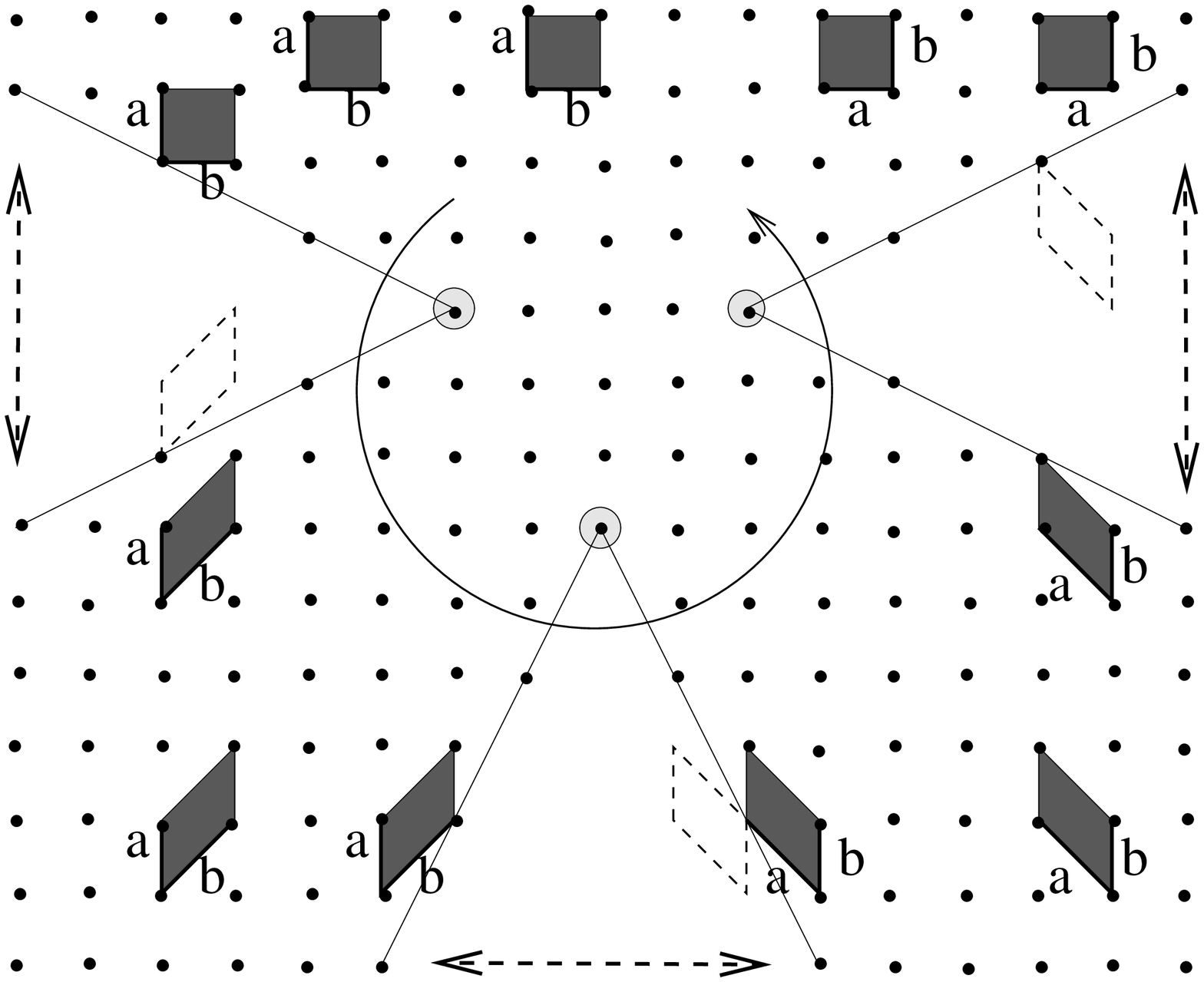} 
\hskip0.5cm
\includegraphics[height=4.5cm]{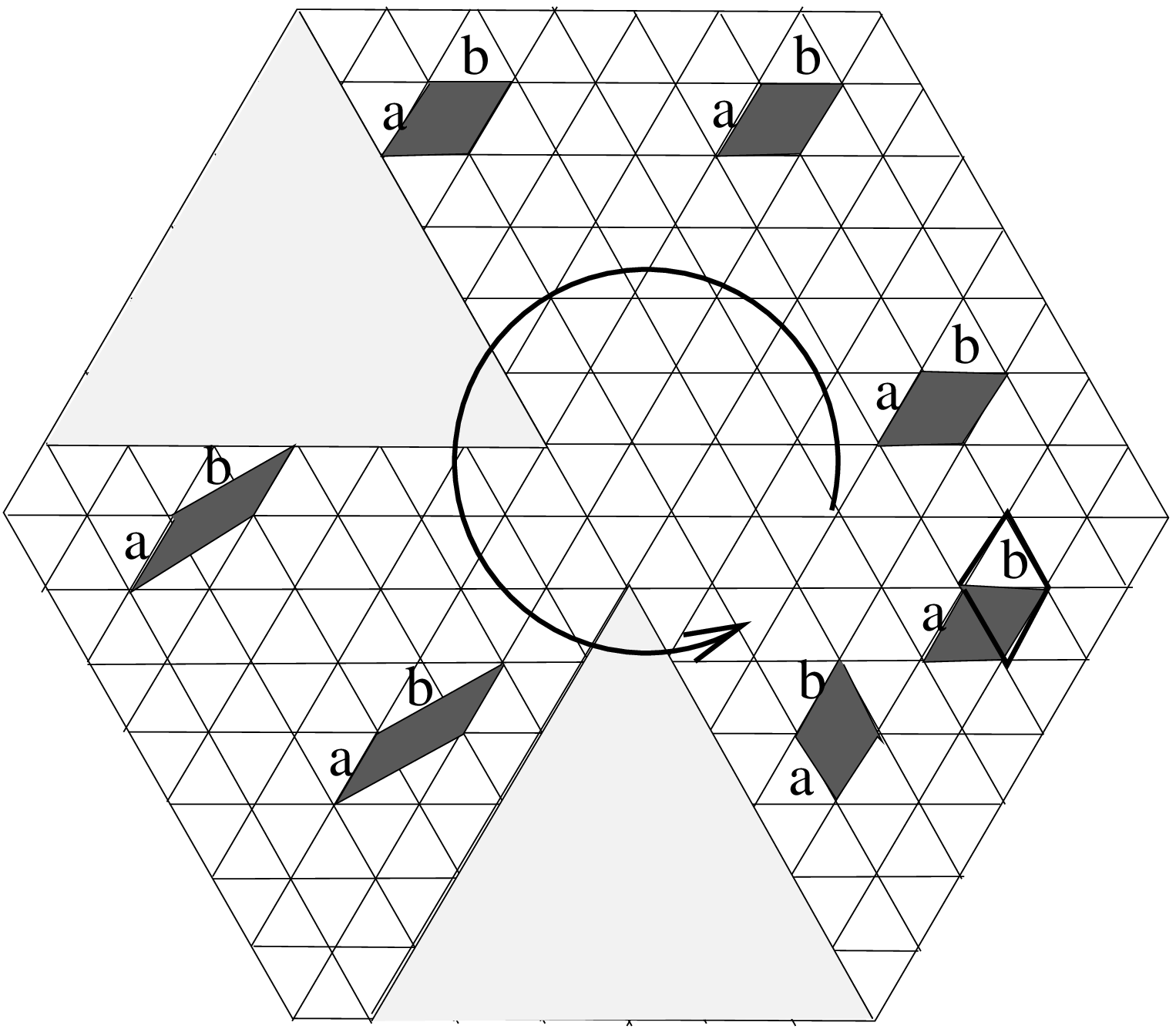}  
\caption{(Left) - Regular square lattice with three  
 elementary monodromy defects.  (Right) - Regular triangular
lattice with two elementary monodromy defects.  }
\label{F:90discl}      
\end{figure}

Figure \ref{F:90discl} shows regular square lattice with three defects
corresponding to elementary monodromy. For two removed angles (sectors
around horizontal lines) the reconstruction of lattice is done 
through sliding points in vertical directions. For the third removed angle
the identification of boundaries is done through the horizontal sliding
of lattice points.  

 The cumulative effect of three elementary cuts is the rotation
of the elementary cell by $\pi/2$. 
The direction of the rotation of the
elementary cell is defined by the direction of the circular path around    
 singularities. Such defect is known in solid
state physics as $\pi/2$ rotational disclination. 
It is easy to see that the same effect takes place if the
three cuts are distributed in another way between vertical and horizontal
directions (two vertical and one horizontal). To see this it is just
sufficient to look at the same figure after rotating it  by $\pi/2$.

Naturally, similar construction can be done with three singular points
corresponding to  adding the
solid angle and reconnecting new boundaries through horizontal or
vertical shift.
The resulting effect on the elementary cell is again the $\pi/2$ rotation,
but now the rotation of the elementary cell is in opposite direction
(as compared to the direction of the circular loop around the
singularity). 

\begin{figure}
\centering
\includegraphics[height=4.5cm]{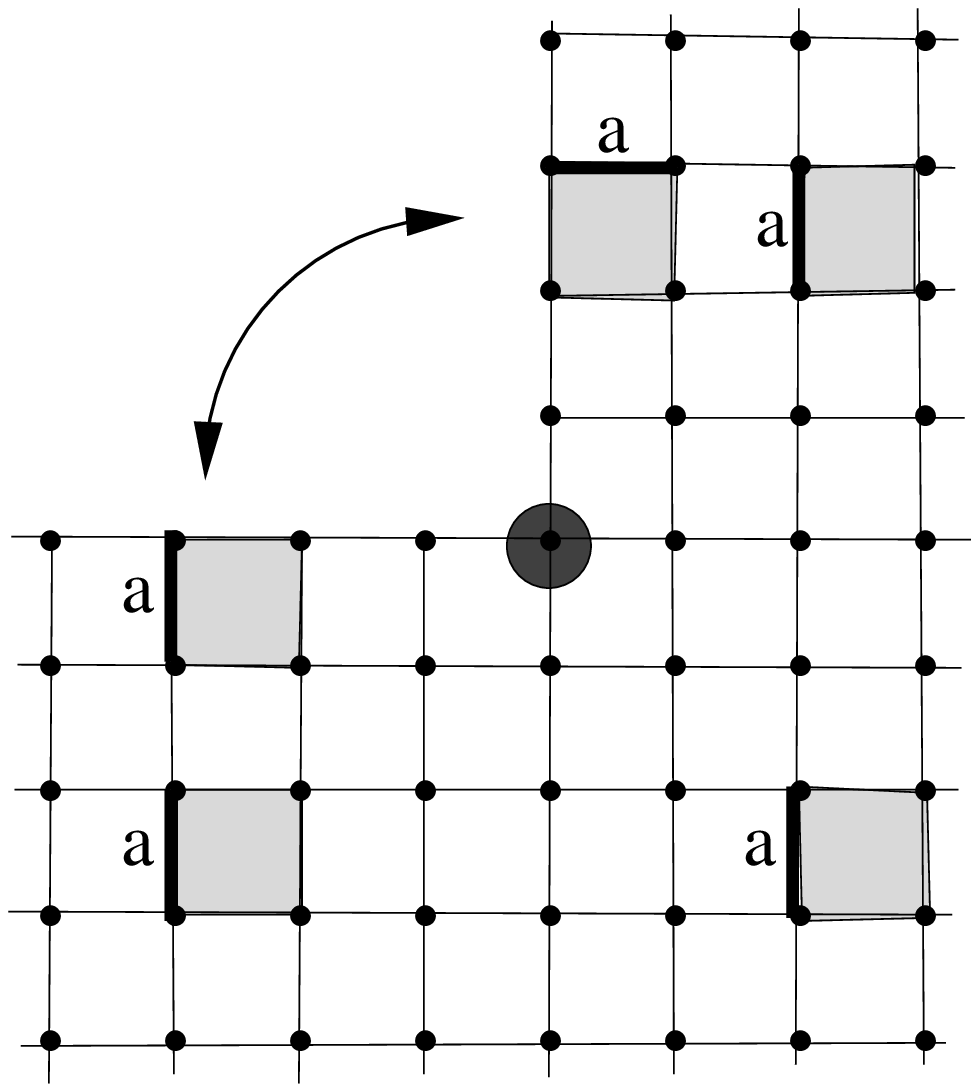}  \hskip1cm
\includegraphics[height=4.5cm]{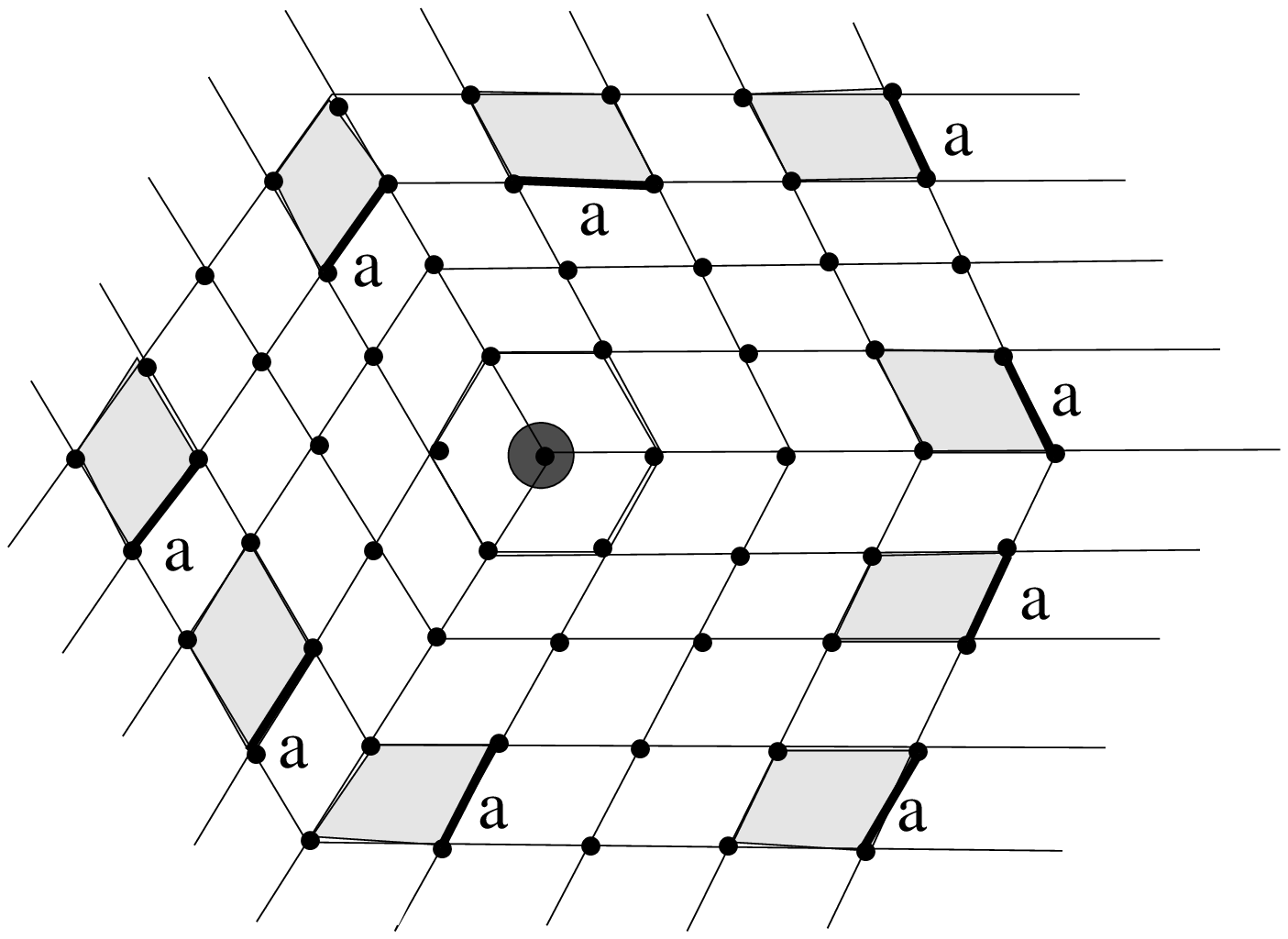}  
\caption{Construction of the rotational disclination
by removing solid angle $\pi/2$ shown on the left picture. }
\label{F:90Rotdiscl}      
\end{figure}

\begin{figure}
\centering
\includegraphics[height=4.5cm]{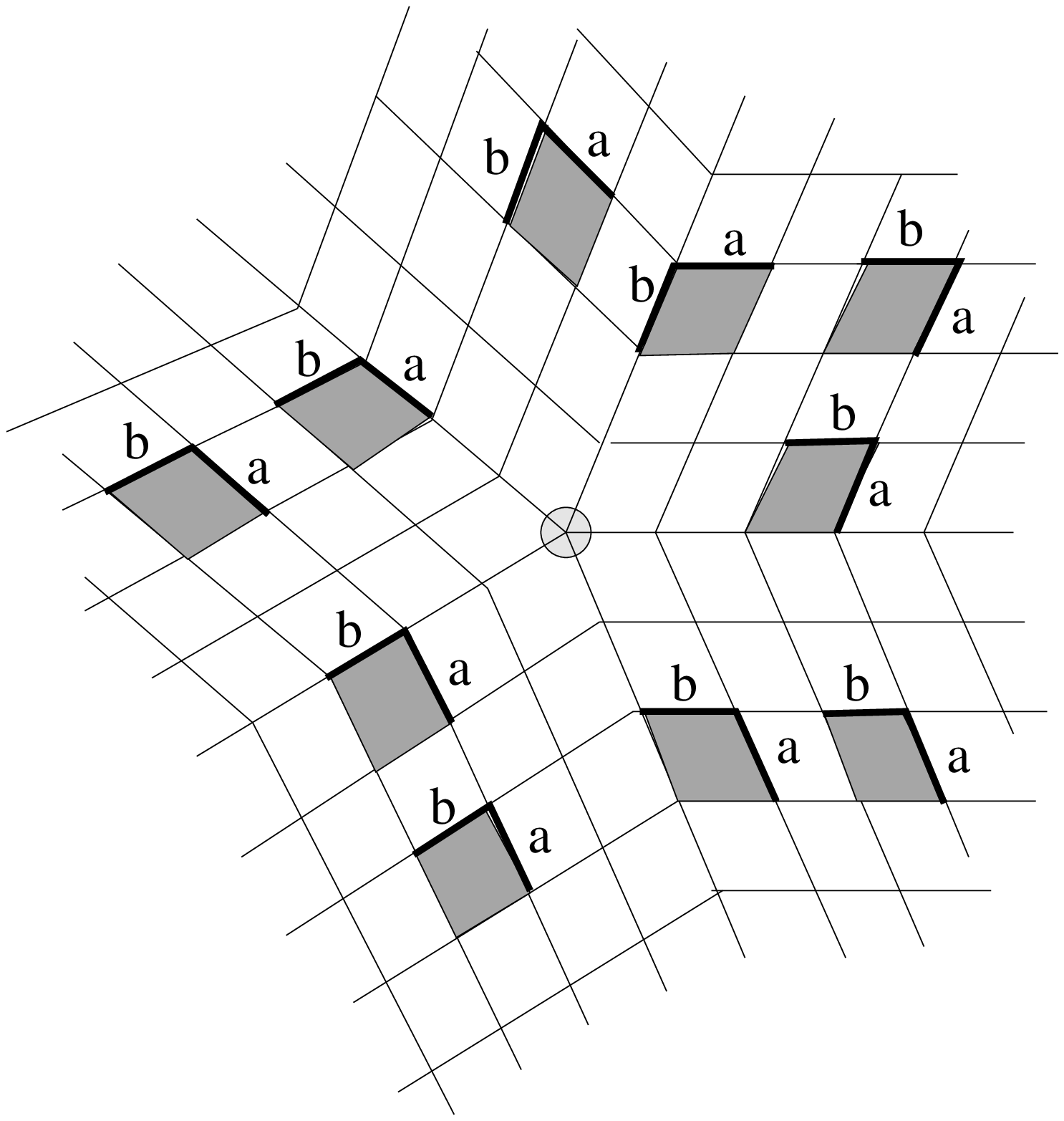}  \hskip1cm
\includegraphics[height=4.5cm]{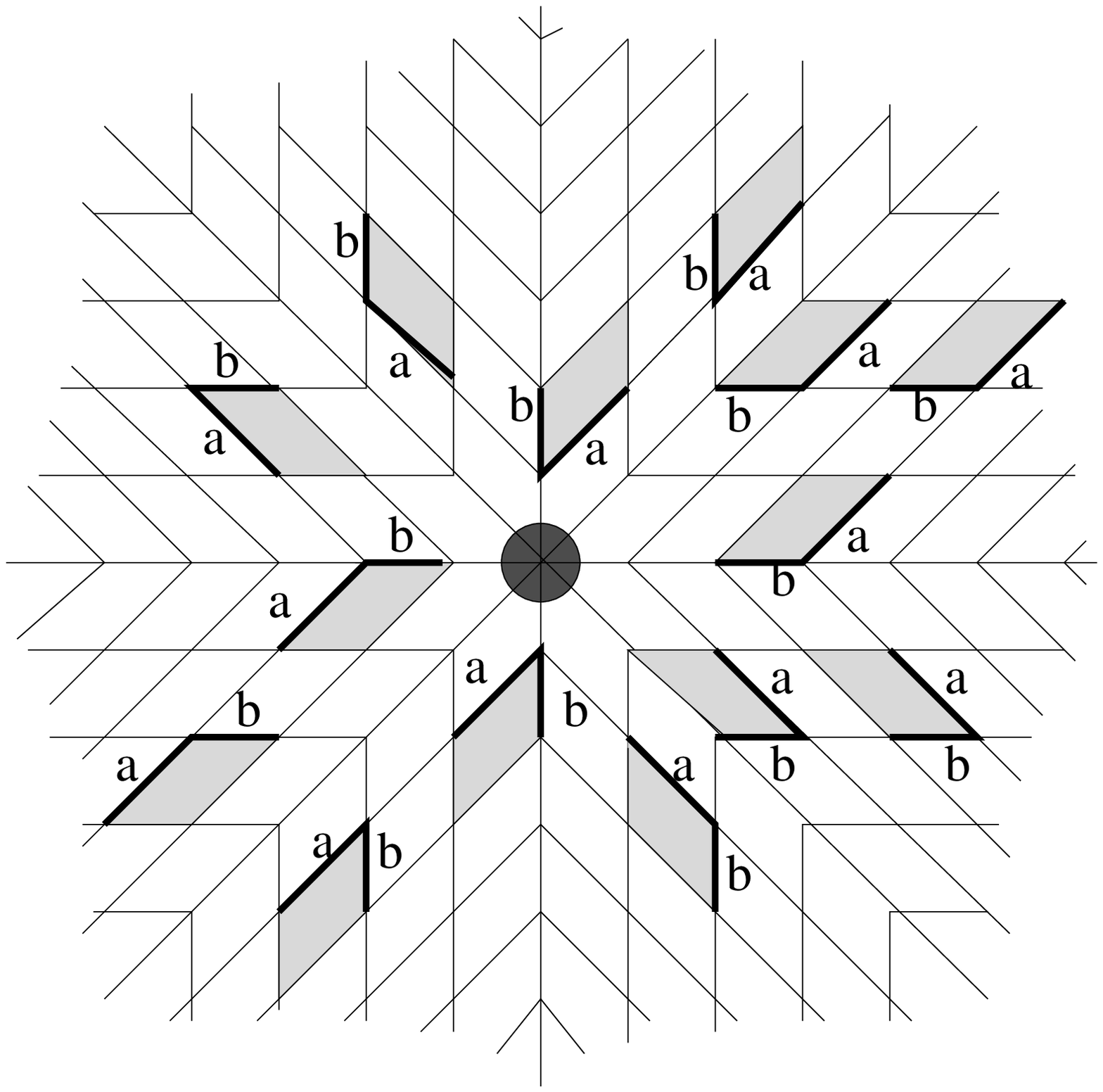}  
\caption{Construction of the rotational dislocation (disclination)
by introducing solid angle $k\pi/2$. $k=1$ on the left
and $k=4$ on the right picture. }
\label{F:addRotdiscl}      
\end{figure}

The global effect in both cases can be reproduced  by removing 
(or adding) the
solid angle $\pi/2$ and by reconstructing lattice through identification
of two boundaries by rotating them  as it is shown in Figure
\ref{F:90Rotdiscl} where  $\pi/2$ solid angle is removed.  
Naturally one can also remove $\pi$ or $3\pi/2$ solid angle  or to
add $\pi/2$ or $k\pi/2$ solid angle as it is shown in 
Figure \ref{F:addRotdiscl}. This gives negative or  positive  rotational 
disclinations.
 
\begin{figure}
\centering
\includegraphics[height=3.0cm]{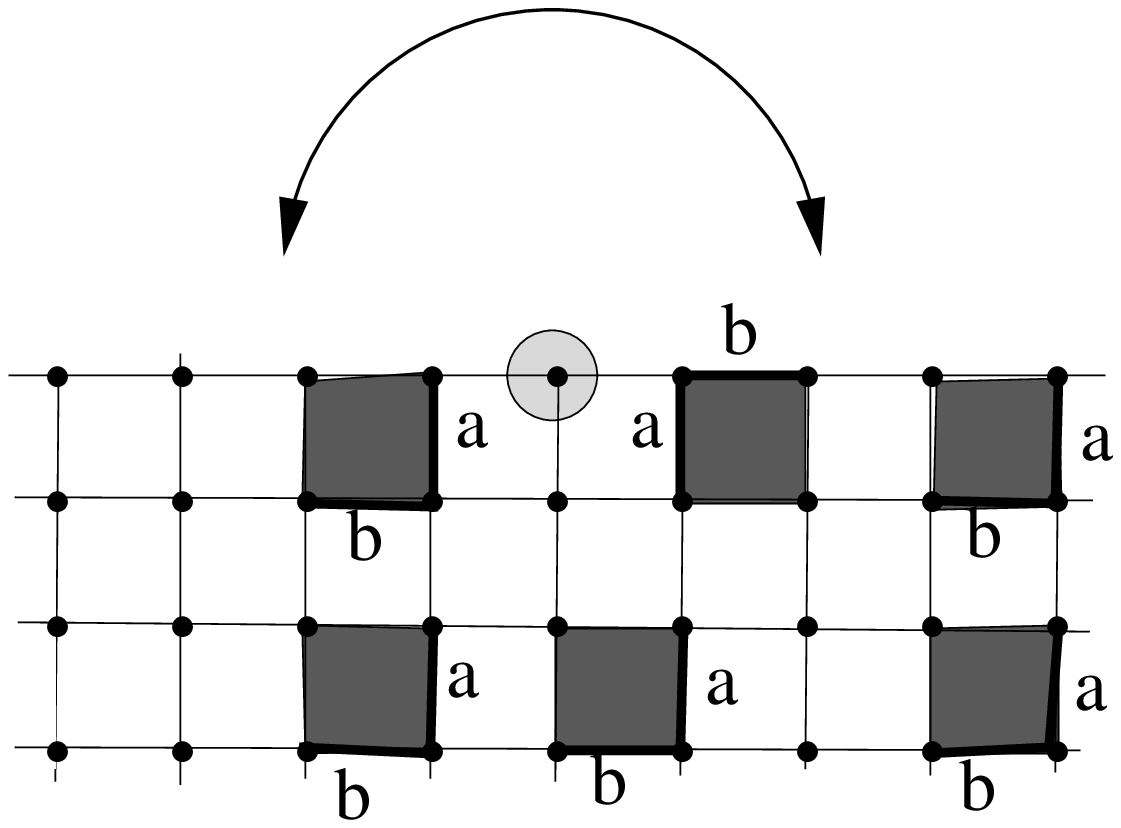}  \hskip0.3cm
\includegraphics[height=3.0cm]{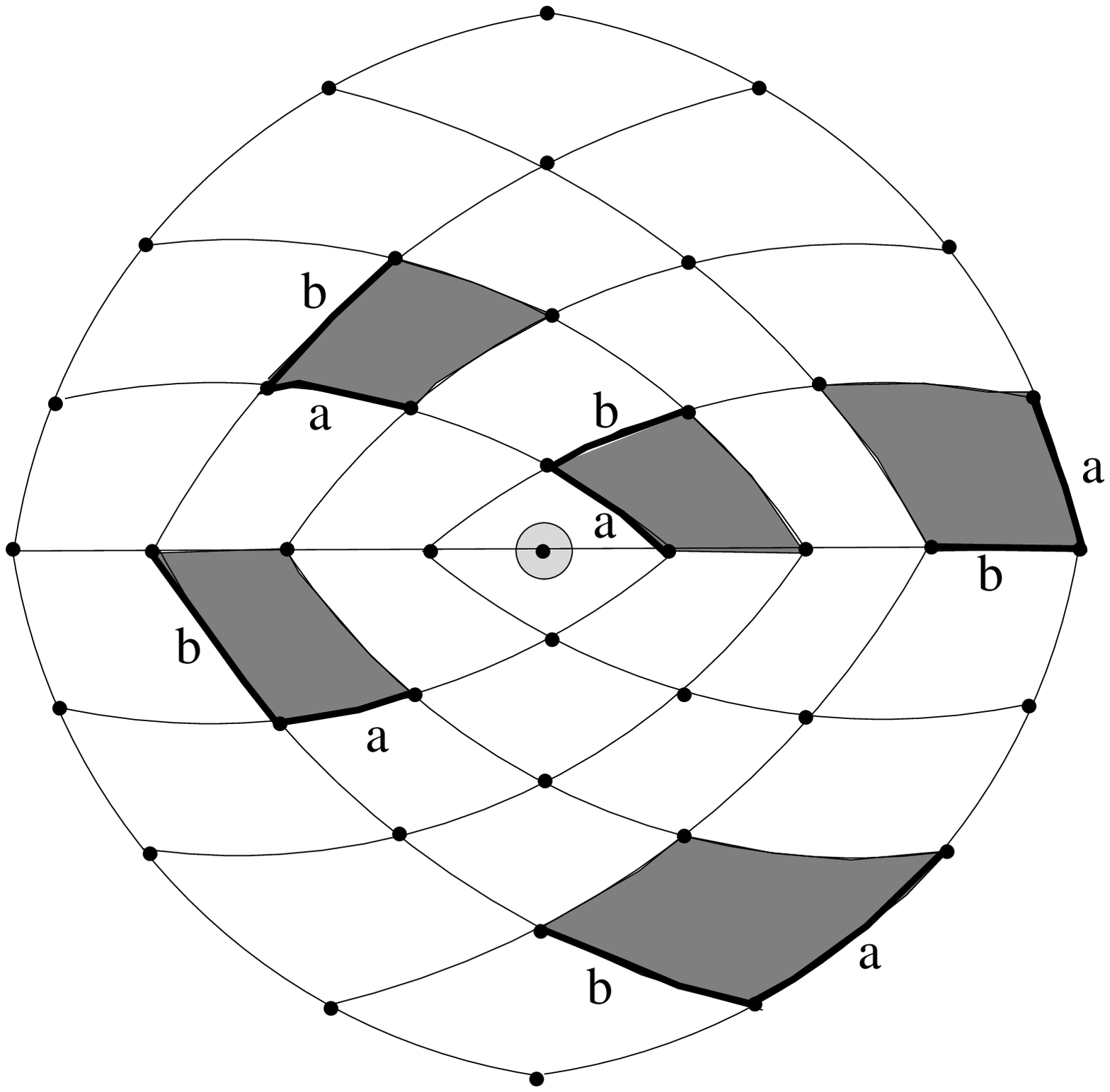}  \hskip0.3cm
\includegraphics[height=3.0cm]{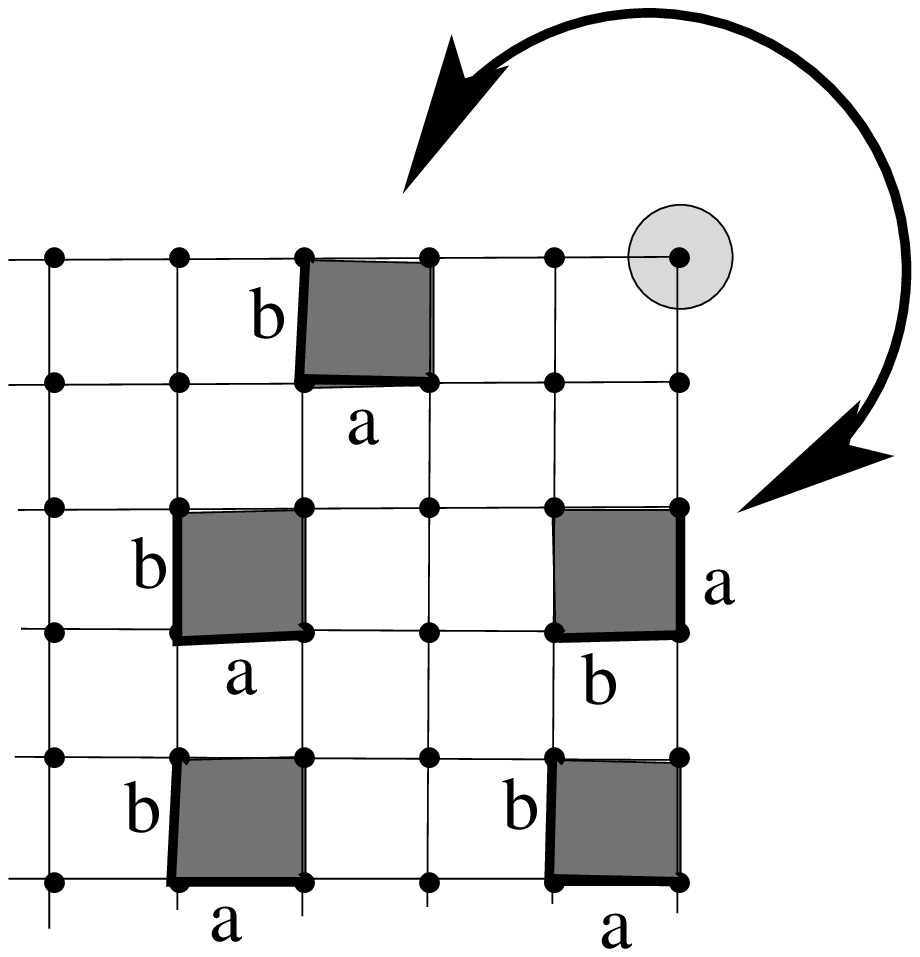}  
\caption{Construction of the rotational disclination
by removing solid angle $\pi$ (Left) and $3\pi/2$ (Right). 
The reconstructed lattice after removing $\pi$ solid angle (center). }
\label{F:180Rotdiscl}      
\end{figure}

The same construction made with triangular lattice and with two elementary
monodromy defects rotated one with respect to another over $2\pi/3$ gives
the cumulative effect consisting in rotation of elementary cell over
$2\pi/6$ after a close path surrounding two elementary defects
(see figure \ref{F:90discl}, right). The cumulative
effects of such two elementary monodromy defects is the $\pi/3$ rotational
disclination. Its multiple, positive or negative analogs can be
immediately constructed. 

Rotational disclinations are well known defects in the solid state physics.
From the point of view of defects of quantum state lattices and classical
Hamiltonian monodromy, the elementary monodromy defects seems to be 
more fundamental.
Any rotational disclination can be constructed as a global effect
in systems with several elementary monodromy defects. 

%%%%%%%%%%%%%%%%%%%%%%%%%%%%%%%%%%%%%%%%%%%%%%%%%%%%%%%%%%%%%%%%%%%%
\subsubsection{Multiple defects with trivial global monodromy}
\label{ssS:Mult Triv}
\begin{figure}
\centering
\includegraphics[height=4cm]{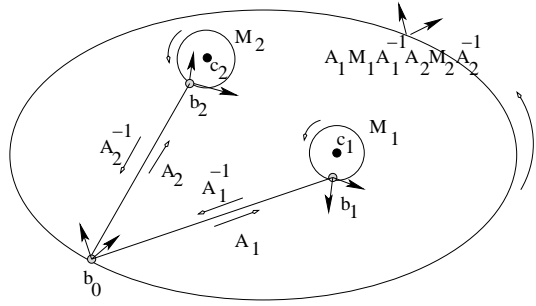} 
\caption{Relation between local monodromy matrices for isolated defects and
global monodromy for the circular path around two defects }
\label{F:prodM}      
\end{figure}

Let us now consider in more general way the correspondence between
local and global monodromy for a system of elementary defects.
 Figure \ref{F:prodM} illustrates this relation.

Suppose we have two defects $c_1$ and $c_2$
characterized by monodromy matrices $M_1$ and $M_2$. These monodromy 
matrices are obtained by going around the defect $c_i$ starting from point
$b_i$ and using local basis associated with point $b_i$. If we are interested
now in global monodromy which corresponds to a close loop going around two
defects and starting at initial point $b_0$ with its own local basis
we can calculate the global monodromy by going first from $b_0$ to $b_1$,
making close loop around $c_1$, returning back by the same way, and repeating
the same for the second defect. The global monodromy calculated in this way
should be the same by homotopy arguments. If the modification of the
basis between $b_0$ and $b_i$ is described by matrices $A_i$ the global
monodromy $M$ can be expressed in terms of $M_1$ and $M_2$ as
$M= A_1 M_1 A_1^{-1} A_2 M_2 A_2^{-1}$.

Naturally for an arbitrary system of elementary defects the global monodromy
matrix can always be represented in the form
$M=\prod_i A_iM_iA_i^{-1}$. As it was already noted the monodromy
matrix is defined up to conjugation with $SL(2,Z)$ matrices, i.e. 
defects with different monodromy are in one to one correspondence with classes
of conjugate elements of $SL(2,Z)$ matrices. 

One can easily verify that arbitrary $SL(2,Z)$ matrix can be represented
in the form of product of matrices, conjugate to elementary monodromy
matrices with one chosen sign \cite{CushZhil}. In particular, 
the identity matrix can also be
represented in the form of product of matrices conjugate to
elementary monodromy matrix. It is obvious that four $\pi/2$ rotational
disclinations (or six $\pi/6$ rotational disclinations)  give trivial 
global monodromy. In fact the elementary cell make a $2\pi$ rotation
when going along close path surrounding these defects and in spite of 
the fact that the monodromy is trivial the close path is not contractible and
the defect exists. An easy consequence of this statement: The monodromy
matrix (defined up to conjugation with $SL(2,Z)$ matrices)
is not sufficient to distinguish defects. Two defects with the same
monodromy matrix can be further labeled by the  number $k$ of $2\pi$
rotations of the elementary cell after a close path around a defect.
This additional number $k$ can be arbitrary integer $k=0, \pm1, \pm2, \ldots$.
Note that an elemenraty $(-)$ monodromy defect can be constructed as a 
cumulative effect of eleven elementary $(+)$ monodromy defects 
\cite{CushZhil}.

\begin{figure}
\centering
\includegraphics[height=2.5cm]{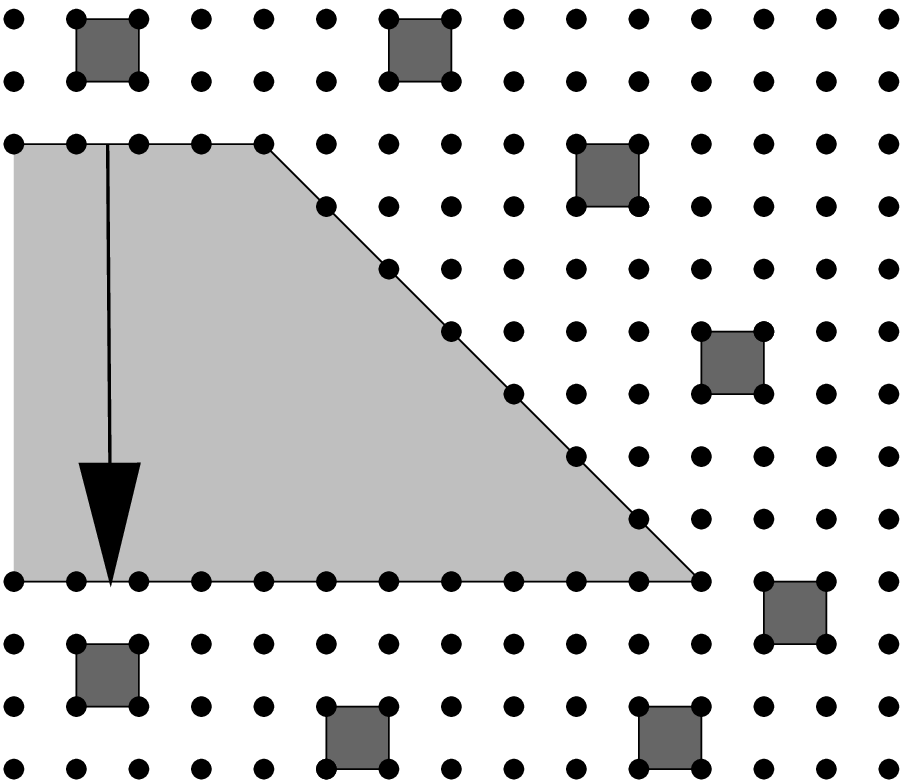} \hskip0.5cm
\includegraphics[height=2.5cm]{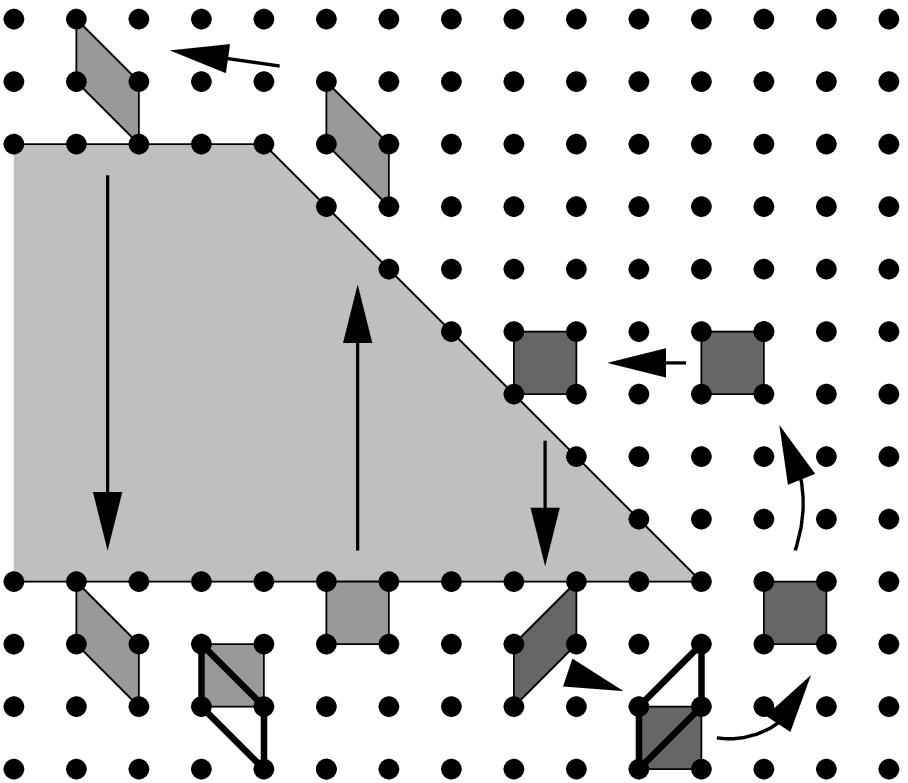} \hskip0.5cm
\includegraphics[height=2.5cm]{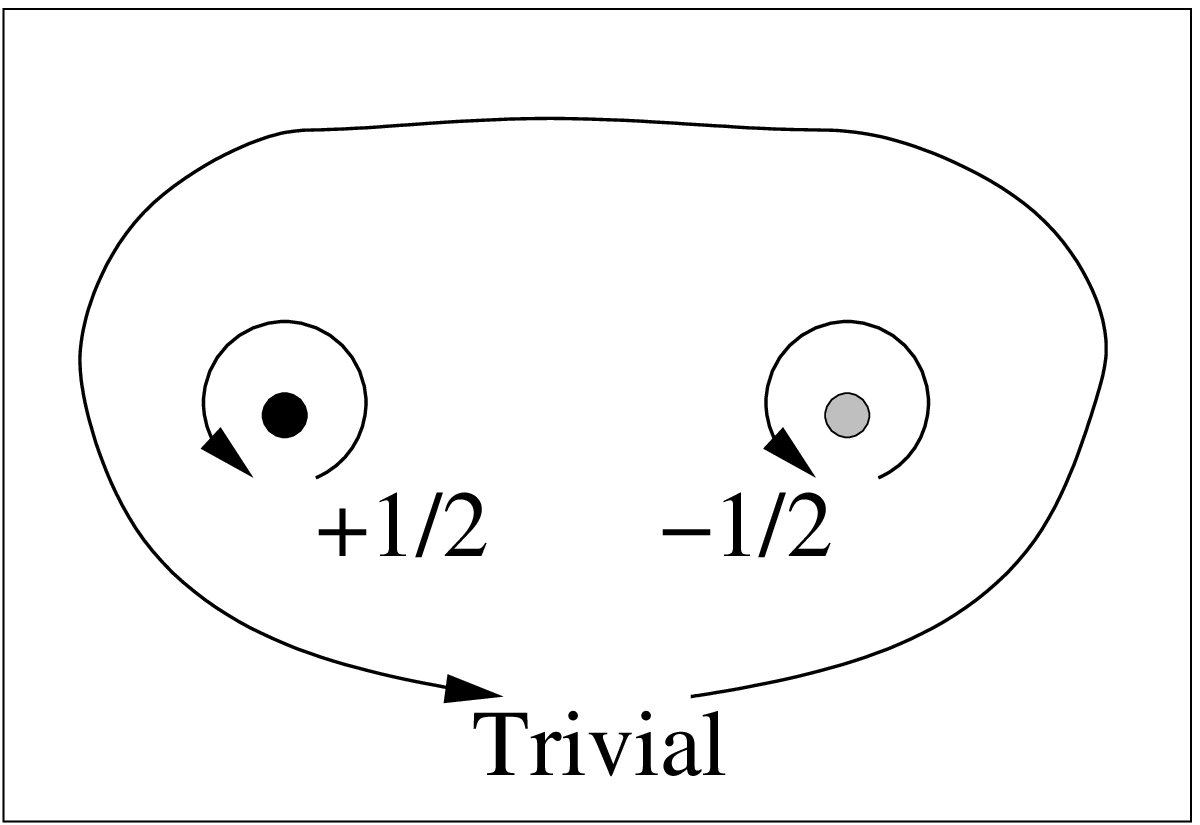}  
\caption{Lattice with two elementary monodromy defects of different sign. }
\label{F:pm11}      
\end{figure}

One can easily obtain trivial global monodromy for a close path around two
defects with different signs. Figure \ref{F:pm11} shows construction of
two elementary monodromy defects with different signs. Signs plus and minus
indicate respectively adding and  removing of the same solid angle.

%%%%%%%%%%%%%%%%%%%%%%%%%%%%%%%%%%%%%%%%%%%%%%%%%%%%%%%%%%%%%%
\subsection{Several rational line defects} \label{sS:MultRatL}
\begin{figure}
\centering
\includegraphics[height=3.cm]{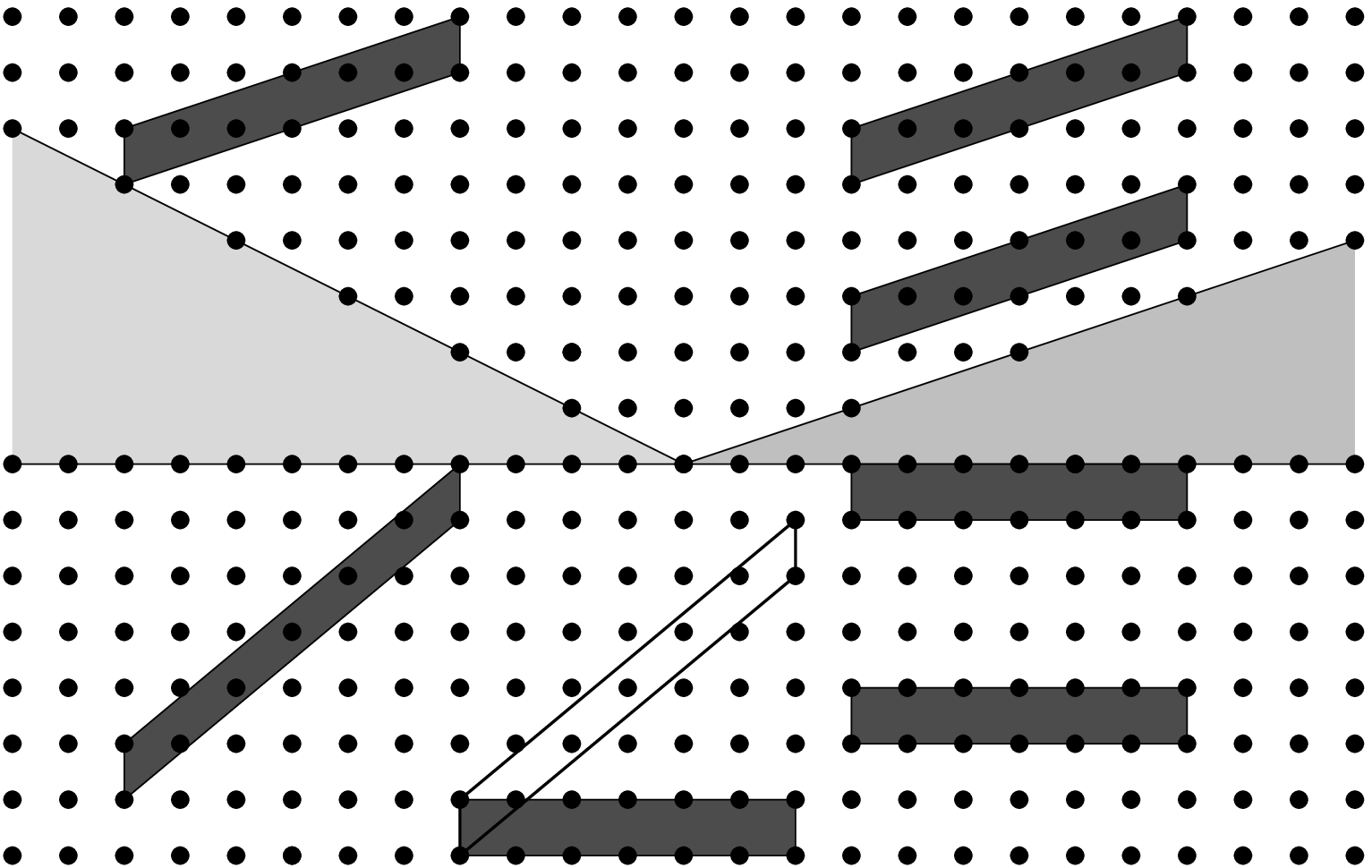} \hskip0.5cm
\includegraphics[height=3.cm]{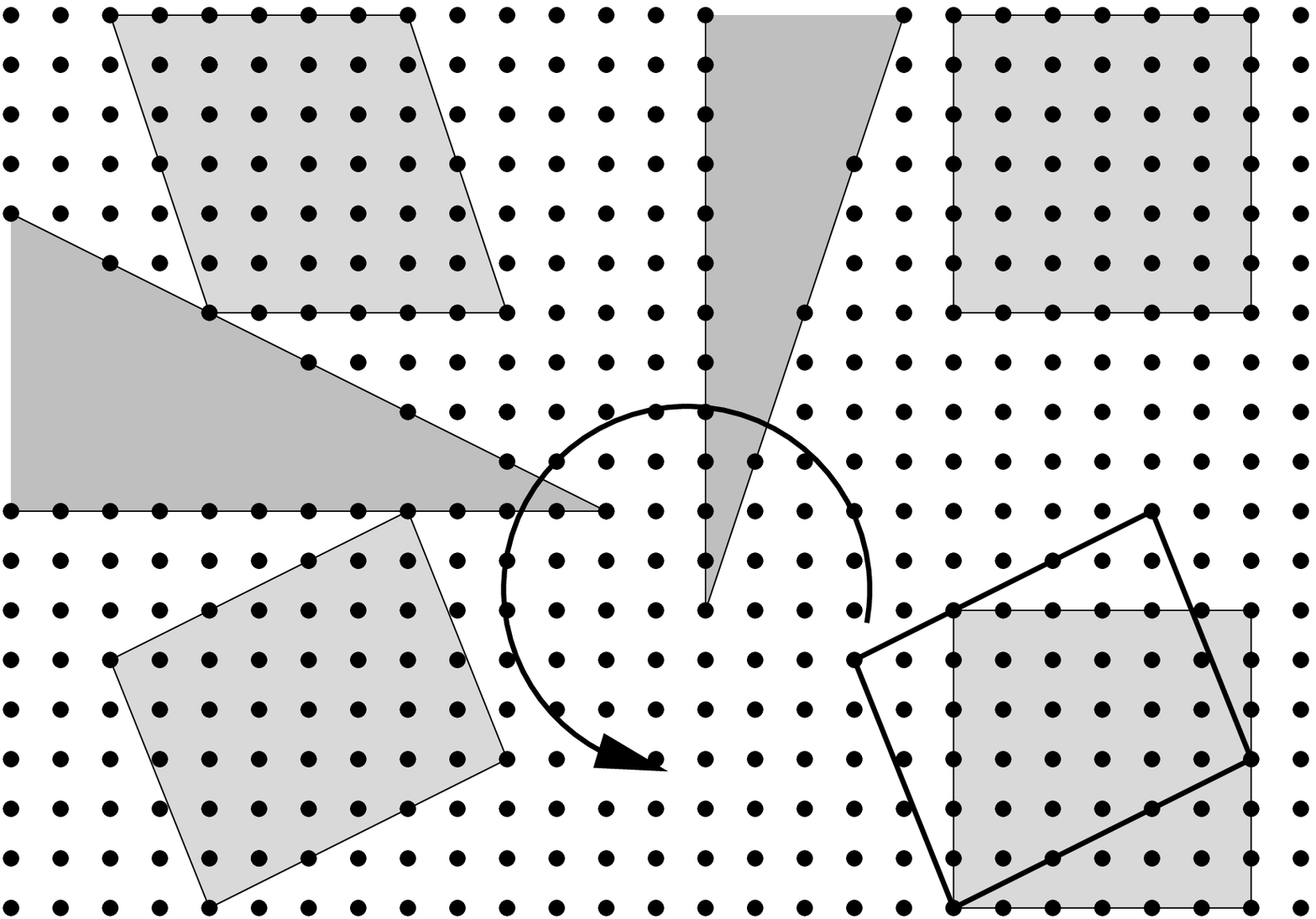}  
\caption{Construction of lattice with two rational defects, $1:2$ and $1:3$
(Left)- Parallel defects which correspond to 
singular one-dimensional strata for $2:(-3)$
resonance oscillator. (Right) - Two orthogonal defects.  }
\label{F:2_3cut}      
\end{figure}

Let us now discuss examples of lattices with multiple rational line
defects. We assume below that all defects are of the same sign, i.e.
obtained by removing solid angle.
We start with example of two defects $1:2$ and $1:3$ which have
similar orientation (see Figure \ref{F:2_3cut}, left). These two defects
model the singularities of integrable toric fibration for $2:-3$
resonant oscillator (\ref{I1-def}). Elementary $1\times 1$ cell can not
cross unambiguously both defect half-lines. The cell should be doubled in
horizontal direction in order to cross unambiguously the $1:2$ defect. In a
similar way the cell should be tripled in the same direction in order to
cross the $1:3$ defect. This means that only $1\times6$ 
cell which is six times larger in the horizontal direction can cross
both defects. Using such cell we can go along a close path surrounding
the singular vertex. After linear extension to elementary lattice vectors
we get the fractional monodromy matrix
$\left(\begin{array}{cc} 1& \ 0\\ 5/6 &\  1\end{array}\right)$. 

If  two rational defects, $1:2$ and $1:3$, have different orientations
the situation becomes quite different. Figure  \ref{F:2_3cut}, right shows
these two defects with orthogonal orientation. In order to pass through
horizontal $1:2$ defect the cell should be doubled in horizontal direction.
In order to pass through $1:3$ vertical defect the cell should be tripled in
vertical direction. Moreover, one should note that all vertices of the cell
should lie on even vertical lines and on horizontal lines having the same number
modulo 3. This means that we need to take at least $6\times6$ cell in order
to cross unambiguously both rational cuts. The resulting monodromy matrix
for a counterclockwise path around two singular points has the form
$\left(\begin{array}{cc} 5/6& \ -1/3\\ 1/2 &\  1\end{array}\right)$. 
This is an elliptic $SL(2,Q)$ matrix.

\subsubsection{Rational defect line with ends and singular points} 
\label{ssS:LineEnds}
\begin{figure}
\centering
\includegraphics[height=1.8cm]{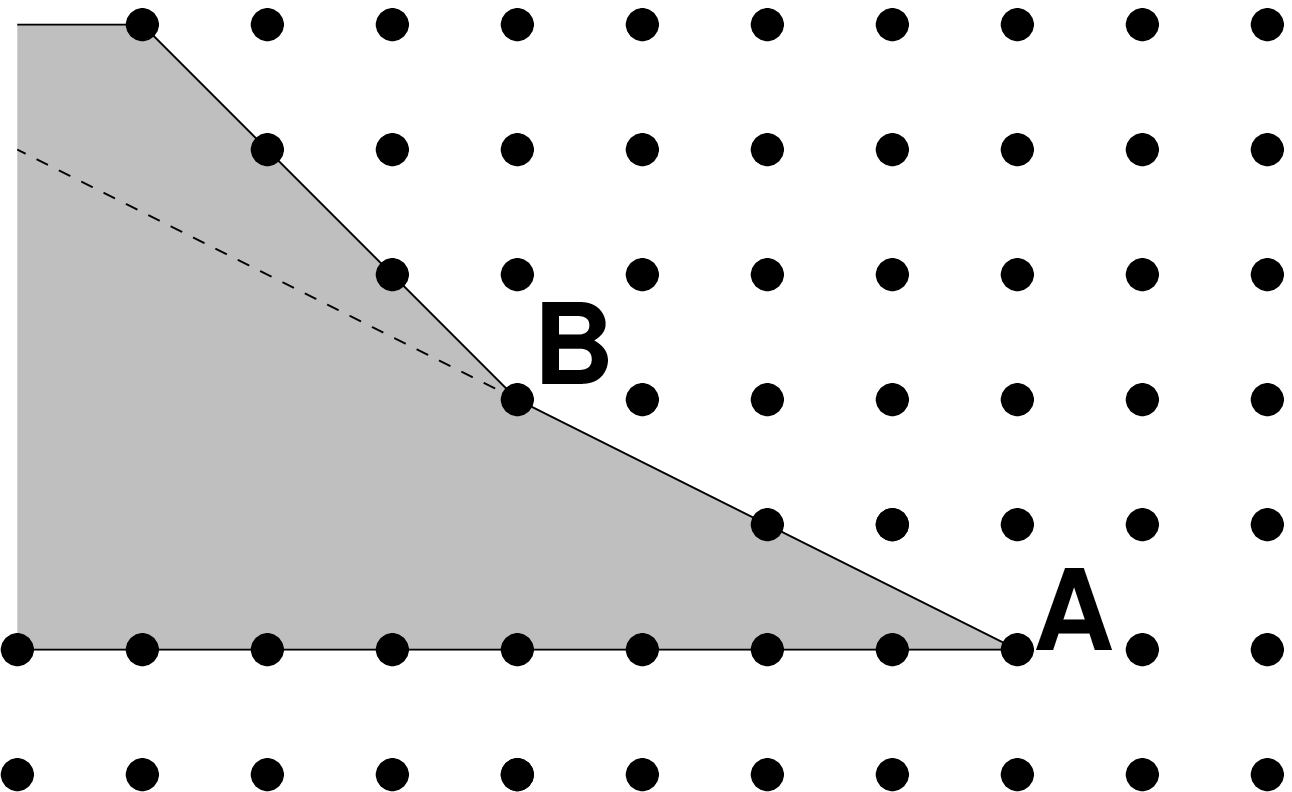} \hskip0.5cm
\includegraphics[height=1.8cm]{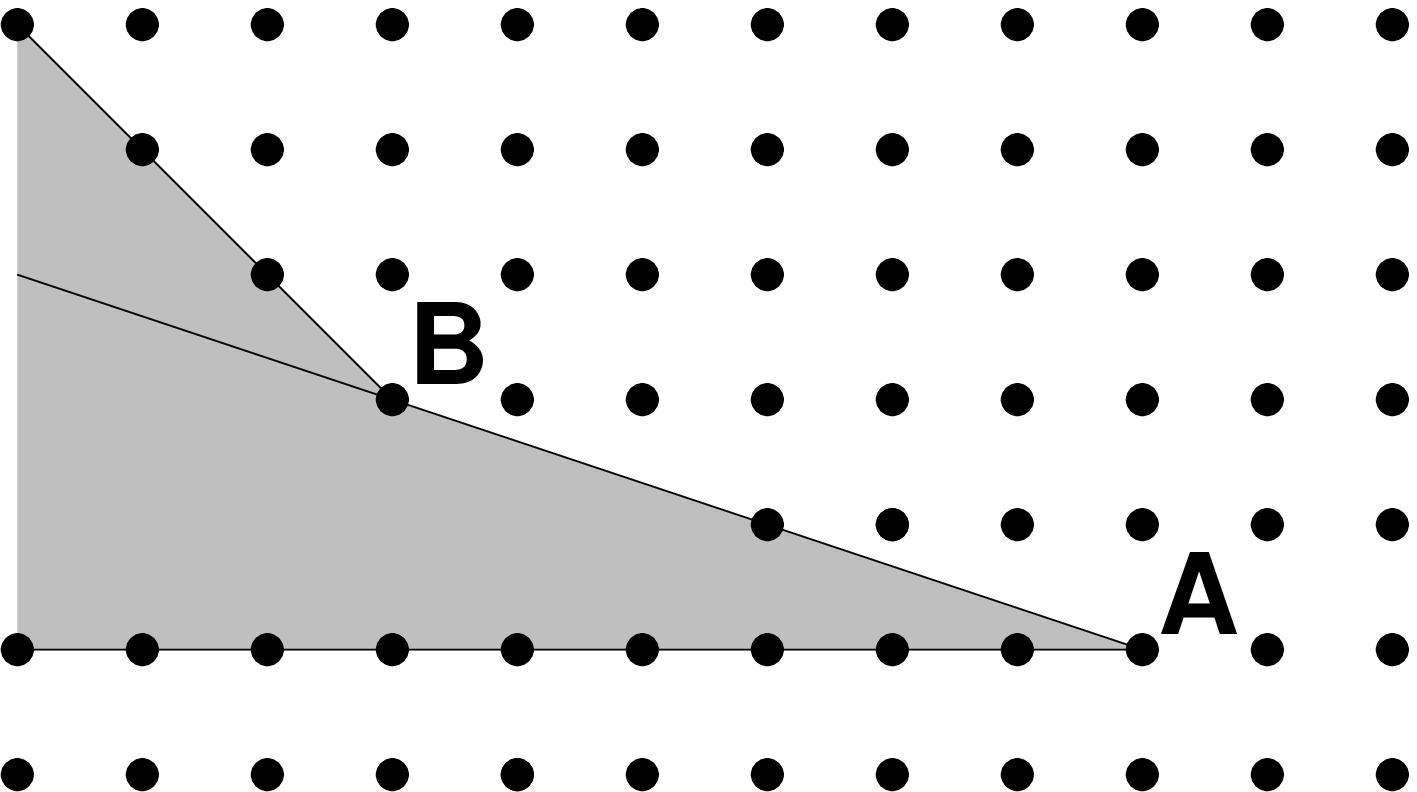} \hskip0.5cm
\includegraphics[height=1.8cm]{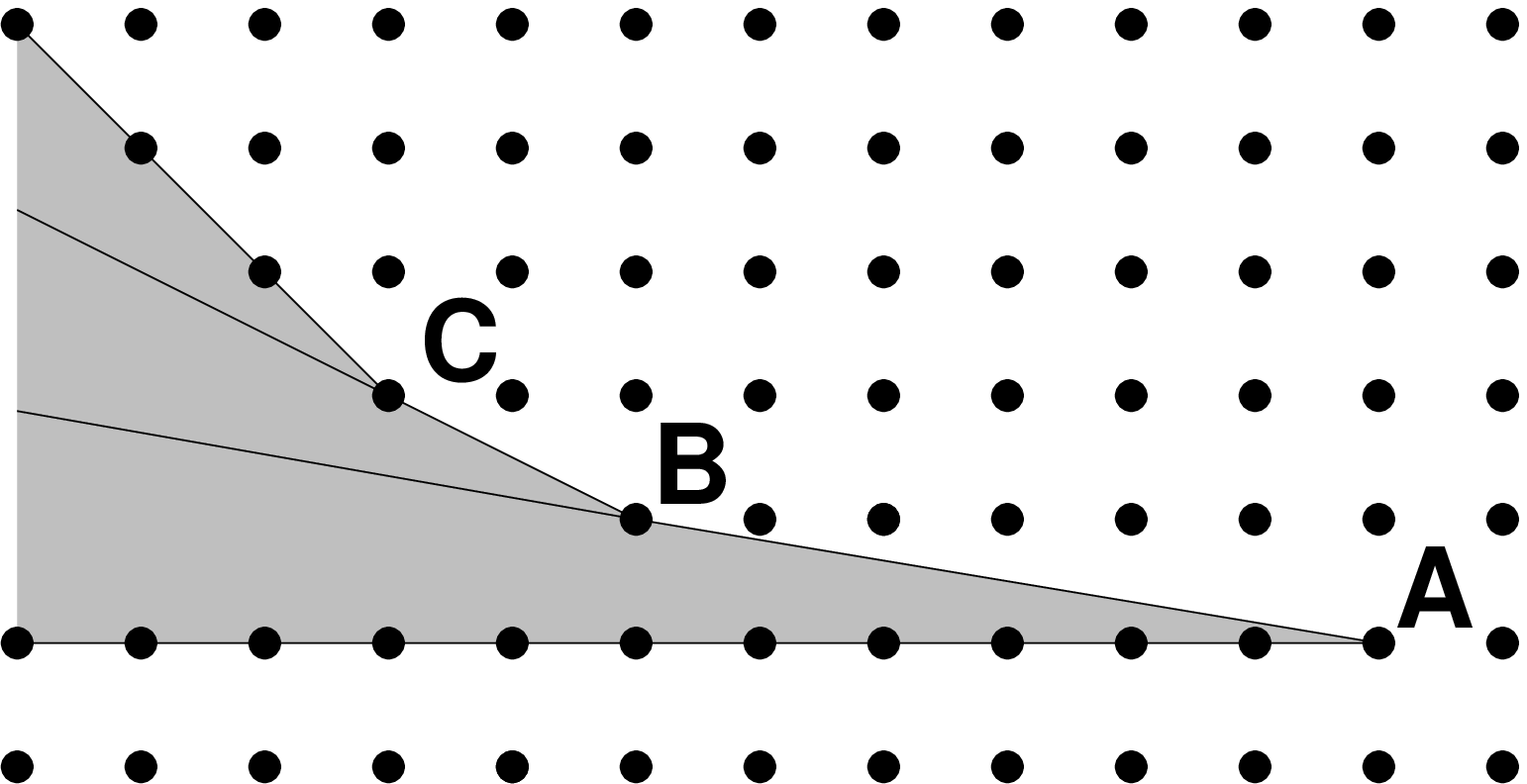} \hskip0.5cm  
\vskip10pt 
\includegraphics[height=1.6cm]{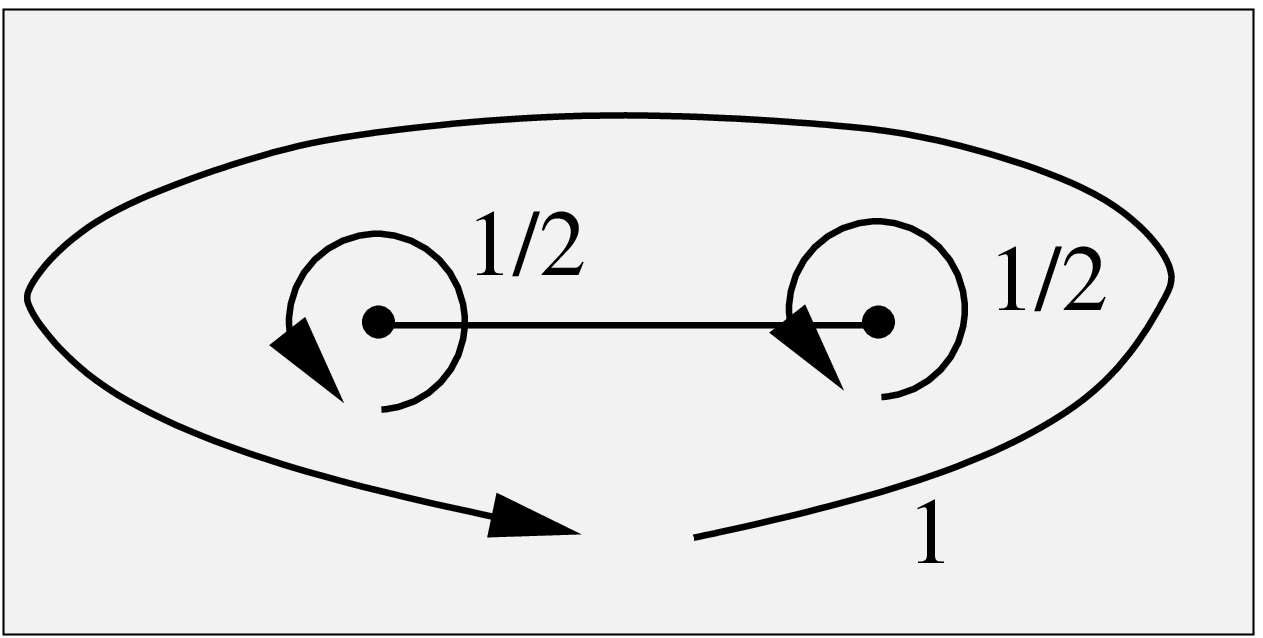} \hskip0.6cm
\includegraphics[height=1.6cm]{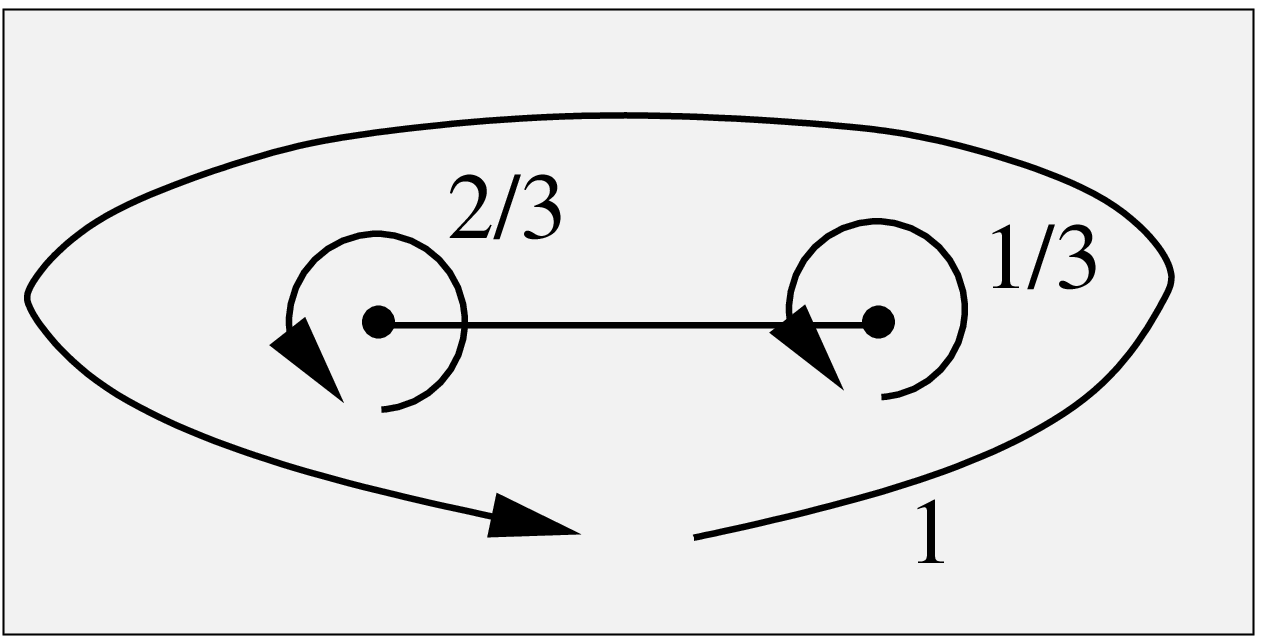} \hskip0.6cm
\includegraphics[height=1.6cm]{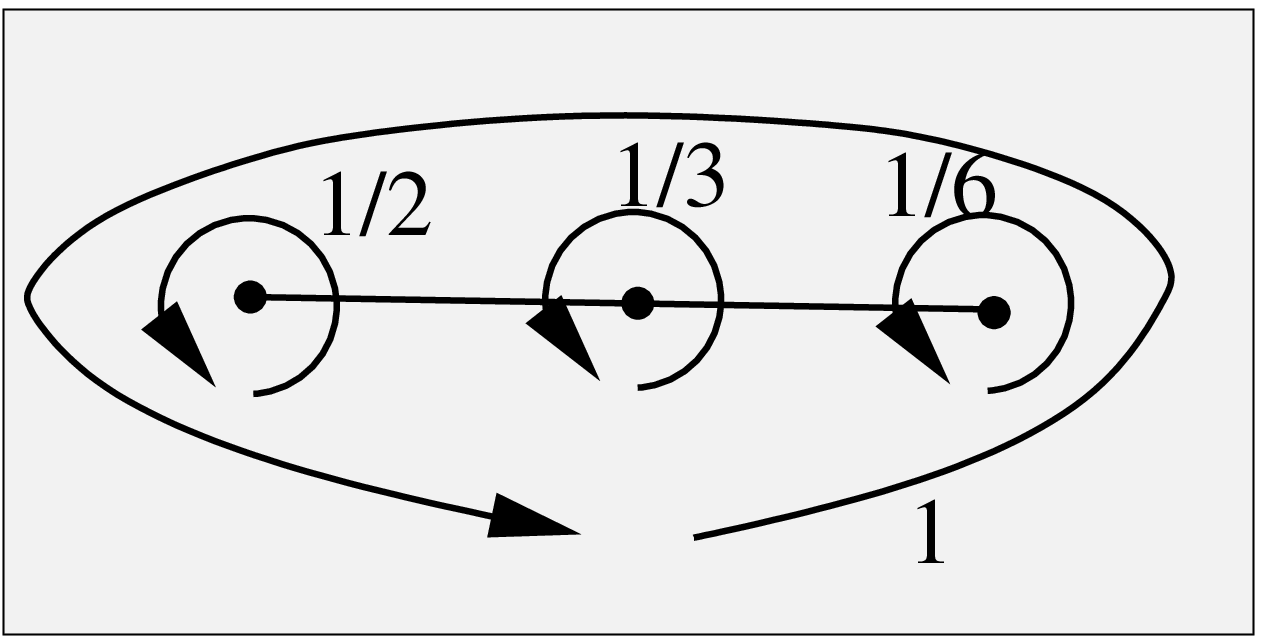}  \hskip0.3cm
 \caption{Construction of the line defects with ends and singular points. 
(Left) - Defect with equivalent ends. (Center) - Defect with inequivalent
ends. (Right) - Defect with inequivalent ends and additional singular point
splitting defect into two fragments.}
\label{F:multsing}      
\end{figure}

It is quite easy to construct rational defect with two ends. It is sufficient
to start to cut solid angle as it was done for rational defect but at
some another point to change the slope and to continue with another
slope related to integer monodromy defect. This situation is shown in 
Figure \ref{F:multsing} on three different examples. Fig. \ref{F:multsing},
left shows $1:2$ cut which starts at point $A$ but at point $B$ the
angle of the cut changes. It becomes equal angle characteristic to
elementary monodromy defect. This means that the line $1:2$ defect on
reconstructed lattice has two ends and the monodromy around each end 
is $\left(\begin{array}{cc} 1& \ 0\\ 1/2 &\  1\end{array}\right)$. At the same
time the global monodromy for close path surrounding $1:2$ defect 
is $\left(\begin{array}{cc} 1& \ 0\\ 1 &\  1\end{array}\right)$.
In a similar way (see  Fig. \ref{F:multsing}, center) we can start 
at point $A$ with $1:3$
cut and change at point $B$ the angle in order to get again on the
reconstructed lattice  elementary monodromy for global close path. 
This means that surrounding point $A$ we get the monodromy
 $\left(\begin{array}{cc} 1& \ 0\\ 1/3 &\  1\end{array}\right)$, while
surrounding point $B$ the monodromy becomes
$\left(\begin{array}{cc} 1& \ 0\\ 2/3 &\  1\end{array}\right)$. Two ends are
not equivalent. Naturally, we can change angle several times. This gives
the line defect with singular points on it. Each singular point corresponds
to modification of value of solid angle removed from the lattice. 
 Fig. \ref{F:multsing}, right shows example with three singular points.
Generalization to more complicated examples is straightforward.

\begin{figure}
\centering
\includegraphics[height=2.3cm]{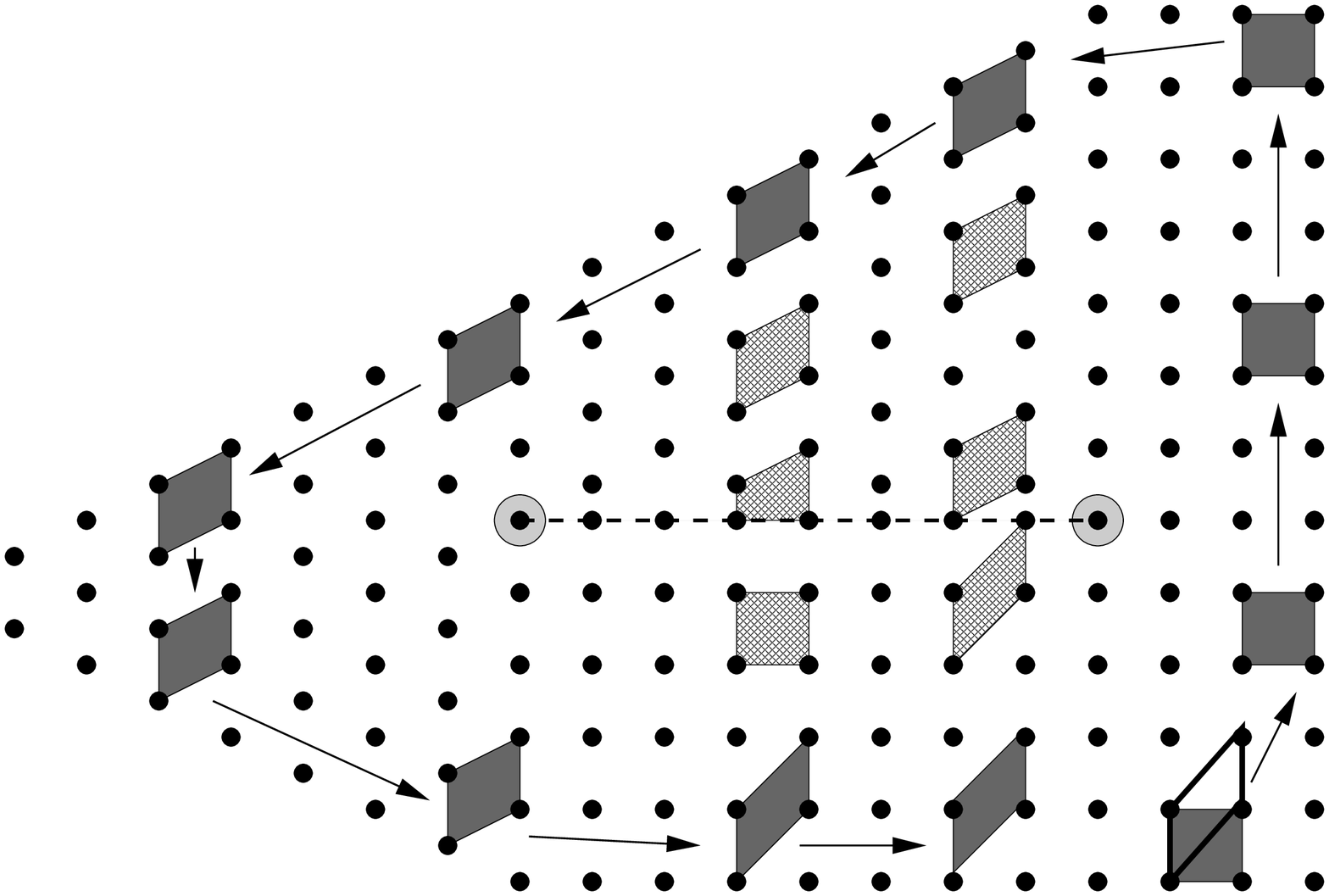} \hskip0.3cm
\includegraphics[height=2.3cm]{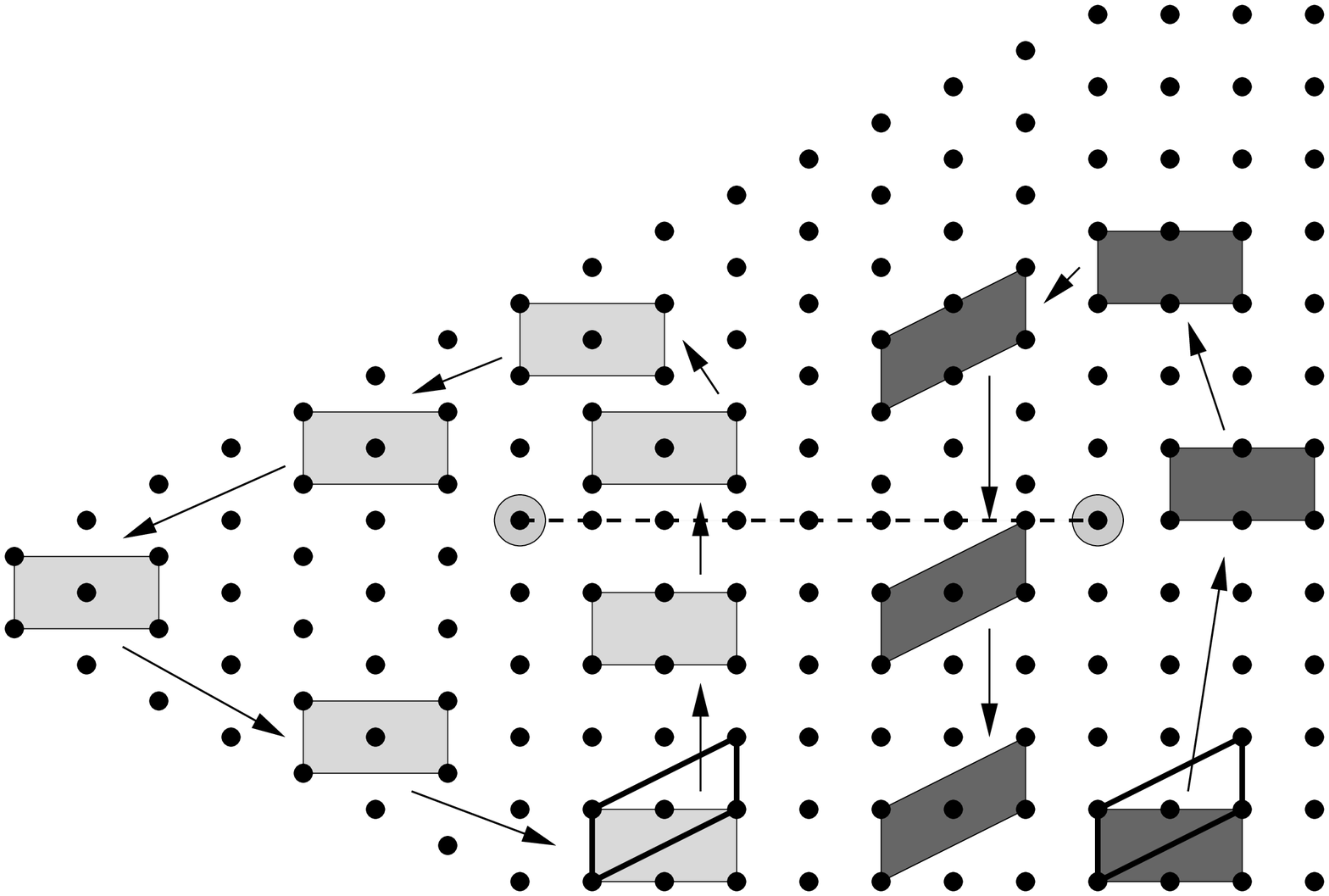}  \hskip0.3cm
\includegraphics[height=2.3cm]{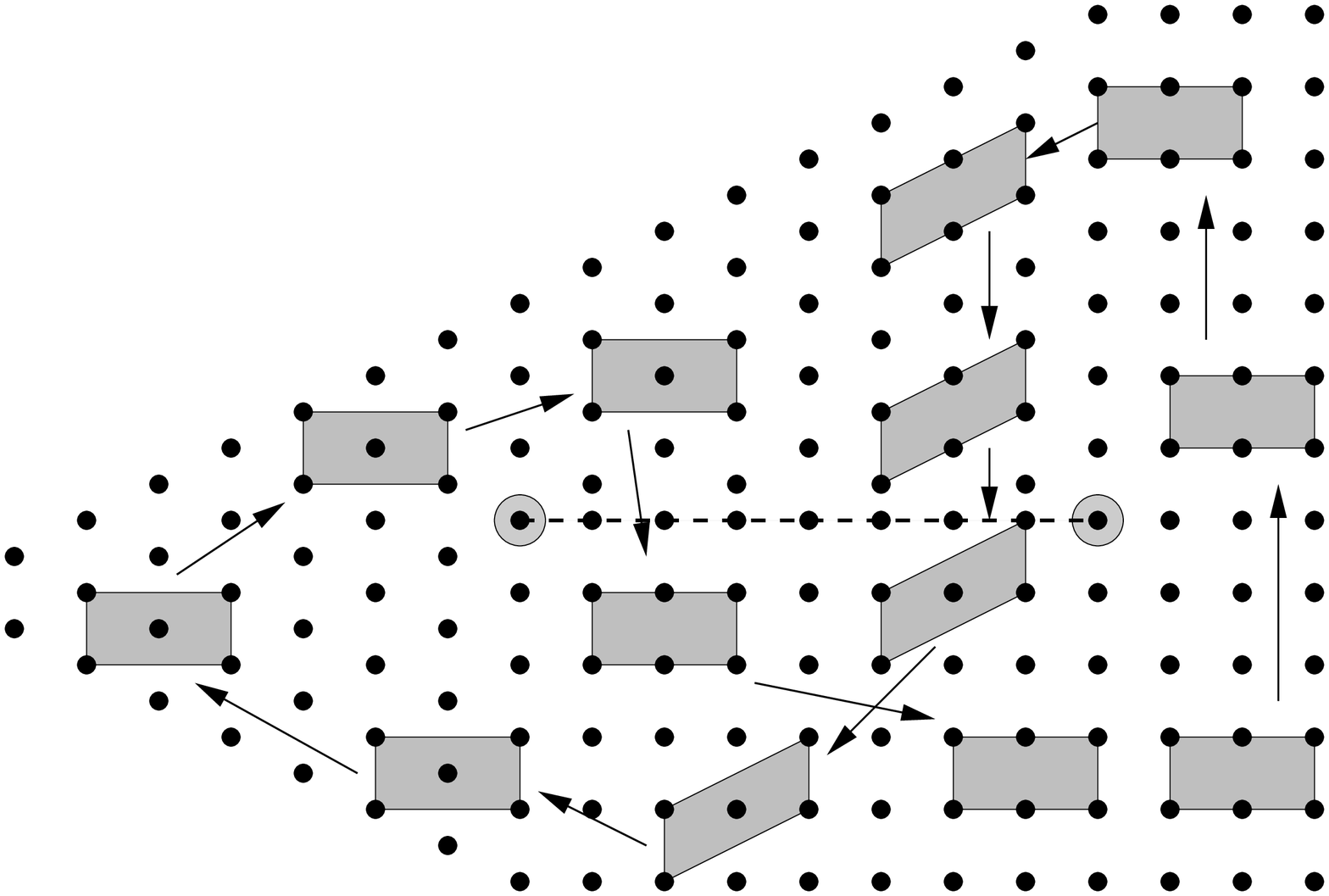}  
\caption{Lattice with two half-integer defects. Defect is an interval with
two end points.
(Left)-  Elementary cell cannot cross unambiguously the defect line.
(Center) - Both ends have the same $(1/2)$ monodromy. (Right) - Figure eight 
close path is not contractible but the monodromy is trivial. }
\label{F:HalfIntcut}      
\end{figure}

Figure \ref{F:HalfIntcut} demonstrates geometrically modifications
which occur with elementary cell after 
traversing different closed paths on the lattice with
$1:2$ defect with two ends. To find the global monodromy one can use
elementary $1\times1$ cell (Fig. \ref{F:HalfIntcut}, left), whereas it is
not possible to cross the line with such a cell because of ambiguity of
cell modifications. 
 
Taking $1\times2$ cell we can easily go around each singular point
and see  (Fig. \ref{F:HalfIntcut}, center)
that the result is exactly the same for both points, namely the half of the
global modifications. 

The last Figure \ref{F:HalfIntcut}, (right) shows the evolution of the
double cell along the figure eight close path which goes around two
centers but in opposite directions. This close path results in trivial
monodromy, the cell has no modifications after returning to the
initial point.

%%%%%%%%%%%%%%%%%%%%%%%%%%%%%%%%%%%%%%%%%%%%%%%%%%%%%%%%%%%%%%%%%%%%
\section{Is there mutual interest in defect - monodromy correspondence?}
Stimulated by analogy between classical and quantum monodromy for
Hamiltonian integrable systems from one side and defects of regular
periodic lattices from another side we have suggested 
construction of ``elementary integer and fractional lattice defects'' 
associated with 
elementary integer and fractional monodromy.  Then we have proposed
how to generate more complicated defects of lattices
by combination of elementary ones. 
Among these more complicated defects there are defects like disclinations
which are well known in solid state physics. At the same time the author
does not know simple examples of dynamical Hamiltonian system with 
similar defects. Reciprocally, many examples of dynamical Hamiltonian systems
with elementary monodromy are known but the ``elementary monodromy defect''
seems not to be individually detected experimentally in periodic
solids.  These observations enables us to formulate below a number of 
problems concerning Hamiltonian dynamical systems with singularities
and periodic lattices with defects. 
Some of these problems have more or less intuitively
evident answers but the strict mathematical proofs are still absent. 
In other cases even the formulation of the problem
is not precise and should be critically
analyzed and corrected before looking for the answer.

\begin{itemize}
\item{}  {\sl About the sign of elementary monodromy defect}. How to
characterize the class of dynamical systems (classical and quantum) 
possessing only elementary monodromy defects of one sign? For Hamiltonian
systems focus-focus singularities correspond to elementary $(-)$ defects. 
Tentative answer is to say that elementary $(+)$ monodromy defects
are generic for $PT$-invariant dynamical systems with non-hermitian
Hamiltonians and real spectra \cite{Bender,G24Zhil}.

\item{} {\sl Correspondence between topology of singular fibers of integrable
toric fibrations and integer and fractional defects of lattices.}
Some simple examples of such correspondence were given. Is it possible
to establish more general correspondence?
In particular it seems natural that elementary $(+)$ and $(-)$ monodromy defects
correspond to pinched tori with different index of transversal self-crossings
\cite{matsumoto}.
 
\item{} {\sl Constructive methods to design Hamiltonian classical and quantum
systems with prescribed
type of monodromy}. Less ambitious task is to propose a list of concrete
examples of classical and quantum systems which show the manifestation of
different elementary and non-elementary defects.

\item{}  {\sl Existence of a topological invariant separating different
singularities (defects) with the same monodromy but with different
numbers of $2\pi$-rotations of elementary cell.}  The analogy
between this problem and the Riemann surfaces description \cite{Cartan}
was pointed out to author on several ocassions.

\item{} {\sl Extension of the correspondence between singularities and defects
from 2D-systems to higher dimensional systems.} This is surely a very wide 
subject and author believes that first steps in 
mathematical generalization should be guided by natural physical examples.

\item{} {\sl Global restrictions on the system of defects and on the
system of
singularities of toric fibrations in the case of compact base space (lattices on
compact spaces).} For example, singular toric fibration over $S^2$ base space
should have 24 elementary focus-focus singularities \cite{zungII} or equivalently
24 elementary $(-)$ monodromy defects or 12 $(\pi/3$)-rotational disclinations
known as pentagonal defects \cite{Nelson83}. This problem has obvious relation
with fullerene-like materials.

\item{} {\sl The relation between number of removed vertices for a defect
and the Duistermaat-Heckman measure for the reduced  Hamiltonian system.}
Hint:   The slope of the function giving the number of removed vertices from
vertical line as a
function of the number of a vertical line coincides with the Chern class of
the integrable fibration used in the Duistermaat-Heckman theorem
\cite{DuistHeck,Guill}.

\item{} {\sl To find macroscopic/mesoscopic physical systems which manifest
presence of elementary $(+/-)$ monodromy defects.} Possible candidates
besides periodic solids or liquid crystals may be membranes 
\cite{Bowick01}, fullerenes and curved carbon surfaces \cite{ConeNature},
viruses \cite{SPID}, colloidal structures \cite{Macromol}, %optical lattices, 
etc.

\item{} {\sl Relation between internal structure of elementary cells
and possible existence of isolated elementary integer and fractional
monodromy defects in real physical systems.} In what kind of systems
(materials) the topological properties are more important than geometric
and steric effects and enable one to see the manifestation of 
elementary monodromy defects?

\item{} {\sl Physical consequences of sign conjecture.} If one accept the
formulated above sign conjecture, i.e. presence of only $(-)$ defects in
generic families of Hamiltonian systems depending on a small number of 
parameters, there is fundamental difference between $(+)$ and $(-)$ 
(or in other terms between ``right'' and ``left'')
in both classical and quantum mechanics.  How to formulate this conjecture in more
precise terms and what kind of physical consequences can be rigorously
deduced? 

\end{itemize}

The author hopes that  Hamiltonian dynamics and  periodic
solids gives  complementary points of view which are useful for both 
fields of scientific interest.
The present article is supposed to stimulate mutual
interest, better understanding and further  cooperation between specialists
working in these fields.

{\bf Acknowledgements.}
{\small The author thanks Drs. R. Cushman, N. Nekhoroshev, D. Sadovskii
 for many stimulating discussions.
This work was essentially done during authors stay in  
IHES, Bures-sur-Yvette, France,
Mathematical Institute, Universitty of Warwick, UK, and 
Max-Planck Institut f\"ur Physik komplexer Systeme, Dresden, Germany.
The support of these instotutions is acknowledged.
This work is a part of the European project MASIE,
contract HPRN-CT-2000-00113.}

\input{MLDref} %%%% references
%%%%%%%%%%%%%%%%%%%%%%%%%%%%%%%%%%%%%%%%%%%%%%%%%%%%%%%%%%%%%%%%%%%%%%  }

\printindex
\end{document}

%% file: MLDref.tex
%
%\title{Hamiltonian monodromy as lattice defect \\ 
%{\rm B. I. Zhilinskii } \\
% 

\bibliographystyle{apsrev}